%% file: main.tex
\pdfoutput=1
\documentclass[twocolumn,twoside]{IEEEtran}

\usepackage{amsmath,amssymb,amsfonts,epsfig,color,rotating,epstopdf,bigstrut,bm,verbatim,epsf,latexsym,graphicx,xspace}
\usepackage{multicol}
\usepackage{amsthm,epsf,graphicx,subcaption}
\usepackage{authblk}
\usepackage{soul}
\usepackage{algorithm}
\usepackage{algorithmic}
\usepackage[dvipsnames]{xcolor}

\input{mysymbol}

\newtheorem{remark}{Remark}

\def \vectorize {{\rm vec}}
\def \unvec {{\rm unvec}}
\def \transpose {\top}
\newcommand{\MATLAB}{\textsc{Matlab}\xspace}

\def \xgen  {\bbx} 
\def \xgrd  {\tilde {\bbx}} 
\def \uGain {g_0}  
\def \shd   {s}    
\def \nshd  {\check s} 
\def \Vnshd {\check {\bbs}_t} 
\def \VnshdIni {\check {\bbs}^{(0)}} 
\def \pathlossexp {\gamma}
\def \interestRegion {\ccalA}

\newcommand \Region[1]{\ccalA_{#1}}
\def \noise {\nu}

\def \mWeight {\bbW} 
\def \vWeight {\bbw} 
\newcommand{\mDiagWeight}[1]{\bbDelta_{w_{#1}}} 
\def \Regularizer {\ccalR} 
\def \RegWeight {\rho_{\field}} 
\def \zSet {\mathcal{Z}}
\def \NumStep {N_{\textrm{Iter}}}

\def \mSet {\mathcal{M}}

\def \numBatch {N_{\textrm{Batch}}}
\def \normEllipMargin {\lambda}

\newcommand \IdenMat[1]{\bbI_{#1}}
\newcommand{\NormDist}[2]{\ccalN({#1},{#2})}

\newcommand{\G}[2]{\mathcal{G}({#1},{#2})}

\newcommand {\Vnshdat}[1]{\check {\bbs}_{#1}}
\newcommand{\EXPwo}[2]{\mathbb{E}_{#2}\left[{#1}\right]}
\newcommand{\diGamma}[1]{\psi\left({#1}\right)}

\def \field {f}
\newcommand {\Meanfield}[1] {\bar{f}_{#1}}
\def \vfield{\bm f}
\def \mfield{\mathbf{F}}
\def \mShadowMap{\mathbf{S}}
\def \mCGmap{\mathbf{G}}

\newcommand {\mCondCovFwith}[1]{\bm{\Sigma}_{f|{#1}}}
\newcommand {\mCondCovFVIwith}[1]{\breve{\bm{\Sigma}}_{f|{#1}}}
\def \Zfield{z}
\def \vZfield{\bm z}
\def \mZfield{\mathbf{Z}}

\def \mDiagFieldPrec {\bbPhi_{{f}|{z}}}
\def \hParameter{\bm \theta}

\def \nVar {\sigma_{\noise}^2} 
\def \granPara {\beta} 

\newcommand {\fMean}[1]{\mu_{{f}_{#1}}}
\def \vfMean {\bm{\mu}_{{f}}}
\newcommand {\fVar}[1]{\sigma_{{f}_{#1}}^2}
\newcommand {\fMeanMean}[1]{m_{#1}}
\newcommand {\fVarVar}[1]{\sigma_{#1}^2}

\newcommand{\neighbor}[1]{\ccalN({#1})}

\def \precNVar {\varphi_{\noise}}
\def \MeanprecNVar {\widetilde{\varphi}_{\noise}}
\def \vfPrec {\bm{\varphi}_{f}}
\newcommand{\fprec}[1]{\varphi_{f_{#1}}}
\newcommand{\Meanfprec}[1]{\widetilde{\varphi}_{f_{#1}}}
\newcommand{\MeanfprecItr}[2]{\widetilde{\varphi}_{f_{#1}}^{(#2)}}
\newcommand{\Meanshd}[1]{\overline{s}_{#1}}

\newcommand {\VIfmean}[2]{\breve{\mu}_{f_{#1}}(\xgrd_{#2})}
\newcommand {\VIfvar}[2]{\breve{\sigma}_{f_{#1}}^2(\xgrd_{#2})}
\newcommand {\VIfmeanItr}[3]{\breve{\mu}_{f_{#1}}^{(#3)}(\xgrd_{#2})}
\newcommand {\VIfvarItr}[3]{\breve{\sigma}_{f_{#1}}^{2(#3)}(\xgrd_{#2})}

\newcommand {\VIfpreca}[1]{\breve{a}_{#1}}
\newcommand {\VIfprecb}[1]{\breve{b}_{#1}}
\newcommand {\VIfprecaItr}[2]{\breve{a}_{#1}^{(#2)}}
\newcommand {\VIfprecbItr}[2]{\breve{b}_{#1}^{(#2)}}
\newcommand {\VIzfieldProb}[2]{\breve{\zeta}_{#1}({#2})}
\newcommand {\VIzfieldProbItr}[3]{\breve{\zeta}_{#1}^{(#3)}({#2})}
\newcommand {\VIzfieldUnnormProbItr}[3]{\breve{\breve{\zeta}}_{#1}^{(#3)}({#2})}
\def \VIaNVar {\breve{a}_{\nu}}

\def \VIbNVar {\breve{b}_{\nu}}
\newcommand {\VIbNVarItr}[1] {\breve{b}_{\nu}^{(#1)}}
\newcommand {\VIfMeanMean}[1] {\breve{m}_{#1}}
\newcommand {\VIfMeanMeanItr}[2] {\breve{m}_{#1}^{(#2)}}
\newcommand {\VIfMeanVar}[1] {\breve{\sigma}_{#1}^2}
\newcommand {\VIfMeanVarItr}[2] {\breve{\sigma}_{#1}^{2(#2)}}

\begin{document}
\title{A Variational Bayes Approach to \\ Adaptive Radio Tomography}

\author{Donghoon Lee,~\IEEEmembership{Student Member,~IEEE,} 
and Georgios B. Giannakis,~\IEEEmembership{Fellow,~IEEE}%
\thanks{Parts of this work were presented at the IEEE International Conference on Acoustics, 
	Speech and Signal Processing, held in Brighton, UK, during May 12-17, 2019~\cite{dg19adaptiveCGcartographyVB}.\\
\indent D.~Lee and G.~B.~Giannakis are with the Department of Electrical
and Computer Engineering and the Digital Technology Center, University
of Minnesota, Minneapolis, MN 55455, USA. Emails: \{leex6962, georgios\}@umn.edu.\\
\indent The work in this paper was supported in part by NSF grants 1508993, 1711471, and 1901134.}
}

\markboth{IEEE TRANSACTIONS ON SIGNAL PROCESSING (submitted, \today)}{Lee and Giannakis: A Variational Bayes Approach to Adaptive Radio Tomography}

\maketitle

\begin{abstract} 
Radio tomographic imaging (RTI) is an emerging technology for localization of physical objects in a geographical area covered by wireless networks. With attenuation measurements collected at spatially distributed sensors, RTI capitalizes on spatial loss fields (SLFs) measuring the absorption of radio frequency waves at spatial locations along the propagation path. These SLFs can be utilized for interference management in wireless communication networks, environmental monitoring, and survivor
localization after natural disasters such as earthquakes. Key to the success of RTI is to accurately model shadowing as the weighted line integral of the SLF. To learn the SLF exhibiting statistical heterogeneity induced by spatially diverse environments, the present work develops a Bayesian framework
entailing a piecewise homogeneous SLF with an underlying hidden Markov random field model. Utilizing variational Bayes techniques, the novel approach yields efficient field estimators at affordable complexity. A data-adaptive sensor selection strategy is also introduced to collect informative measurements for effective reconstruction of the SLF. Numerical tests using synthetic and real datasets demonstrate the capabilities of the proposed approach to radio tomography and channel-gain estimation.
\end{abstract}


\begin{IEEEkeywords}
Radio tomography, channel-gain estimation, variational Bayes, active learning, Bayesian inference
\end{IEEEkeywords}

\section{Introduction}
\label{sec:intro}

Tomography is imaging by sectioning through the use of a penetrating wave, and has been widely appreciated by natural sciences, notably in medical imaging~\cite{smith2010}. The principles underpinning radio tomographic methods have been carried over to construct what are termed~\emph{spatial loss fields} (SLFs), which are maps quantifying the attenuation experienced by electromagnetic waves in radio frequency (RF) bands at every
spatial position~\cite{patwari2008correlated}. To this end, pairs of collaborating sensors are deployed over the area of interest to estimate the attenuation introduced by the channel between those pairs of sensors. 
Different from conventional methods, radio tomography relies on~\emph{incoherent} measurements containing no phase information, e.g., the received signal strength (RSS). Such simplification saves costs for
synchronization needed to calibrate phase differences among waveforms received at different sensors.

SLFs are instrumental in several tasks including radio tomography~\cite{wilson2009regularization} and channel-gain cartography~\cite{kim2011kalmansensing}. Absorption captured by the SLF allows one to discern objects located in the area of interest, thus enabling radio tomographic imaging (RTI). Benefiting from the ability of RF waves to penetrate physical structures such as trees and buildings, RTI provides a means of device-free passive localization~\cite{woyach2005sensorless,youssef07devicefree}, and has found diverse applications in disaster response for e.g., detecting individuals trapped in buildings or smoke~\cite{wilson2011throughwalls}. SLFs are also useful in channel-gain cartography to provide channel-state information (CSI) for a link between any two locations even where no sensors are present~\cite{kim2011kalmansensing}. Such maps can be employed by cognitive radios to control the interference that a secondary network inflicts to primary users that do not transmit--a setup encountered with television broadcast systems~\cite{zhao2007survey,fcc2011order,KDBRG13}. The non-collaborative nature of primary users precludes training-based channel estimation between a secondary transmitter and a primary receiver, and vice versa. Note that channel-gain cartography is also instrumental for interference
management in the Internet-of-things (IoT)~\cite{krr17CRIoT}.

The key premise behind RTI is that spatially close radio links exhibit similar shadowing due to the presence of common obstructions. This shadowing correlation is related to the geometry of objects present in the area that waves propagate through~\cite{patwari2008correlated,AgP09}. As a result, shadowing is modeled as the weighted line integral of the underlying two-dimensional SLF. The weights in the integral are determined by a function depending on the transmitter-receiver locations~\cite{patwari2008correlated,hamilton2014modeling,Daniel18BlindTomography}, which models the SLF effect on shadowing over a link. Inspired by this SLF model, various tomographic imaging methods were  proposed~\cite{wilson2009regularization,wilson2011throughwalls,wilson2010tomography,Kaltio12ChannelDiversity}.
To detect locations of changes in the propagation environment, one can use the difference between the SLF across consecutive time slots~\cite{wilson2009regularization,wilson2010tomography}. To cope with multipath in a cluttered environment, multi-channel measurements can be utilized to enhance localization accuracy~\cite{Kaltio12ChannelDiversity}. Although these are calibration-free approaches, they cannot reveal
static objects in the area of interest. It is also possible to replace the SLF with a label field
indicating presence (or absence) of objects in motion on each voxel~\cite{wilson2011throughwalls}, and leverage the influence that moving objects on the propagation path have, on the variance of a RSS measurement. On the other hand, the SLF itself was reconstructed in~\cite{mostofi13CalibrationWithLineFitting,hamilton2014modeling,dbg18adaptiveRT,lee2017lowrank} to depict static objects in the area of interest, but calibration was necessary by using extra measurements (e.g., collected in free space). One can avoid extra data for calibration by estimating the SLF together with pathloss components~\cite{brian16jointCalibration,Daniel18BlindTomography}.

Another body of work leveraging the SLF model is that of channel-gain cartography when employing tomogaphy based approaches~\cite{kim2011kalmansensing,Daniel18BlindTomography,dallanese2011kriging,lee2017lowrank}.
Linear interpolation techniques such as kriging were further employed to estimate shadowing based on spatially correlated measurements~\cite{dallanese2011kriging}, while spatio-temporal dynamics were tracked via Kalman filtering~\cite{kim2011kalmansensing}. SLFs with regular patterns of objects have also been modeled as a superposition of a low-rank matrix plus a sparse matrix capturing structure irregularities~\cite{lee2017lowrank}. While the aforementioned methods rely on heuristic criteria to choose the weight function, \cite{Daniel18BlindTomography} provides blind algorithms to learn the weight function using a non-parametric kernel regression. 

Conventionally, the SLF is learned via least-squares (LS) estimation regularized by the propagation environment~\cite{hamilton2014modeling,lee2017lowrank,wilson2010tomography}. The resultant ridge-regularized LS solution can be interpreted as a maximum a posteriori (MAP) estimator when  
the SLF is statistically homogeneous and modeled as a zero-mean Gaussian random field. However, these estimators are less effective when the propagation environment is spatially diverse due to a combination of free space and objects in different sizes and materials (e.g., as in urban areas), which subsequently induces statistical heterogeneity in the SLF. To account for environmental heterogeneity, we proposed in~\cite{dbg18adaptiveRT} a Bayesian approach to learn a piecewise homogeneous SLF through a binary hidden Markov random field (MRF) model~\cite{higdon94Potts} via Markov chain Monte Carlo
(MCMC)~\cite{gilks96MCMC}. But this approach does not scale because MCMC is computationally demanding. 

Aiming at efficient field estimators at affordable complexity, we propose a variational Bayes (VB) framework for radio tomography to approximate the analytically intractable minimum mean-square error (MMSE) or MAP estimators. Instead of considering the binary hidden MRF to model statistical heterogeneity in the SLF~\cite{dbg18adaptiveRT}, we further generalize the SLF model by considering~\emph{$K$-ary} piecewise 
homogeneous regions for $K \geq 2$, to address a richer class of environmental heterogeneity. 
Besides developing efficient and affordable solutions for RTI, another contribution here is a data-adaptive sensor selection technique, with the goal of reducing uncertainty in the SLF, by cross-fertilizing ideas from the fields of experimental design~\cite{fedorov72experimentalDesign} and active learning~\cite{Mackay92AL}. 
The conditional entropy of the SLF is considered as an uncertainty measure, giving rise to a novel sensor selection criterion. Although this criterion is intractable especially when the size of the SLF is large, its efficient proxy can be obtained thanks to the availability of an approximate posterior model from the proposed VB algorithm.

The rest of the paper is organized as follows. Sec.~\ref{sec:model}
reviews the radio tomography model and states the problem. The Bayesian model 
and the resultant field reconstruction are the subjects of Sec. III, together with the proposed
sensor selection method. Numerical tests with synthetic as well as real measurements
are provided in Sec.~\ref{sec:numerical}. Finally, Sec.~\ref{sec:conc} summarizes the main
conclusions.

\textit{Notation}. Bold uppercase (lowercase) letters denote matrices (column vectors). 
Calligraphic fonts are used for sets; $\IdenMat{n}$ is the $n \times n$ identity matrix.
Operator $(\cdot)^{\transpose}$ represents the transpose a matrix $\mathbf{X} \in 
\mathbb{R}^{N_x \times N_y}$; $|\cdot|$ is used for the cardinality of a set, the magnitude 
of a scalar, or the determinant of a matrix; and $\vectorize(\mathbf{X})$ produces a column vector 
$\bbx \in\mathbb{R}^{N_xN_y}$ by stacking the columns of a matrix one below the other 
($\unvec(\bbx)$ denotes the reverse process).

\section{Background and Problem Statement}
\label{sec:model}

Consider a set of sensors deployed over a two-dimensional geographical area $\interestRegion \subset \mathbb{R}^2$. After averaging out small-scale fading effects, the channel-gain measurement over a link between a transmitter located at $\xgen \in \mathcal{A}$ and a receiver located at $\xgen' \in \mathcal{A}$ can be represented (in dB) as
\begin{equation}
g(\xgen,\xgen') = \uGain - \pathlossexp 10\log_{10} d(\xgen,\xgen') -  \shd(\xgen, \xgen') \label{eq:cg}
\end{equation}
where $\uGain$ is the path gain at unit distance; $d(\xgen,\xgen') := \|\xgen - \xgen'\|$ is the Euclidean distance between the transceivers at $\xgen$ and $\xgen'$; $\pathlossexp$ is the pathloss exponent; 
and $\shd(\xgen, \xgen')$ is the attenuation due to shadow fading. 

A tomographic shadow fading model is \cite{patwari2008correlated,hamilton2014modeling,lee2017lowrank}
\begin{align}
s(\bbx,\bbx') = \int_{\interestRegion} w(\bbx,\bbx',\xgrd) f(\xgrd) d\xgrd \label{eq:SLFmodel} 
\end{align}
where $f:\interestRegion\rightarrow \mathbb{R}$ denotes the {\em spatial loss field} (SLF) capturing 
the attenuation at location $\xgrd$, and $w:\interestRegion \times \interestRegion \times \interestRegion\rightarrow \mathbb{R}$ is a weight function describing how the SLF at $\tilde \bbx$ contributes to the shadowing experienced over the link $\bbx$--$\bbx'$. Typically, $w$ confers a greater weight $w(\bbx,\bbx',\xgrd)$ to those locations $\xgrd$ lying closer to the link $\xgen$--$\xgen'$. Examples of the weight function include the \emph{normalized ellipse model}~\cite{wilson2010tomography}  
\begin{align}
w(\bbx,\bbx',\tilde{\bbx}):=\begin{cases}
1/\sqrt{d(\bbx,\bbx')}, & \textrm{if}~d(\bbx,\tilde\bbx) + d(\bbx',\tilde\bbx) \\
& \hspace{0.6cm} < d(\bbx,\bbx')+\normEllipMargin/2 \\ 
0, & \textrm{otherwise}
\end{cases}\label{eq:ellipse_model}
\end{align}
where $\normEllipMargin > 0$ is a tunable parameter. The value of $\normEllipMargin$ is commonly set to 
the wavelength to assign non-zero weights only within the first Fresnel zone. In radio tomography, 
the integral in~\eqref{eq:SLFmodel} is approximated by a finite sum as
\begin{align}
s(\bbx,\bbx') &\simeq c\sum_{i=1}^{N_g} w(\bbx,\bbx',\xgrd_{i}) f(\xgrd_{i}) \label{eq:aproxSLFmodel} 
\end{align}
where $\{\xgrd_{i}\}_{i=1}^{N_g}$ is a grid of points over $\interestRegion$ and $c$ is a constant that
can be set to unity without loss of generality by absorbing
any scaling factor in $\field$. Clearly,~\eqref{eq:aproxSLFmodel} shows that
$s(\bbx,\bbx')$ depends on $f$ only through its values at the grid points.

The model in~\eqref{eq:SLFmodel} describes how the spatial distribution
of obstructions in the propagation path influences the attenuation between a pair of locations.
The usefulness of~\eqref{eq:SLFmodel} is twofold: i) as $\field$ represents absorption across space, 
it can be used for imaging; and ii) once $\field$ and $w$ are known, the gain
between any two points $\xgen$ and $\xgen'$ can be recovered through
\eqref{eq:cg} and~\eqref{eq:SLFmodel}, which is precisely the objective of channel-gain cartography.

The goal of radio tomography is to obtain a tomogram by estimating $\field$.
To this end, $N$ sensors located at $\{\xgen_1,\ldots,\xgen_N\} \in \interestRegion$
collaboratively obtain channel-gain measurements. At time slot $\tau$, the radios indexed by $n(\tau)$ and $n'(\tau)$
measure the channel-gain $\check{g}_\tau:=g(\xgen_{n(\tau)},\xgen_{n'(\tau)}) + \noise_\tau$ by 
exchanging training sequences known to both transmitting and receiving radios, where
$n(\tau),n'(\tau)\in \{1,\ldots,N\}$ and $\noise_\tau$ denotes measurement noise. 
It is supposed that $\uGain$ and $\gamma$ have been estimated during a calibration stage. 
After subtracting known components from $\check{g}_\tau$, the shadowing estimate is found as
\begin{align}
	\nshd_\tau&:=  \uGain - \pathlossexp 10\log_{10} d(\xgen_{n(\tau)}, \xgen_{n'(\tau)}) - \check{g}_\tau  \nonumber \\
	&=\shd(\xgen_{n(\tau)}, \xgen_{n'(\tau)}) - \noise_\tau. \label{eq:calibratedMea}
\end{align}
Having available $\Vnshd:=[\nshd_1,\ldots,\nshd_t]^{\transpose}\in \mathbb{R}^{t}$ 
along with the known set of links $\{(\xgen_{n(\tau)}, \xgen_{n'(\tau)})\}_{\tau=1}^{t}$
and the weight function $w$ at the fusion center, the problem is to estimate $\field$, and thus
$\vfield:=[\field(\xgrd_{1}),\ldots,\field(\xgrd_{N_g})]^{\transpose}\in \mathbb{R}^{N_g}$
using~\eqref{eq:aproxSLFmodel}.

Conventional regularized LS estimators of $\vfield$ solve~\cite{hamilton2014modeling,wilson2010tomography}
\begin{equation}
	\min_{\vfield}~\sum_{\tau=1}^{t}\bigg(\nshd_\tau - \sum_{i=1}^{N_g} w(\xgen_{n(\tau)}, \xgen_{n'(\tau)},
	\xgrd_{i}) f(\xgrd_{i}) \bigg)^2 + 
	\RegWeight\Regularizer(\vfield) \label{eq:regLS}
\end{equation}
where $\Regularizer:\mathbb{R}^{N_g} \rightarrow \mathbb{R}$ is a generic regularizer to promote a 
known attribute of $\vfield$, and $\RegWeight \geq 0$ is a regularization scalar to reflect compliance 
of $\vfield$ with this attribute. Although~\eqref{eq:regLS} has been successfully applied to radio tomography
after customizing the regularizer to the propagation environment, how accurate approximation is provided by a 
regularized solution of~\eqref{eq:regLS} is unclear, especially when the propagation environment exhibits inhomogeneous characteristics.  
  	
To overcome this and improve the SLF estimator performance, prior knowledge
on the heterogeneous structure of $\field$ will be exploited next, using a Bayesian approach.	

\section{Adaptive Bayesian Radio Tomography}
\label{sec:BayesModel}

In this section, we view $\vfield$ as random, and introduce a two-layer Bayesian SLF model, along with 
a VB-based approach to inference. We further develop a data-adaptive sensor selection 
method to collect informative measurements. 

\subsection{Bayesian model and problem formulation}
\label{sec:BayesianModel}

\begin{figure}[t!]
	\centering
	\includegraphics[width=.32\linewidth]{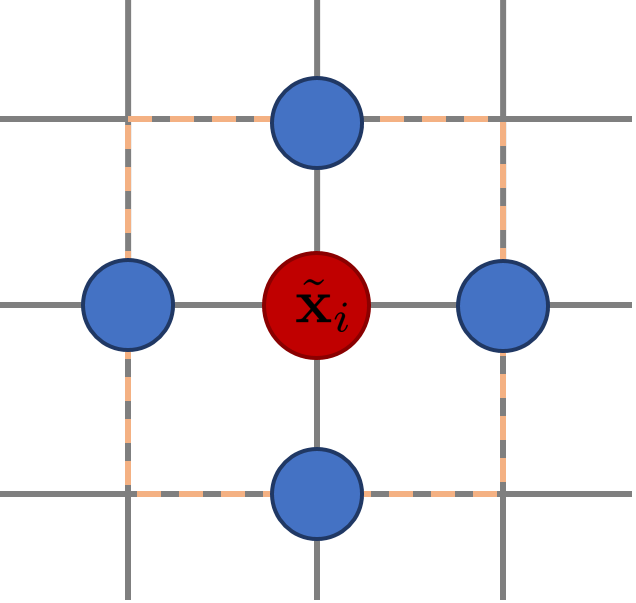}
	\caption{Four-connected MRF with $\Zfield(\xgrd_{i})$ marked red and its
		neighbors in $\neighbor{\xgrd_{i}}$ marked blue.}\label{fig:IsingPrior}
\end{figure}

\begin{figure}[t!]
	\centering
	\includegraphics[width=.6\linewidth]{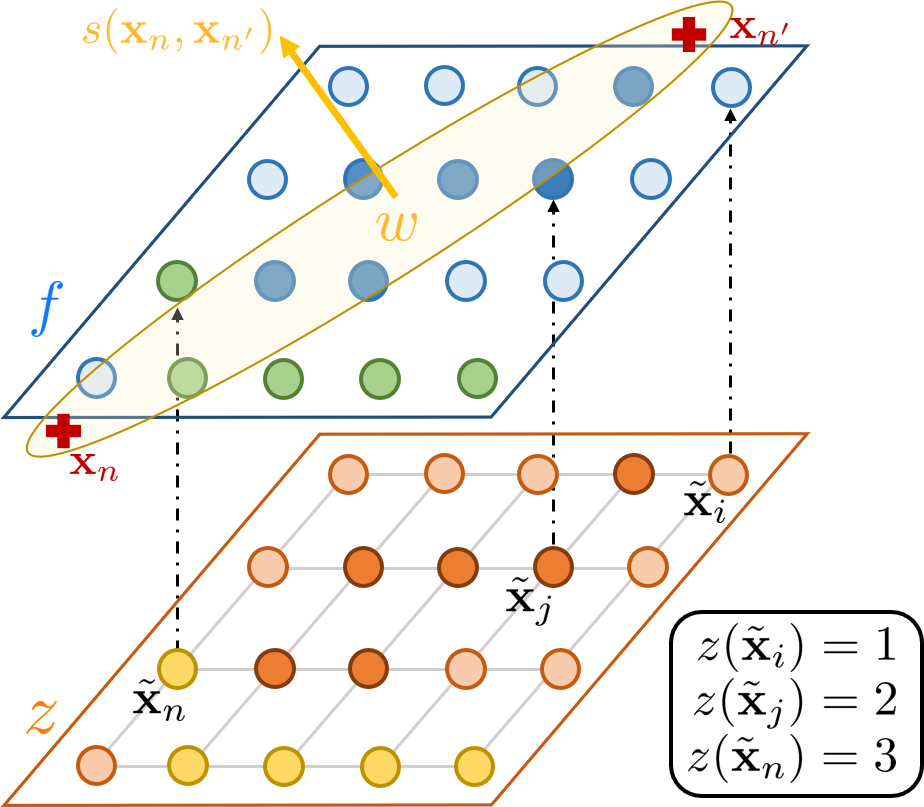}
	\caption{Gauss-Markov-Potts radio tomography model with $K=3$, together with 
measuring sensors located at $(\xgen_n,\xgen_{n'})$.}\label{fig:GP_model}
\end{figure}

Let $\interestRegion$ consist of $K$ disjoint homogeneous regions $\Region{k}:=\{\xgen|\mathbb{E}[\field(\xgen)] = \fMean{k},\textrm{Var}[\field(\xgen)] = \fVar{k} \}$ for $k=1,\ldots,K$, giving rise to a latent random label field 
$\vZfield := [\Zfield(\xgrd_1),\ldots,\Zfield(\xgrd_{N_g})]^{\transpose}\in \{1,\ldots,K\}^{N_g}$ with 
$K$-ary entries $\Zfield(\xgrd_{i}) = k$ if $\xgrd_{i} \in \interestRegion_{k}~\forall i,k$. 
The $K$ separate regions will model heterogeneous environments. With $K=2$ and $\mathcal{A}$ corresponding to an urban area, $\Region{2}$ may include densely populated regions with buildings, while $\Region{1}$ with $\fMean{1} < \fMean{2} $ may capture the less obstructive open spaces. For such a paradigm, we model the conditional distribution of $\field(\xgrd_{i})$ as
\begin{equation}
	p(\field(\xgrd_{i})|\Zfield(\xgrd_{i}) = k) = \mathcal{N}(\fMean{k},\fVar{k})~\forall k\:. \label{eq:GaussianField}
\end{equation}
We further assign the Potts prior to $\vZfield$ in order to capture the dependency among spatially correlated labels. By the Hammersley-Clifford theorem~\cite{HC71Gibbs2}, the Potts prior of $\vZfield$ follows a Gibbs distribution 
\begin{equation}
	p(\vZfield;\granPara) 
	= \frac{1}{C(\granPara)}\exp\left[ \sum_{i=1}^{N_g}\sum_{j \in \neighbor{\xgrd_{i}}}  \granPara\delta(\Zfield(\xgrd_{j})-\Zfield(\xgrd_{i}))\right] \label{eq:PottsPrior}
\end{equation}
where $\neighbor{\xgrd_{i}}$ is a set of indices comprising 1-hop neighbors of $\xgrd_{i}$ on the rectangular grid in Fig.~\ref{fig:IsingPrior}, $\granPara$ is a granularity coefficient controlling the degree of homogeneity in $\Zfield$, $\delta(\cdot)$ is Kronecker's delta, and the normalization constant
\begin{equation}
C(\granPara) := \sum_{\vZfield\in\zSet} \exp\left[ \sum_{i=1}^{N_g}\sum_{j \in \neighbor{\xgrd_{i}}}  \granPara\delta(\Zfield(\xgrd_{j})-\Zfield(\xgrd_{i}))\right]
\end{equation}
is the partition function with $\zSet:=\{1,\ldots,K\}^{N_g}$.
To ease exposition, $\granPara$ is assumed known or fixed a priori; see e.g.,~\cite{dmzb99granEstMCMC,mtrp09granEstVB,dbg18adaptiveRT} for a means of estimating $\granPara$.
If $\{\field(\xgrd_{i})\}_{i=1}^{N_g}$ are conditionally independent given $\vZfield$, 
the model reduces to the Gauss-Markov-Potts model~\cite{ayasso10pott}. 
Such a model with $K=3$ is depicted in Fig.~\ref{fig:GP_model} 
with the measurement model in \eqref{eq:aproxSLFmodel}.

Noise $\noise_t$ in \eqref{eq:calibratedMea} is assumed independent and identically distributed (i.i.d.) Gaussian with zero mean and variance $\nVar$. Here, we correspondingly consider precisions of $\noise_t$ and $\{\field_k\}_{k=1}^{K}$ that are denoted as $\precNVar:=1/\nVar$ and $\fprec{k}:=1/\fVar{k}\forall k$, respectively. Let also $\hParameter$ be a hyperparameter vector comprising $\precNVar$ and $\hParameter_f:=[\vfMean^{\transpose},\vfPrec^{\transpose}]^{\transpose}$ with  $\vfMean:=[\fMean{1},\ldots,\fMean{K}]^{\transpose}\in\mathbb{R}^{K}$ and $\vfPrec:=[\fprec{1},\ldots,\fprec{K}]^{\transpose}\in\mathbb{R}^{K}$.  
Assuming the independence among entries of $\hParameter$, we deduce that 
\begin{equation}
p(\hParameter)=p(\precNVar)p(\vfMean)p(\vfPrec)=p(\precNVar)\prod_{k=1}^{K}p(\fMean{k}) p(\fprec{k})
\end{equation}
where the priors $p(\precNVar)$, $p(\vfMean)$, and $p(\vfPrec)$ are as follows.

\begin{figure}[t!]
	\centering
	\includegraphics[width=.6\linewidth]{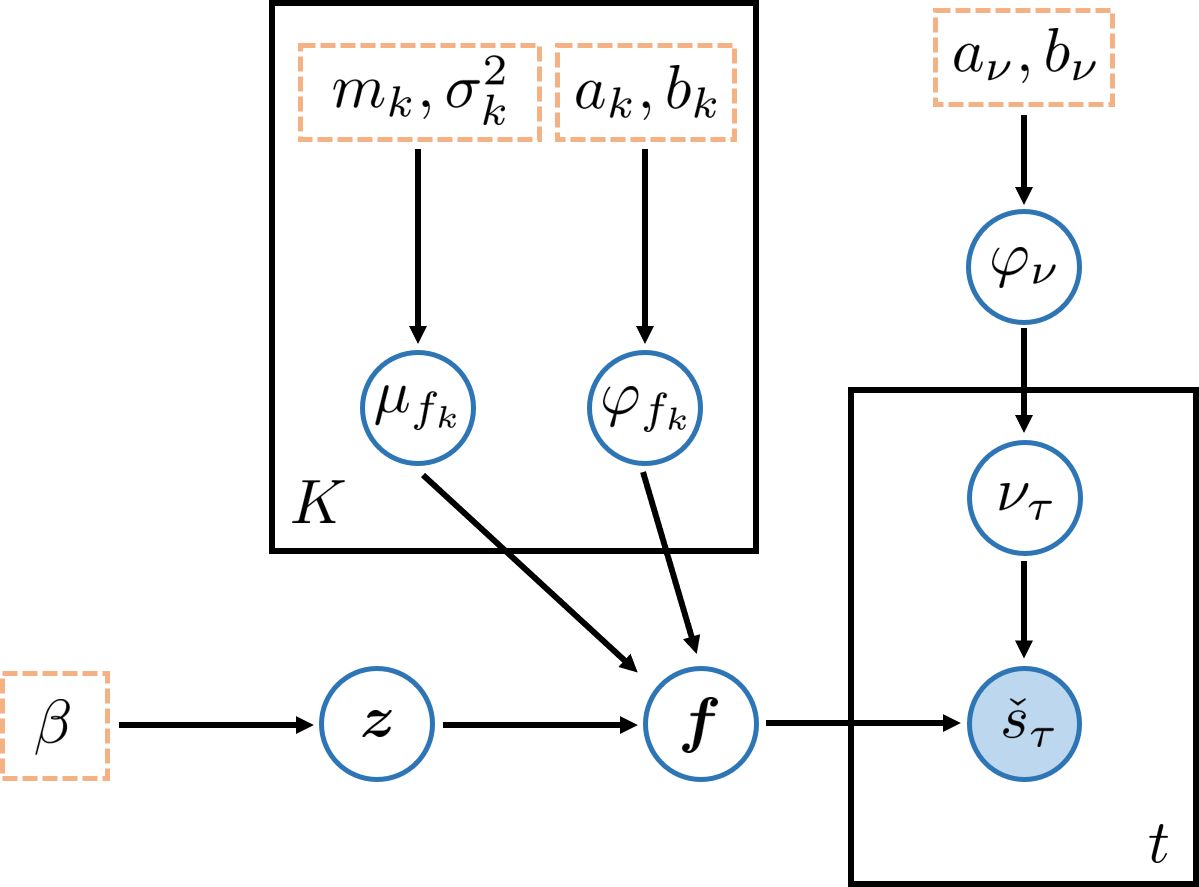}
	\caption{Graphical model representation of the hierarchical Bayesian model for (hyper) 
		parameters (those in dashed boxes are fixed).}\label{fig:Graph_Bayes_model}
\end{figure}

\subsubsection{Noise precision $\precNVar$}
With additive Gaussian noise having fixed mean, it is common to assign a conjugate prior to $\precNVar$ that can reproduce a posterior in the same family of its prior. The gamma distribution for $\precNVar\in \mathbb{R}^{+}$ serves this purpose, as 
\begin{align}
p(\precNVar)=\G{a_{\nu}}{b_{\nu}}:= \frac{1}{\Gamma(a_{\nu})b_{\nu}^{a_{\nu}}}
(\precNVar)^{a_{\nu}-1}e^{-\precNVar / b_\nu}  \label{eq:VarPrior}
\end{align}
where $a_{\nu}$ is referred to as the shape parameter, $b_{\nu}$ as the scale parameter, 
and $\Gamma(\cdot)$ denotes the gamma function.

\subsubsection{Hyperparameters $\hParameter_f$ of the SLF }
While the prior for $\fMean{k}$ is assumed to be Gaussian with mean $\fMeanMean{k}$ and variance $\fVarVar{k}$ (see also~\cite{ayasso10pott}), similar to the noise, the prior for $\fprec{k}\in \mathbb{R}^{+}$
is the Gamma distribution parameterized by $\{a_k, b_k\}$; that is,
\begin{align}
	p(\fMean{k}) & = \NormDist{\fMeanMean{k}}{\fVarVar{k}},~k = 1,\ldots,K\label{eq:fMeanPrior} \\
	p(\fprec{k}) & = \G{a_k}{b_k},~~~k = 1,\ldots,K. \label{eq:fVarPrior}
\end{align}
We stress that analytical tractability is the main motivation behind selecting the conjugate
priors in~\eqref{eq:VarPrior}--\eqref{eq:fVarPrior}.

Our goal of inferring $\vfield$, relies on the following posterior distribution that can be factored (within a constant) as 
\begin{equation}
	p(\vfield,\vZfield,\hParameter|\Vnshd)\propto p(\Vnshd|\vfield,\precNVar)p(\vfield|\vZfield,\hParameter_f) 
	p(\vZfield;\granPara) p(\hParameter) \label{eq:BayesPosterior}
\end{equation}
where $p(\Vnshd|\vfield,\precNVar) \sim \mathcal{N}(\mWeight_t^{\transpose}\vfield,\nVar\IdenMat{t})$ is 
the data likelihood with the weight matrix $\mWeight_t \in \mathbb{R}^{N_g \times t}$ formed from
columns $\vWeight_{\tau}^{(n,n')}:=[w(\xgen_{n(\tau)},\xgen_{n'(\tau)},\xgrd_{1}),
\ldots,w(\xgen_{n(\tau)},\xgen_{n'(\tau)},\xgrd_{N_g})]^\transpose:=[w_{\tau,1},\ldots,w_{\tau,N_g}]^{\transpose} 
\in \mathbb{R}^{N_g}$ of the link $\xgen_{n(\tau)}$--$\xgen_{n'(\tau)}$ for $\tau=1,\ldots,t$.
Fig.~\ref{fig:Graph_Bayes_model} depicts our hierarchical Bayesian model for 
$\{\Vnshd,\vfield,\vZfield,\hParameter\}$ as a directed acyclic graph, where the dependence between 
(hyper) parameters is indicated with an arrow.

Given the posterior in~\eqref{eq:BayesPosterior}, the conditional minimum mean-square error (MMSE) estimator of the field is 
\begin{equation}
	\hat{\vfield}_{\textrm{MMSE}}:=\mathbb{E}[\vfield | \vZfield = \hat{\vZfield}_{\textrm{MAP}},\Vnshd] \label{est_1}
\end{equation}
where the maximum a posteriori (MAP) label estimator is
\begin{equation}
	\hat{\vZfield}_{\textrm{MAP}} :=\arg\max_{\vZfield} p(\vZfield|\Vnshd)
\end{equation}
and the MMSE estimators of $\hParameter$ entries are   
\begin{align} 
	\widehat{\precNVar}_{\textrm{MMSE}}&:=\mathbb{E}[\precNVar | \Vnshd] \\\label{est_2}
	\widehat{\fMean{k}}_{\textrm{MMSE}}&:=\mathbb{E}[\fMean{k} | \Vnshd],~k=1,\ldots,K \\\label{est_4}
	\widehat{\fprec{k}}_{\textrm{MMSE}}&:=\mathbb{E}[\fprec{k} | \Vnshd],~k=1,\ldots,K.
\end{align}

\subsection{Radio tomography using variational Bayes}
\label{sec:VB}

Although the estimator forms in \eqref{est_1}-\eqref{est_4} have been considered also in~\cite{kail12BlindDecon}, obtaining estimates in practice is not tractable because the complex posterior in~\eqref{eq:BayesPosterior} is not amenable to marginalization or maximization. To overcome this hurdle, one can resort to approximate Bayesian inference methods such as MCMC~\cite{gilks96MCMC} that relies on samples of $\{\vfield,\vZfield,\hParameter\}$ drawn from a complex distribution. Although MCMC can asymptotically approach an exact target distribution, such as the sought one in~\eqref{eq:BayesPosterior}, it can be computationally demanding and thus does not scale well. Aiming at a scalable alternative, we will adopt the so-termed variational Bayes (VB) approach.

VB is a family of techniques to approximate a complex distribution by a tractable one termed variational distribution. A typical choice of an approximation criterion is to find the variational distribution $q$ minimizing the Kullback-Leibler (KL) divergence ($D_{\rm KL}(q\|p)$) to a target distribution $q$. The variational distribution $q$ is further assumed to belong to a certain family $\mathcal{Q}$ of distributions possessing a simpler form of dependence between variables than the original one; see also~\cite{opper01meanfield} for the so-termed mean-field approximation. 

Tailored to the posterior in~\eqref{eq:BayesPosterior} the variational one, solves 
\begin{align}
	\min_{q(\vfield,\vZfield,\hParameter) \in \mathcal{Q}} D_{\textrm{KL}}\left(q(\vfield,\vZfield,\hParameter)\|
	p(\vfield,\vZfield,\hParameter|\Vnshd)\right)~\label{eq:KL}
\end{align}  
Using that $D_{\textrm{KL}}\left(q \| p \right):= - \mathbb{E}_{q}[\ln(p/q)] $, the latter reduces to   
\begin{align}
\textrm{(P1)}~\max_{q(\vfield,\vZfield,\hParameter) \in \mathcal{Q}} 
\underbrace{\mathbb{E}_{q(\vfield,\vZfield,\hParameter)}\left[\ln\left(
	\frac{p(\vfield,\vZfield,\hParameter,\Vnshd)}{q(\vfield,\vZfield,\hParameter)}\right)\right]}_{
	=:\textrm{ELBO}(q(\vfield,\vZfield,\hParameter))} \label{eq:P1}
\end{align}
where we drop the constant $p(\Vnshd)$ from the posterior that resulted in the so-termed~\emph{evidence lower bound} 
(ELBO) in (P1), which involves the joint $p(\vfield,\vZfield,\hParameter,\Vnshd)$ factored as in the right-hand side (RHS) of~\eqref{eq:BayesPosterior}. We choose the family $\mathcal{Q}$ as 
\begin{align}
\mathcal{Q}:=\bigg\{q:q(\vfield,\vZfield,\hParameter)&:=q(\vfield|\vZfield)q(\vZfield)q(\hParameter)\nonumber \\ &=\prod_{i=1}^{N_g}q(\field_i|\Zfield_i) \prod_{i=1}^{N_g}q(\Zfield_i)q(\hParameter) \bigg\}\label{eq:factPosterior}
\end{align}
where $\field_i:=\field(\xgrd_{i})$ and $\Zfield_i:=\Zfield(\xgrd_{i})$ $\forall i$ for simplicity, and 
\begin{equation}
q(\hParameter):=q(\precNVar)q(\vfMean)q(\vfPrec)=q(\precNVar)\prod_{k=1}^{K}q(\fMean{k})\prod_{k=1}^{K}q(\fprec{k}).
\label{eq:variationalDist}
\end{equation}
Following the general VB steps~\cite{murphy12ML}, we will solve our (P1) here via coordinate minimization among factors of $q(\vfield,\vZfield,\hParameter)$. Within a constant $c$, the optimal solutions have the form 
\begin{align}
&\ln q^{*}(\field_i|\Zfield_i) = \mathbb{E}_{-q(\field_i|\Zfield_i)}\left[ \ln p(\vfield,\vZfield,\hParameter,\Vnshd)\right] 
+ c~\forall i\label{eq:variationalProbSLF}\\
&\ln q^{*}(\Zfield_i) = \mathbb{E}_{-q(\Zfield_i)}\left[ \ln p(\vfield,\vZfield,\hParameter,\Vnshd)\right] + c~\forall i \label{eq:variationalProbHidden} \\
&\ln q^{*}(\hParameter) = \mathbb{E}_{-q(\hParameter)}\left[ \ln p(\vfield,\vZfield,\hParameter,\Vnshd)\right] 
+ c\label{eq:variationalProbHyper}
\end{align}
where the expectation in~\eqref{eq:variationalProbSLF} is over the variational pdf of $\vfield_{-i}$, $\vZfield$, and $\hParameter$, that is $\prod_{j\neq i}q(\field_j|\Zfield_{j})q(\vZfield)q(\hParameter)$. Similar expressions are available for \eqref{eq:variationalProbHidden} and \eqref{eq:variationalProbHyper}. The solutions in~\eqref{eq:variationalProbSLF}--\eqref{eq:variationalProbHyper} are intertwined since 
the evaluation of one requires the others. We show in Appendices A-E that the optimal solutions 
can be obtained iteratively; that is, per iteration $\ell=1,2,\ldots$, we have 
\begin{align}
q^{(\ell)}(\field_{i}|\Zfield_{i}=k) &= \NormDist{\VIfmeanItr{k}{i}{\ell}}{\VIfvarItr{k}{i}{\ell}}
\forall k \label{eq:VD_field}\\
q^{(\ell)}(\Zfield_{i}=k) &=:\VIzfieldProbItr{k}{\xgrd_{i}}{\ell} = \frac{\VIzfieldUnnormProbItr{k}{\xgrd_{i}}{\ell}}{\sum_{k=1}^{K}\VIzfieldUnnormProbItr{k}{\xgrd_{i}}{\ell}} 
\forall k\label{eq:VP_hidden} \\
q^{(\ell)}(\precNVar) &= \G{\VIaNVar}{\VIbNVarItr{\ell}} \\
q^{(\ell)}(\fMean{k}) &= \NormDist{\VIfMeanMeanItr{k}{\ell}}{\VIfMeanVarItr{k}{\ell}}~\forall k \\
q^{(\ell)}(\fprec{k}) &= \G{\VIfprecaItr{k}{\ell}}{\VIfprecbItr{k}{\ell}}~\forall k
\end{align}
with variational parameters 
\begin{align}
\VIfvarItr{k}{i}{\ell} &= \left({\MeanprecNVar}^{(\ell-1)} \sum_{\tau=1}^{t}w_{\tau,i}^{2} 
+ \MeanfprecItr{k}{\ell-1}\right)^{-1}~\forall k \label{eq:VPara_update_fvar}\\
\VIfmeanItr{k}{i}{\ell} &= \Meanfield{i}^{(\ell-1)} + \VIfvarItr{k}{i}{\ell}\bigg[\big(\VIfMeanMeanItr{k}{\ell-1} 
- \Meanfield{i}^{(\ell-1)}  \big)\MeanfprecItr{k}{\ell-1} \nonumber \\
&+{\MeanprecNVar}^{(\ell-1)}\sum_{\tau=1}^{t} w_{\tau,i}\left(\nshd_{\tau}-\Meanshd{\tau}^{(\ell-1)}\right)\bigg]~\forall k \label{eq:VPara_update_fmean}\\
\VIzfieldUnnormProbItr{k}{\xgrd_{i}}{\ell} &=\exp\bigg\{-\frac{\MeanfprecItr{k}{\ell-1}}{2}\bigg[\VIfvarItr{k}{i}{\ell} +\left(\VIfmeanItr{k}{i}{\ell}\right)^{2}  \nonumber \\
& - 2\VIfMeanMeanItr{k}{\ell-1}\VIfmeanItr{k}{i}{\ell} +\VIfMeanVarItr{k}{\ell-1}
+\left(\VIfMeanMeanItr{k}{\ell-1}\right)^{2} \bigg]\nonumber \\
& + \frac{1}{2}\left(\diGamma{\VIfprecaItr{k}{\ell-1}}+\ln \VIfprecbItr{k}{\ell-1} \right) \nonumber\\
& + \sum_{j \in \neighbor{\xgrd_{i}}} \granPara  \VIzfieldProbItr{k}{\xgrd_{j}}{\ell-1} \bigg\}~\forall k  
\label{eq:VP_hidden_unnorm}\\
\VIaNVar  &= a_{\nu} + \frac{t}{2} \label{eq:VPara_aNVar} \\
\VIbNVarItr{\ell}  &= \Bigg\{\frac{1}{b_{\nu}} + \frac{1}{2}\sum_{\tau=1}^{t} \nshd_{\tau}^2
-2\nshd_{\tau}\Meanshd{\tau}^{(\ell)} +\sum_{i=1}^{N_g}w_{\tau,i}^2\nonumber \\
&\times \Bigg[\sum_{k=1}^{K}\VIzfieldProbItr{k}{\xgrd_{i}}{\ell}\left(
\VIfvarItr{k}{i}{\ell} +\left(\VIfmeanItr{k}{i}{\ell}\right)^{2}\right) \nonumber \\
&- \left({\Meanfield{i}}^{(\ell)} \right)^2\Bigg] + \left(\Meanshd{\tau}^{(\ell)}\right)^2 \Bigg\}^{-1} 
\label{eq:VPara_bNVar}\\
\VIfMeanVarItr{k}{\ell} &= \left(\frac{1}{\sigma_{k}^2} + \sum_{i=1}^{N_g}\VIzfieldProbItr{k}{\xgrd_{i}}{\ell}\MeanfprecItr{k}{\ell-1}\right)^{-1}
\forall k \label{eq:VPara_fMeanVar}\\
\VIfMeanMeanItr{k}{\ell} &= \VIfMeanVarItr{k}{\ell}\left(\frac{m_k}{\sigma_{k}^2} + \sum_{i=1}^{N_g}\VIzfieldProbItr{k}{\xgrd_{i}}{\ell}\MeanfprecItr{k}{\ell-1}\VIfmeanItr{k}{i}{\ell}\right) 
\forall k \label{eq:VPara_fMeanMean}\\
\VIfprecaItr{k}{\ell} &= a_k + \frac{1}{2}\sum_{i=1}^{N_g} \VIzfieldProbItr{k}{\xgrd_{i}}{\ell}
\forall k \label{eq:VPara_fpreca}
\end{align}
\begin{align}
&\VIfprecbItr{k}{\ell} =\Bigg[ \frac{1}{b_k} + \frac{1}{2}\sum_{i=1}^{N_g} \VIzfieldProbItr{k}{\xgrd_{i}}{\ell} 
\bigg(\VIfvarItr{k}{i}{\ell} +\left(\VIfmeanItr{k}{i}{\ell}\right)^{2}  \nonumber \\
&- 2\VIfmeanItr{k}{i}{\ell}\VIfMeanMeanItr{k}{\ell-1} +\VIfMeanVarItr{k}{\ell-1}
+\left(\VIfMeanMeanItr{k}{\ell-1}\right)^{2} \bigg)\Bigg]^{-1} \forall k\label{eq:VPara_fprecb}
\end{align}
where $\diGamma{\cdot}$ is the digamma function, ${\Meanfield{i}}^{(\ell)}:=\sum_{k=1}^{K}\VIzfieldProbItr{k}{\xgrd_{i}}{\ell}\VIfmeanItr{k}{i}{\ell}~\forall i$, 
$\Meanshd{\tau}^{(\ell)}:=\sum_{i=1}^{N_g} w_{\tau,i}{\Meanfield{i}}^{(\ell)}~\forall \tau$, $\MeanprecNVar^{(\ell)}:=\mathbb{E}_{q^{(\ell)}(\precNVar)}[\precNVar]=\VIaNVar\VIbNVarItr{\ell}$, 
and $\MeanfprecItr{k}{\ell}:=\mathbb{E}_{q^{(\ell)}(\fprec{k})}[\fprec{k}]=\VIfprecaItr{k}{\ell}\VIfprecbItr{k}{\ell}~\forall k$;
see Appendices~\ref{sec:CondMeanVar4SLF}--\ref{sec:VIClassPrecsPosterior} for detailed derivation of the variational factors and parameters in~\eqref{eq:VD_field}--\eqref{eq:VPara_fprecb}.

Upon convergence of the iterative solvers, the (approximate) MAP estimator of $\vZfield$ can be obtained as 
\begin{equation}
	\hat{\Zfield}_{\textrm{MAP},i} =  \arg \max_{\Zfield_i \in {1,\ldots,K}} q^{*}(\Zfield_i)~\forall i,
\end{equation} 
and then the (approximate) MMSE estimator of $\vfield$ as 
\begin{equation}
	\hat{\field}_{\textrm{MMSE},i} \simeq \mathbb{E}_{q^{*}(\field_i|\hat{\Zfield}_{\textrm{MAP},i})}\left[\field_i\right]
	={\breve{\mu}}_{f_{\hat{\Zfield}_{\textrm{MAP},i}}}^{*}(\xgrd_{i})~\forall i
\end{equation}
while $\hParameter$ is estimated using the marginal MMSE estimators 
\begin{align}
	\widehat{\precNVar}_{\textrm{MMSE}} &\simeq \mathbb{E}_{q^{*}(\precNVar)}\left[\precNVar\right]
	=\VIaNVar^{*} \VIbNVar^{*} \label{eq:MMSE_PrednVar}\\
	\widehat{\fMean{k}}_{\textrm{MMSE}} &\simeq \mathbb{E}_{q^{*}(\fMean{k})}\left[\fMean{k}\right]
	=\VIfMeanMean{k}^{*}~\forall k \label{eq:MMSE_fMean}\\
	\widehat{\fprec{k}}_{\textrm{MMSE}} &\simeq \mathbb{E}_{q^{*}(\fprec{k})}\left[\fprec{k}\right]
	=\VIfpreca{k}^{*}\VIfprecb{k}^{*}~\forall k. \label{eq:MMSE_fPrec}
\end{align}

The VB algorithm to obtain $\{\hat{\field}_{\textrm{MMSE},i}\}_{i=1}^{N_g}$, $\{\hat{\Zfield}_{\textrm{MAP},i}\}_{i=1}^{N_g}$,
$\hat{\hParameter}_{\textrm{MMSE}}$,  and $q^{*}(\vfield,\vZfield,\hParameter)$ is tabulated in Alg.~\ref{alg:VB}.  

\noindent{\remark \bf{(Assessing convergence)}.}
The steps of Alg.~\ref{alg:VB} guarantee that the ELBO monotonically increases across iterations $\ell$~\cite{beal03}. Hence, convergence of the solution can be assessed by monitoring the change in the ELBO of (P1) in~\eqref{eq:P1}, which for a preselected threshold $\xi>0$ suggests stopping at iteration $\ell$ if $\textrm{ELBO}\left(q^{(\ell)}(\vfield,\vZfield,\hParameter)\right) 
- \textrm{ELBO}\left(q^{(\ell-1)}(\vfield,\vZfield,\hParameter)\right)\leq \xi$. 

\noindent{\remark \bf{(Computational complexity)}.}
For Alg.~\ref{alg:VB}, the complexity order to
update $q(\field_{i}|\Zfield_{i}=k)~\forall i,k$ per iteration $\ell$ is $\mathcal{O}(tKN_g)$
to compute $\VIfmean{k}{i}$ in~\eqref{eq:VPara_update_fmean}, while updating 
$\VIzfieldProb{k}{\xgrd_{i}}~\forall i,k$ via~\eqref{eq:VP_hidden} incurs complexity 
$\mathcal{O}(KN_g)$. In addition, updating $q(\hParameter)$ has complexity 
$\mathcal{O}(tKN_g)$ that is dominated by the computation of $\VIbNVar$ in~\eqref{eq:VPara_bNVar}. 
Overall, the per-iteration complexity of Alg.~\ref{alg:VB} is $\mathcal{O}\left((2t+1)KN_g\right)$. 

Note that a sample-based counterpart of Alg.~\ref{alg:VB} via MCMC in~\cite{dbg18adaptiveRT} 
incurs complexity in the order of $\mathcal{O}(N_g^3)$. For conventional methods to estimate $\vfield$, 
the ridge regularized LS~\cite{hamilton2014modeling} has a one-shot (non-iterative) complexity 
of $\mathcal{O}(N_g^{3})$, while the total variation (TV) regularized LS via the alternating 
direction method of multipliers (ADMM) in~\cite{QGM15ADMMTV} incurs complexity of $\mathcal{O}(N_g^{3})$ 
per iteration $\ell$; see also~\cite{lee2017lowrank,Daniel18BlindTomography} for details.
This means that Alg.~\ref{alg:VB} incurs the lowest per-iteration complexity, which becomes more critical as 
$N_g$ increases.

\begin{algorithm}[t!]
	\caption{Field estimation via variational Bayes}
	\begin{algorithmic}[1]
	\renewcommand{\algorithmicrequire}{\textbf{Input:}}
	\REQUIRE $\Vnshd$, $\mWeight_{t}$, $\bigg\{a_{\nu}, b_{\nu}, 
	\big\{\fMeanMean{k},\fVarVar{k},a_k,b_k\big\}_{k=1}^{K}\bigg\}$, and $\NumStep$.
	\STATE Initialize $q^{(0)}(\vfield,\vZfield,\hParameter)$ and set $\ell=0$
	\STATE Obtain $\VIaNVar$ with~\eqref{eq:VPara_aNVar}
	\WHILE {\textrm{ELBO} has not converged and $\ell\leq \NumStep$ }
		\STATE Set $\ell \leftarrow \ell + 1$
		\STATE Obtain $\VIfvarItr{k}{i}{\ell}~\forall i,k$ via~\eqref{eq:VPara_update_fvar} 
		\STATE Obtain $\VIfmeanItr{k}{i}{\ell}~\forall i,k$ via~\eqref{eq:VPara_update_fmean} 
		\STATE Obtain $q^{(\ell)}(\Zfield_i=k)~\forall i,k$ via~\eqref{eq:VP_hidden}
		\STATE Obtain $\VIbNVarItr{\ell}$ via~\eqref{eq:VPara_bNVar}
		\STATE Obtain $\VIfMeanVarItr{k}{\ell}~\forall k$ via~\eqref{eq:VPara_fMeanVar}
		\STATE Obtain $\VIfMeanMeanItr{k}{\ell}~\forall k$ via~\eqref{eq:VPara_fMeanMean}
		\STATE Obtain $\VIfprecaItr{k}{\ell}~\forall k$ via~\eqref{eq:VPara_fpreca}
		\STATE Obtain $\VIfprecbItr{k}{\ell}~\forall k$ via~\eqref{eq:VPara_fprecb}
	\ENDWHILE
	\STATE Set $q^{*}(\field_i|\Zfield_i)=q^{(\ell)}(\field_i|\Zfield_i)$ and $q^{*}(\Zfield_i)=q^{(\ell)}(\Zfield_i)$~$\forall i$
	\STATE Set $q^{*}(\hParameter) = q^{(\ell)}(\hParameter)$
	\STATE Estimate $\hat{\Zfield}_{\textrm{MAP},i} = \arg \max_{\Zfield_{i} \in \{1,\ldots,K\}} q^{*}(\Zfield_i)~\forall~i$
	\STATE Estimate $\hat{\field}_{i,\textrm{MMSE}}={\breve{\mu}}_{f_{\hat{\Zfield}_{\textrm{MAP},i}}}^{*}(\xgrd_{i})~\forall~i$
	\STATE Estimate $\hat{\hParameter}_{\textrm{MMSE}}=\mathbb{E}_{q^{*}(\hParameter)}[\hParameter]$ via~\eqref{eq:MMSE_PrednVar}--\eqref{eq:MMSE_fPrec}
	\RETURN $\hat{\vfield}_{\textrm{MMSE}}$, $\hat{\vZfield}_{\textrm{MAP}}$, 
	$\hat{\hParameter}_{\textrm{MMSE}}$, $q^{*}(\vfield|\vZfield)$, $q^{*}(\vZfield)$, and $q^{*}(\hParameter)$
\end{algorithmic} \label{alg:VB} 
\end{algorithm} 

\subsection{Data-adaptive sensor selection via uncertainty sampling}
\label{sec:adaptiveSampling}
Here we deal with cost-effective radio tomography as new data are collected by interactively querying the location of sensing radios to acquire a minimal but most informative measurements. To this end, 
a measurement (or a mini-batch of measurements) can be adaptively collected using a set of available sensing radio pairs, with the goal of reducing the uncertainty of $\vfield$. Since the proposed Bayesian framework accounts for the uncertainty through $\VIfvar{k}{i}$ in~\eqref{eq:VPara_fMeanMean}, we adopt the conditional entropy~\cite{cover91IT} to serve as an uncertainty measure of $\vfield$ at time slot $\tau$, namely,
\begin{align}
	H(\vfield|\vZfield,\Vnshdat{\tau};\hat{\hParameter}_{\tau})  &= 
	\sum_{\vZfield'\in\zSet}\int  p(\vZfield',\Vnshdat{\tau}';\hat{\hParameter}_{\tau})  \nonumber \\
	&\times H(\vfield|\vZfield=\vZfield',\Vnshdat{\tau}=\Vnshdat{\tau}';\hat{\hParameter}_{\tau}) d\Vnshdat{\tau}', \label{eq:conditionalEntropy} 
\end{align} 
where $\hat{\hParameter}_{\tau}$ is the estimate obtained via~\eqref{eq:MMSE_PrednVar}--\eqref{eq:MMSE_fPrec}
per slot $\tau$, and 
\begin{align}
	& H(\vfield|\vZfield=\vZfield',\Vnshdat{\tau}=\Vnshdat{\tau}';\hat{\hParameter}_{\tau}) \nonumber \\
	& \hspace{-0.1cm} := - \hspace{-0.12cm}\int \hspace{-0.12cm} p(\vfield|\vZfield=\vZfield'\hspace{-0.05cm},\Vnshdat{\tau}=\Vnshdat{\tau}';\hat{\hParameter}_{\tau}) 
	\ln p(\vfield|\vZfield=\vZfield'\hspace{-0.05cm},\Vnshdat{\tau}=
	\Vnshdat{\tau}';\hat{\hParameter}_{\tau}) d\vfield \nonumber \\
	& \hspace{-0.1cm} = \frac{1}{2}\ln \left| \mCondCovFwith{z',\Vnshdat{\tau}';\hat{\hParameter}_{\tau}}  
	\right| + \frac{N_g}{2}\bigg(1 + \ln 2\pi \bigg) \label{eq:condEntropyGivenParameters}
\end{align}
since $p(\vfield|\vZfield,\Vnshdat{\tau};\hat{\hParameter}_{\tau})$ is Gaussian with covariance matrix $\mCondCovFwith{z,\Vnshdat{\tau};\hat{\hParameter}_{\tau}}:=\left(\widehat{\precNVar}\mWeight_\tau\mWeight_{\tau}^{\transpose} 
+ \widehat{\mDiagFieldPrec} \right)^{-1}$ with $\widehat{\mDiagFieldPrec}:=\textrm{diag}\left(\{ \widehat{\fprec{\Zfield_{i}}} \}_{i=1}^{N_g}\right)$~\cite{dbg18adaptiveRT}. Then, using the matrix determinant identity lemma~\cite[Chap. 18]{harville97MatrixDeterminantLemma}, it is not hard to show that \begin{align}
&H(\vfield|\vZfield,\Vnshdat{\tau+1};\hat{\hParameter}_{\tau}) = H(\vfield|\vZfield,\Vnshdat{\tau};
\hat{\hParameter}_{\tau}) -\frac{1}{2}\hspace{-0.05cm}\sum_{\vZfield'\in\zSet}\int\hspace{-0.08cm}  
p(\vZfield',\Vnshdat{\tau}';\hat{\hParameter}_{\tau}) \nonumber \\
&\hspace{1cm}  \times\ln\left(1 + \widehat{\precNVar} 
{\vWeight_{\tau+1}^{(n,n')}}^{\transpose} \mCondCovFwith{z',\Vnshdat{\tau}';\hat{\hParameter}_{\tau}} \vWeight_{\tau+1}^{(n,n')} 
\right)\hspace{-0.06cm} d\Vnshdat{\tau}'.\label{eq:condEntropyAtNextTimeSlot} 
\end{align} 
To obtain $\nshd_{\tau+1}$, we choose a pair of sensors $(n^{*},n'^{*})$, or equivalently 
find $\vWeight_{\tau+1}^{(n^{*},n'^{*})}$ minimizing $H(\vfield|\vZfield,\Vnshdat{\tau+1};\hat{\hParameter}_{\tau})$. 

Given $\Vnshdat{\tau}$, we then find $\vWeight_{\tau+1}^{(n^{*},n'^{*})}$ by solving
\begin{equation}
	\textrm{(P2)}~~~\max_{\stackrel{\vWeight_{\tau+1}^{(n,n')}:}{(n,n')\in \mathcal{M}_{\tau+1}}} \mathbb{E}_{p(\vZfield|\Vnshdat{\tau};\hat{\hParameter}_{\tau})}
	\left[ h(\vZfield,\Vnshdat{\tau},\vWeight_{\tau+1}^{(n,n')};\hat{\hParameter}_{\tau})\right] \label{eq:P2}  
\end{equation}
where $h(\vZfield,\Vnshdat{\tau},\vWeight;\hat{\hParameter}_{\tau}):=\ln
\left(1 + \widehat{\precNVar} \vWeight^{\transpose} \mCondCovFwith{z,\Vnshdat{\tau};\hat{\hParameter}_{\tau}} 
\vWeight \right)$ and $\mathcal{M}_{\tau} :=\{(n,n')|\exists (\xgen_n\textrm{--}\xgen_{n'})
\textrm{~at~}\tau,(n,n')\in\{1,\ldots,N\}^{2}\}$ denotes the set of available sensing radio pairs at slot $\tau$. 

Clearly, (P2) in~\eqref{eq:P2} cannot be directly solved because  $p(\vZfield|\Vnshdat{\tau};\hat{\hParameter}_{\tau})$ is not tractable e.g., by marginalizing the posterior in~\eqref{eq:BayesPosterior}. Hence, evaluating the cost of (P2) is intractable for large $N_g$ as $|\zSet|=2^{N_g}$. Fortunately, we show next how (P2) can be approximately reformulated using the variational 
distribution $q(\vfield,\vZfield,\hParameter)$. Consider first that  
\begin{equation}
	p(\vfield|\vZfield,\Vnshdat{\tau},\hParameter) = \frac{p(\vfield,\vZfield,\hParameter|\Vnshdat{\tau})}{p(\vZfield,\hParameter|\Vnshdat{\tau})}\approx
	\frac{q(\vfield,\vZfield,\hParameter)}{q(\vZfield,\hParameter)}=q(\vfield|\vZfield),
\end{equation}
which yields the approximation of $H$ in~\eqref{eq:condEntropyGivenParameters}, as
\begin{align}
	& H(\vfield|\vZfield=\vZfield',\Vnshdat{\tau}=\Vnshdat{\tau}';\hat{\hParameter}_{\tau}) \nonumber \\
	& \hspace{1cm}\approx \frac{1}{2}\ln \left| \mCondCovFVIwith{z',\Vnshdat{\tau}';\hat{\hParameter}_{\tau}}  \right| 
	+ \frac{N_g}{2}\bigg(1 + \ln 2\pi \bigg) \label{eq:condEntropyGivenParametersVI}
\end{align}
with $\mCondCovFVIwith{z,\Vnshdat{\tau};\hat{\hParameter}_{\tau}}:=\textrm{diag}\left(\{ 
\VIfvar{z_i}{i} \}_{i=1}^{N_g}\right)$; and subsequently, that of $H(\vfield|\vZfield,
\Vnshdat{\tau};\hat{\hParameter}_{\tau})$ by substituting~\eqref{eq:condEntropyGivenParametersVI} 
into~\eqref{eq:conditionalEntropy}. 

Similar to~\eqref{eq:condEntropyAtNextTimeSlot}, we then show in 
Appendix~\ref{sec:approxEntropyDerivation} that 
\begin{align}
&H(\vfield|\vZfield,\Vnshdat{\tau+1};\hat{\hParameter}_{\tau}) \approx H(\vfield|\vZfield,\Vnshdat{\tau}
;\hat{\hParameter}_{\tau}) \label{eq:ApproxCondEntropyAtNextTimeSlot} \\
&\hspace{-0.1cm} -\frac{1}{2}\sum_{\vZfield'\in\zSet}\int p(\vZfield',\Vnshdat{\tau}'
;\hat{\hParameter}_{\tau}) \ln\bigg| \bbI_{N_g}  + \MeanprecNVar\mDiagWeight{\tau+1} \mCondCovFVIwith{z',\Vnshdat{\tau}';\hat{\hParameter}_{\tau}}\bigg| d\Vnshdat{\tau}'\nonumber 
\end{align} 
where $\mDiagWeight{\tau+1}:=\textrm{diag}\left(\vWeight_{\tau+1}^{(n,n')} \circ \vWeight_{\tau+1}^{(n,n')}\right)$,
with $\circ$ denoting the Hadamard product. Given $\Vnshdat{\tau}$, and using the approximation  $p(\vZfield|\Vnshdat{\tau};\hat{\hParameter}_{\tau})\approx q(\vZfield)$, we can reformulate (P2) 
as (cf.~\eqref{eq:ApproxCondEntropyAtNextTimeSlot})
\begin{equation*}
	\textrm{(P2')}~~\max_{\stackrel{\vWeight_{\tau+1}^{(n,n')}:}{(n,n')\in \mathcal{M}_{\tau+1}}}  \underbrace{\sum_{i=1}^{N_g}\mathbb{E}_{q(\Zfield_i)}\left[\ln \left( 1 +  \MeanprecNVar\VIfvar{z_i}{i}w_{\tau+1,i}^2\right)\right]}_{  =:\bar{h}(\vWeight_{\tau+1}^{(n,n')})}.
\end{equation*}
Solving (P2') using a greedy search, we obtain the pair of sensors $(n^{*},n'^{*})$ 
associated with $\vWeight_{\tau+1}^{(n^{*},n'^{*})}$, based on which we collect the informative measurement $\nshd_{\tau+1}$.

The overall algorithm for adaptive radio tomography via VB is tabulated in Alg.~\ref{alg:AdapBayesCG}. 

\noindent {\remark \textbf{(Mini-batch setup).}}
The proposed data-adaptive sensor selection scheme can be easily 
extended to a mini-batch setup of size $\numBatch$ per time slot $\tau$ as follows: i) find 
weight vectors $\big\{\vWeight_{\tau+1}^{\left(n^{(m)},n'^{(m)}\right)}\big\}_{m=1}^{\numBatch}$ 
for $\big\{\left(n^{(m)},n'^{(m)}\right)\big\}_{m=1}^{\numBatch} \subset \mathcal{M}_{\tau+1}$
associated with $N_{\textrm{Batch}}$ largest values of~$\bar{h}(\vWeight_{\tau+1}^{(n,n')})$ in~(P2'),
and collect $\{\nshd_{\tau+1}^{(m)}\}_{m=1}^{\numBatch}$ from pairs of sensors revealed 
from those weight vectors (steps $4$--$5$ in Alg.~\ref{alg:AdapBayesCG}); and ii) 
aggregate those measurements below $\Vnshdat{\tau}$ to construct $\Vnshdat{\tau+1}:=[
\Vnshdat{\tau}^{\transpose},\nshd_{\tau+1}^{(1)},\ldots,\nshd_{\tau+1}^{(\numBatch)}]^{\transpose}$ 
(step $6$ in Alg.~\ref{alg:AdapBayesCG}). Numerical tests are presented next to 
assess the mini-batch operation of Alg.~\ref{alg:AdapBayesCG}. 

\begin{algorithm}[t]
	\caption{Adaptive radio tomography via variational Bayes}
	\begin{algorithmic}[1]
		\renewcommand{\algorithmicrequire}{\textbf{Input:}}
		\REQUIRE $\check{\mathbf{s}}^{(0)}$, $\mWeight^{(0)}$, $\bigg\{a_{\nu}, b_{\nu}, 
		\big\{\fMeanMean{k},\fVarVar{k},a_k,b_k\big\}_{k=1}^{K}\bigg\}$, and $\NumStep$.
		\STATE Set $\check{\mathbf{s}}_0=\check{\mathbf{s}}^{(0)}$ and 
		$\mWeight_0 = \mWeight^{(0)}$ 
		\FOR {$\tau = 0,1,\ldots$}
		\STATE Obtain $\hat{\vfield}_{\textrm{MMSE}}$, $\hat{\hParameter}_{\textrm{MMSE}}$, 
		and $q^{*}(\vfield,\vZfield,\hParameter)$
		via Alg.~\ref{alg:VB}$\left(\check{\mathbf{s}}_\tau,\mWeight_{\tau},\bigg\{a_{\nu}, b_{\nu}, 
		\big\{\fMeanMean{k},\fVarVar{k},a_k,b_k\big\}_{k=1}^{K}\bigg\},\NumStep\right)$
		\STATE Evaluate $\bar{h}(\vWeight_{\tau+1}^{(n,n')})$ in~(P2') $\forall \{n,n'\} \in \mathcal{M}_{\tau+1}$
		\STATE Collect $\nshd_{\tau+1}$ from $(n^{*},n'^{*})$ with $\max \bar{h}(\vWeight_{\tau+1}^{(n,n')})$
		\STATE Set $\Vnshdat{\tau+1}=[\Vnshdat{\tau}^{\transpose},\nshd_{\tau+1}]^{\transpose}$ and 
		$\mathbf{W}_{\tau+1} \hspace{-0.05cm}= \hspace{-0.05cm} [\mathbf{W}_{\tau},\mathbf{w}_{\tau+1}^{(n^{*},n'^{*})}]$
		\ENDFOR
		\RETURN $\hat{\vfield}_{\textrm{MMSE}}$ 
	\end{algorithmic} \label{alg:AdapBayesCG} 
\end{algorithm} 

\section{Numerical Tests}
\label{sec:numerical}
Performance of the proposed algorithms was assessed through numerical tests using $\MATLAB$ on synthetic and real datasets. Comparisons were carried out with existing methods, including the ridge-regularized SLF estimate given by $\hat{\vfield}_{\textrm{LS}}=(\mWeight_{t}\mWeight_{t}^{\transpose}+\RegWeight \bbC_{\field}^{-1})^{-1} \mWeight_{t}\Vnshd$~\cite{hamilton2014modeling}, where $\bbC_{\field}$ is a spatial covariance matrix modeling the similarity between points $\xgrd_i$ and $\xgrd_j$ in area $\interestRegion$.
We further tested the TV-regularized LS scheme in~\cite{QGM15ADMMTV}, which solves the problem in~\eqref{eq:regLS} with 
\begin{equation}
\Regularizer(\vfield)= \sum_{i=1}^{N_x-1}\sum_{j=1}^{N_y}|F_{i+1,j}-F_{i,j}| 
+ \sum_{i=1}^{N_x}\sum_{j=1}^{N_y-1}|F_{i,j+1}-F_{i,j}|
\end{equation}
where $\mfield := \unvec(\vfield)\in\mathbb{R}^{N_x \times N_y}$ and $F_{i,j}:=[\mfield]_{i,j}$.
We also tested an MCMC-based counterpart of Alg.~\ref{alg:AdapBayesCG} for estimating the posterior in~\eqref{eq:BayesPosterior}, and solving (P2) in~\eqref{eq:P2}; see e.g.,
\cite{dbg18adaptiveRT,pereyra13GranularityEstimation} for details.

We further compared the proposed data-adaptive sensor selection with simple \emph{random sampling} for both regularized LS estimators, by selecting $\big\{\left(n^{(m)},n'^{(m)}\right)\big\}_{m=1}^{\numBatch}$ 
uniformly at random to collect $\{\nshd_{\tau+1}^{(m)} \}_{m=1}^{\numBatch}~\forall \tau$. 
Alg.~\ref{alg:AdapBayesCG} after replacing steps $4$--$5$ with random sampling is termed \emph{non-adaptive VB algorithm}, and will be compared with the proposed method throughout synthetic and real data tests.

\begin{figure}[t]
	\qquad
	\begin{minipage}[h]{.4\linewidth}
		\centering
		\includegraphics[width=\linewidth]{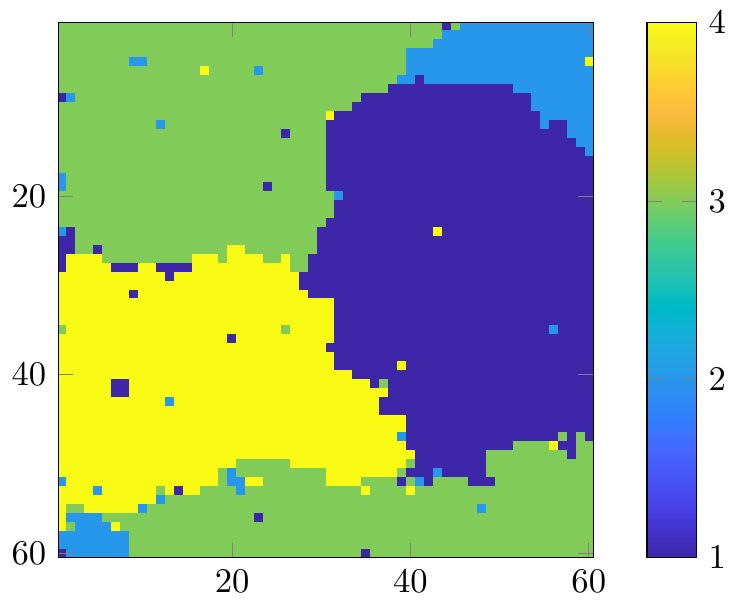}
		\vspace{-10pt}
		\subcaption{}
		\label{fig:true_z_0}
	\end{minipage} 
	\quad 
	\begin{minipage}[h]{.4\linewidth}
		\centering
		\includegraphics[width=\linewidth]{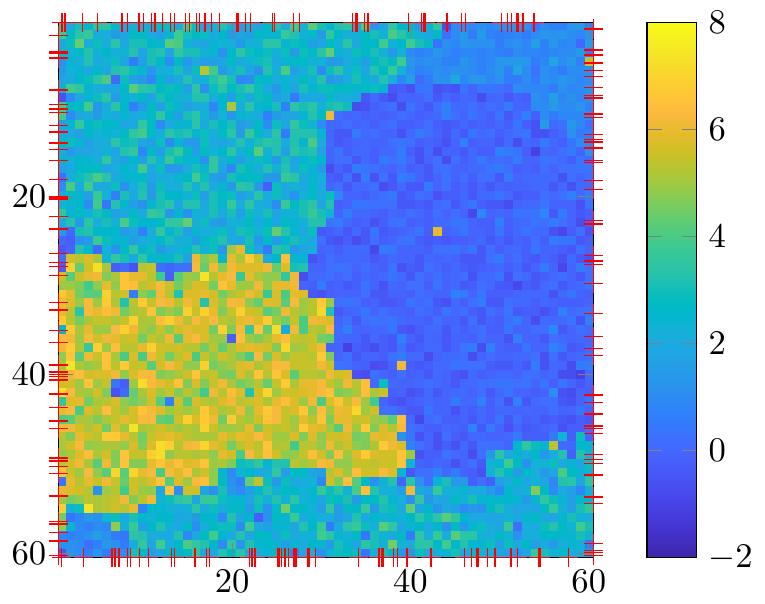}
		\vspace{-10pt}
		\subcaption{}
		\label{fig:true_f_0}
	\end{minipage} 
	\caption{True fields for synthetic tests: (a) hidden label field $\mZfield_{0}$ 
		and (b) spatial loss field $\mfield_{0}$ with $N=200$ sensor locations 
		marked with crosses.} \label{fig:trueSLFandHidden}
\end{figure}

\subsection{Test with synthetic data}
This section validates the proposed algorithm using synthetic datasets. Random tomographic measurements were collected from $N=200$ sensors uniformly deployed on the boundary of $\mathcal{A} := [0.5, 60.5] \times [0.5, 60.5]$. Using these measurements, the SLF was reconstructed over the grid $\{\xgen_{i}\}_{i=1}^{3,600}:=\{1,\ldots,60\}^2$ . To generate the ground-truth SLF $\vfield_0$, the ground-truth label field $\vZfield_0$ was generated via Gibbs sampling~\cite{Geman84GibbsSampler} by using the Potts prior of $\vZfield$ in~\eqref{eq:PottsPrior} with $\granPara=1.5$ and $K=4$. Given 
$\hParameter_f := [\vfMean^{\transpose},\vfPrec^{\transpose}]^{\transpose}$
with $\vfMean=[0,1,2.5,5.5]^{\transpose}$ and $\vfPrec=[10,10,2,2]^{\transpose}$, 
vector $\vfield_0$ was constructed to have $f(\xgen_i)\sim\NormDist{\fMean{k}}{\fprec{k}^{-1}}
~\forall \xgen_i \in \mathcal{A}_k,~\forall k$ conditioned on the labels in $\vZfield_0$. The resulting 
hidden label field $\mZfield_{0}:=\unvec(\vZfield_0)\in\{1,2,3,4\}^{60\times 60}$, and the true SLF $\mfield_{0}:=\unvec(\vfield_0)\in\mathbb{R}^{60\times 60}$ are depicted in Fig.~\ref{fig:trueSLFandHidden} with sensor locations marked by crosses. The effects of calibration are not accounted for, meaning that $\uGain$ and $\pathlossexp$ are assumed to be known, and the fusion center directly uses shadowing measurements $\Vnshdat{\tau}$. Under the mini-batch operation, each measurement $\nshd_{\tau}^{(m)}~\forall \tau,m$ was generated according to~\eqref{eq:calibratedMea}, where $\shd_{\tau}$ was obtained using \eqref{eq:aproxSLFmodel} with $w$ set to the normalized ellipse model in~\eqref{eq:ellipse_model} with $\normEllipMargin=0.39$, while $\noise_\tau$ was set to follow a zero-mean Gaussian pdf with $\precNVar=20$. To construct $\mSet_{\tau+1}$ per time slot $\tau$,  $|\mSet_{\tau+1}|=200$ pairs of sensors were uniformly selected at random with replacement. Then, $\numBatch=100$ shadowing measurements were collected at  $\big\{\left(n^{(m)},n'^{(m)}\right)\big\}_{m=1}^{\numBatch} \subset \mSet_{\tau+1}$ to run Alg.~\ref{alg:AdapBayesCG} for $\tau = 0,1,\ldots,8$.

In all synthetic tests, the simulation parameters were set to $\NumStep=3,000$ and $\xi = 10^{-6}$; 
hyper-hyper parameters of $\noise_t$ were set to $a_{\nu}=1,300$ and $b_{\nu}=2$; and 
those of $\hParameter_{f}$ were set as listed in Table.~\ref{tab:parameter_setup_synthetic}.
To execute Alg.~\ref{alg:VB}, variational parameters of $q^{(0)}(\vfield,\vZfield,\hParameter)$ 
were initialized as follows:
$\left\{\VIfmeanItr{k}{i}{0}\right\}_{i=1}^{N_g}~\forall k$, $\VIbNVarItr{0}$, 
$\left\{\VIfMeanVarItr{k}{0}\right\}_{k=1}^{4}$, and 
$\left\{\VIfprecaItr{k}{0},\VIfprecbItr{k}{0}\right\}_{k=1}^{4}$ were drawn 
from the uniform distribution $\mathcal{U}(0,1)$,
while $\VIfMeanMeanItr{k}{0} = m_k~\forall k$; and it was set to 
$\VIzfieldProbItr{k}{\xgrd_{i}}{0}=1/4~\forall i,k$.
Furthermore, $\VnshdIni$ was collected from $800$ pairs of sensors selected at random,
which determined $\mWeight^{(0)}$. 
To find $\RegWeight$ of the competing alternatives, the L-curve~\cite[Chapter 26]{lawson74Lcurev} 
was used for the ridge regularization, while the generalized cross-validation~\cite{golub79gcv} 
was adopted for the TV regularization. The hyper-hyper parameters of $\hParameter$
used for the proposed algorithm were also adopted to run its MCMC-based counterpart.

\begin{figure}[t!]
	\begin{minipage}[h]{.24\linewidth}
		\centering
		\includegraphics[width=\linewidth]{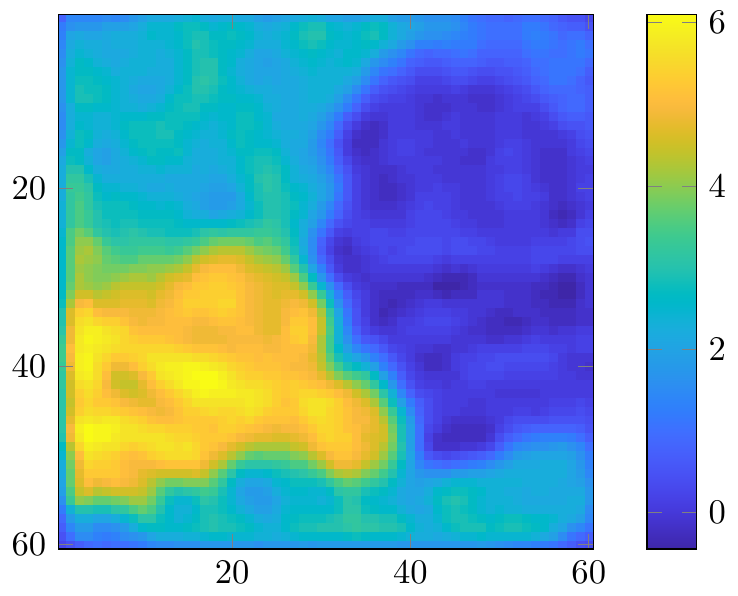}
		\vspace{-10pt}
		\subcaption{}
		\label{fig:Ridge_LS_syn}
	\end{minipage}  
	\begin{minipage}[h]{.24\linewidth}
		\centering
		\includegraphics[width=\linewidth]{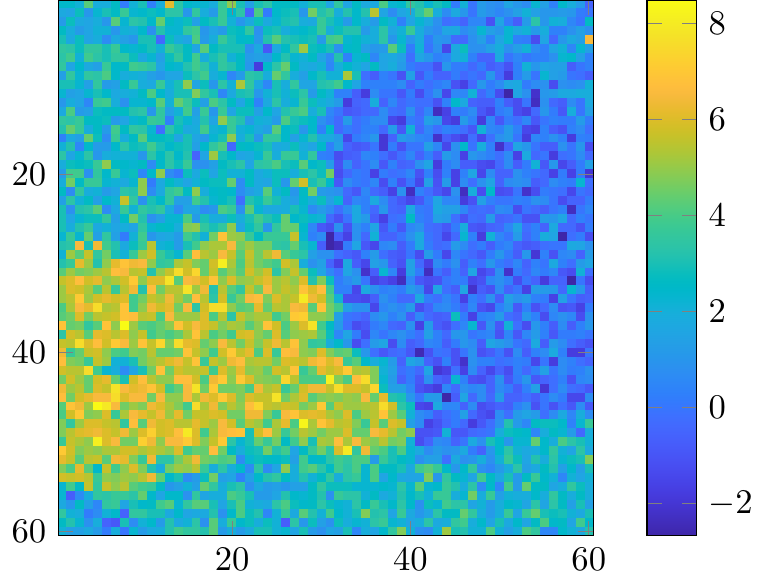}
		\vspace{-10pt}
		\subcaption{}
		\label{fig:TV_LS_syn}
	\end{minipage} 
	\begin{minipage}[h]{.24\linewidth}
		\centering
		\includegraphics[width=\linewidth]{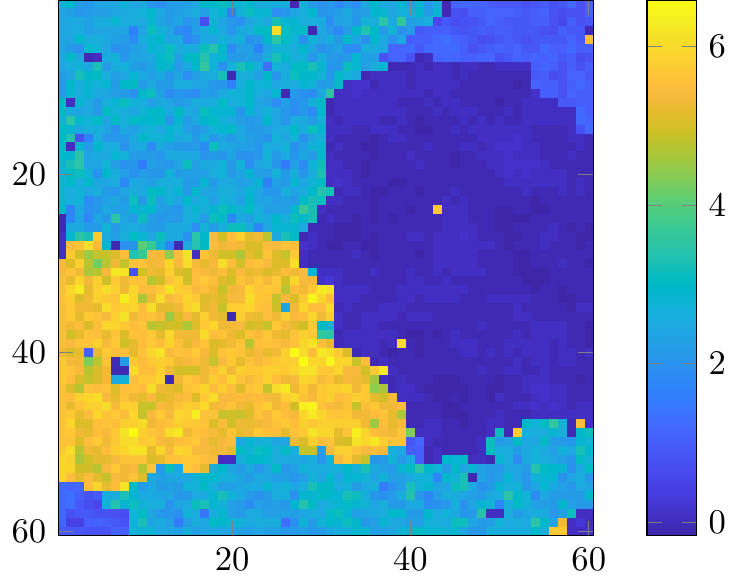}
		\vspace{-10pt}
		\subcaption{}
		\label{fig:VB_adap_syn}
	\end{minipage}  
	\begin{minipage}[h]{.24\linewidth}
		\centering
		\includegraphics[width=\linewidth]{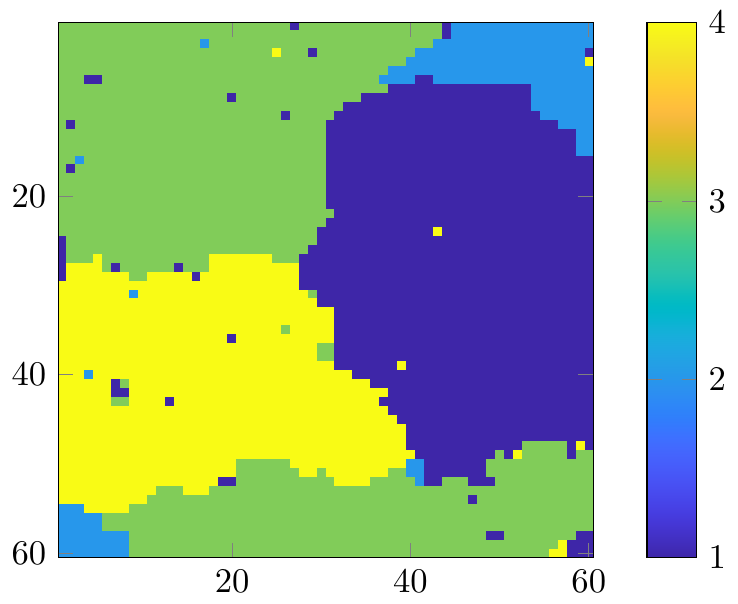}
		\vspace{-10pt}
		\subcaption{}
		\label{fig:VB_adap_hidden_syn}
	\end{minipage} 

	\begin{minipage}[h]{.24\linewidth}
		\centering
		\includegraphics[width=\linewidth]{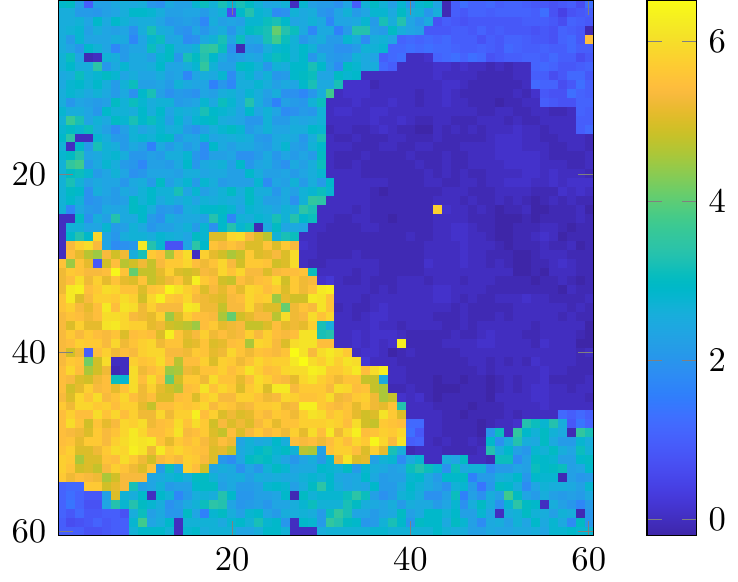}
		\vspace{-10pt}
		\subcaption{}
		\label{fig:VB_random_syn}
	\end{minipage}  
	\begin{minipage}[h]{.24\linewidth}
		\centering
		\includegraphics[width=\linewidth]{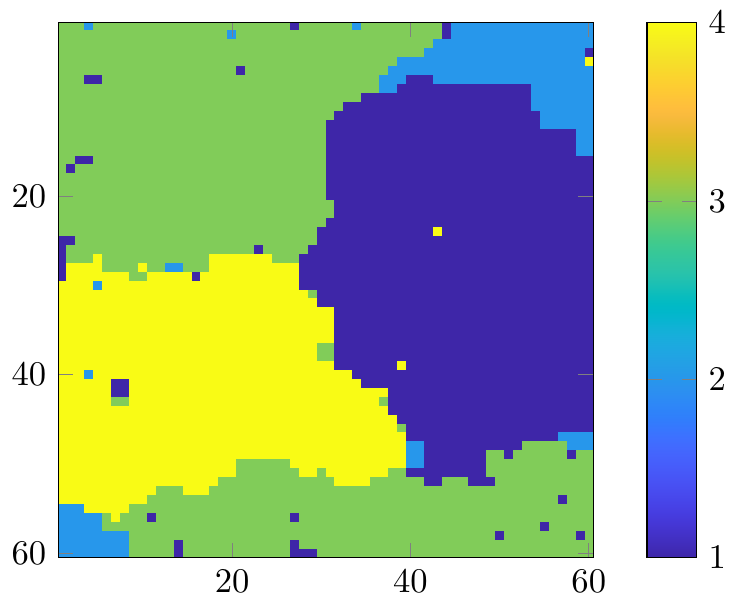}
		\vspace{-10pt}
		\subcaption{}
		\label{fig:VB_random_hidden_syn}
	\end{minipage} 
	\begin{minipage}[h]{.24\linewidth}
		\centering
		\includegraphics[width=\linewidth]{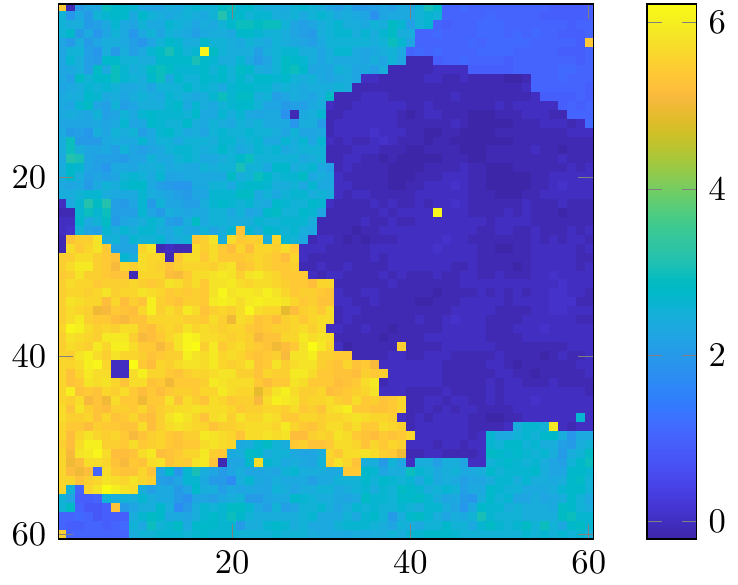}
		\vspace{-10pt}
		\subcaption{}
		\label{fig:MCMC_adap_syn}
	\end{minipage}  
	\begin{minipage}[h]{.24\linewidth}
		\centering
		\includegraphics[width=\linewidth]{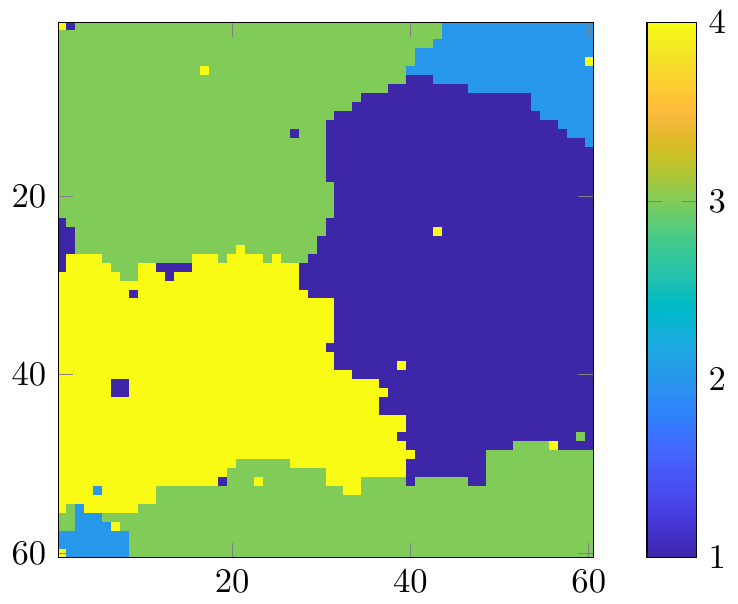}
		\vspace{-10pt}
		\subcaption{}
		\label{fig:MCMC_adap_hidden_syn}
	\end{minipage} 

	\begin{minipage}[h]{.24\linewidth}
		\centering
		\includegraphics[width=\linewidth]{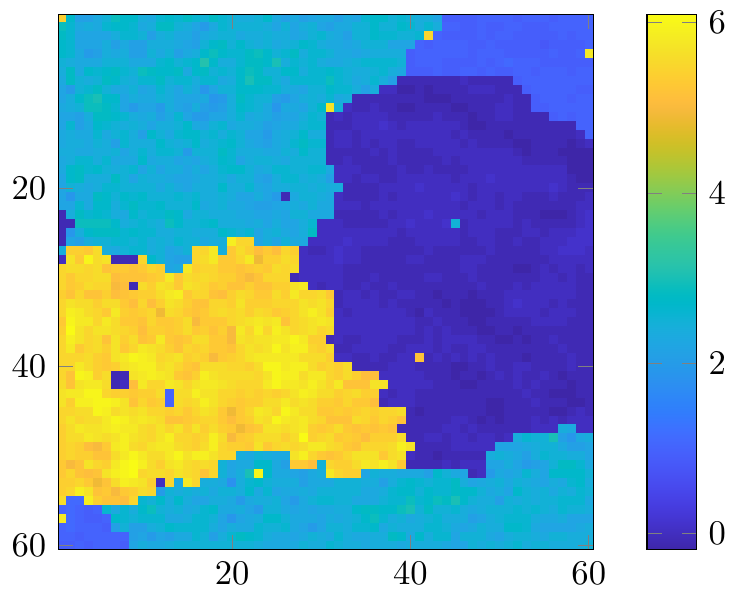}
		\vspace{-10pt}
		\subcaption{}
		\label{fig:MCMC_random_syn}
	\end{minipage}  
	\begin{minipage}[h]{.24\linewidth}
		\centering
		\includegraphics[width=\linewidth]{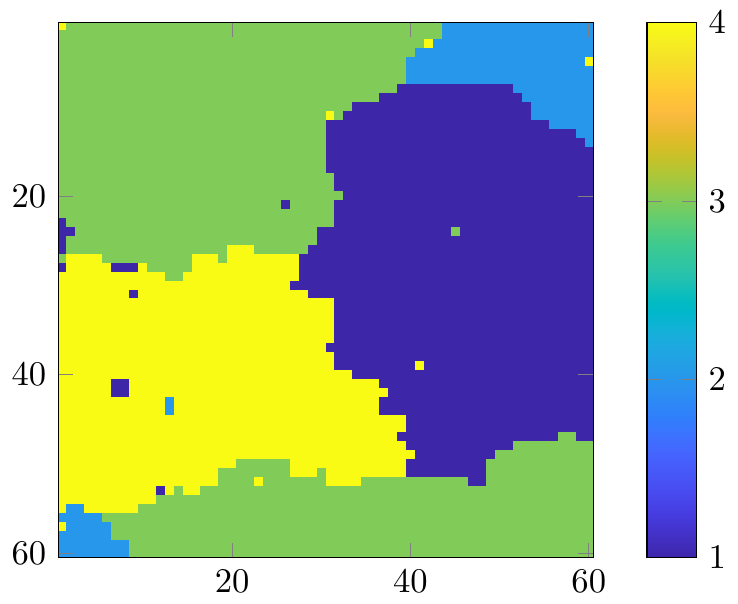}
		\vspace{-10pt}
		\subcaption{}
		\label{fig:MCMC_random_hidden_syn}
	\end{minipage}
	\begin{minipage}[h]{.24\linewidth}
		\centering
		\includegraphics[width=\linewidth]{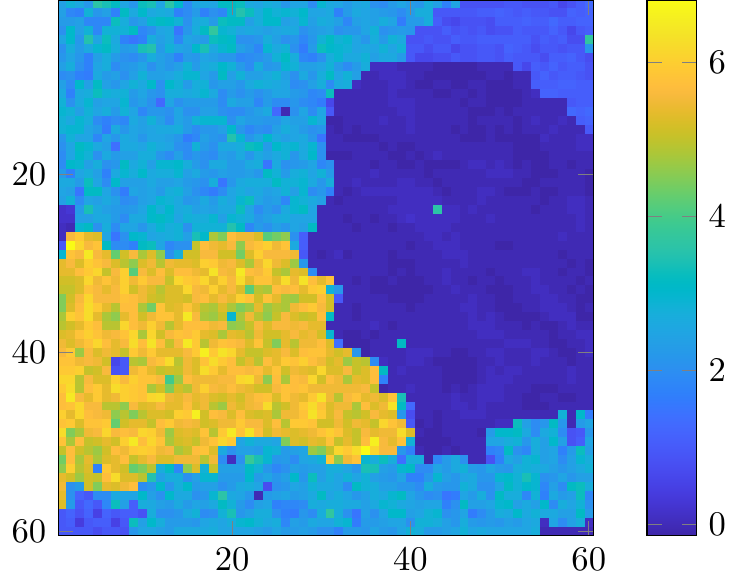}
		\vspace{-10pt}
		\subcaption{}
		\label{fig:VB_full_syn}
	\end{minipage}  
	\begin{minipage}[h]{.24\linewidth}
		\centering
		\includegraphics[width=\linewidth]{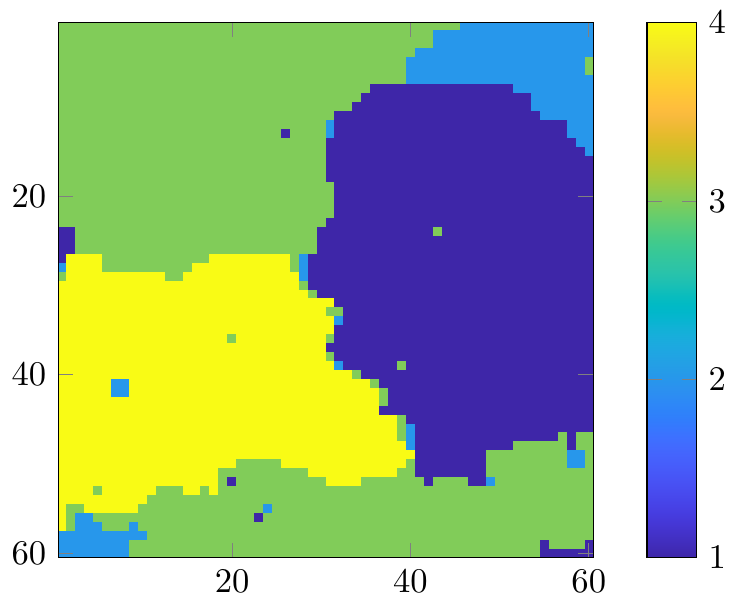}
		\vspace{-10pt}
		\subcaption{}
		\label{fig:VB_full_hidden_syn}
	\end{minipage}  
	\caption{SLF estimates $\hat{\mfield}$ at $\tau=8$ (with $1,600$ measurements) via; 
		(a) ridge-regularized LS ($\RegWeight=0.015$ and
		$\bbC_{\field}=\bbI_{3,600}$); (b) TV-regularized LS ($\RegWeight=10^{-11}$);
		(c) Alg.~\ref{alg:AdapBayesCG} through (d) estimated hidden field $\hat{\mZfield}$; 
		(e) non-adaptive VB algorithm through (f) $\hat{\mZfield}$; 
		(g) adaptive MCMC algorithm through (h) $\hat{\mZfield}$;
		(i) non-adaptive MCMC algorithm through (j) $\hat{\mZfield}$; and
		(k) $\hat{\mfield}_{\textrm{full}}$ and (l) $\hat{\mZfield}_{\textrm{full}}$ 
		obtained by using the full data
		(with $2,400$ measurements) via Alg.~\ref{alg:AdapBayesCG}.} \label{fig:est_Synthetic_SLFs}
\end{figure}

The first experiment is performed to validate Alg.~\ref{alg:AdapBayesCG}. Estimates of SLFs $\hat{\mfield}:=\unvec(\hat{\vfield})$ and the associated hidden label fields 
$\hat{\mZfield}:=\unvec(\hat{\vZfield})$ at time slot {$\tau=8$ obtained via Alg.~\ref{alg:AdapBayesCG}, and the competing alternatives, are depicted in Figs.~\ref{fig:Ridge_LS_syn}}--\ref{fig:MCMC_random_hidden_syn}. One-shot estimates of the SLF and associated hidden field, denoted as $\hat{\mfield}_{\textrm{full}}$ and $\hat{\mZfield}_{\textrm{full}}$, respectively, are also displayed in Figs.~\ref{fig:VB_full_syn}
and~\ref{fig:VB_full_hidden_syn}, which were obtained via Alg.~\ref{alg:AdapBayesCG} by using the entire set of $2,400$ measurements collected till $\tau=8$. Clearly, satisfactory results were obtained only by
teh approximate Bayesian inference methods including MCMC and VB because every piecewise homogeneous region was accurately classified through the hidden label field. As discussed in Remark $2$ however, the proposed algorithm is computationally much more efficient than the ones using MCMC. Per-iteration execution time was $0.04$ (sec) for Alg.~\ref{alg:AdapBayesCG} on average, while that was $3.64$ (sec) for the MCMC method. On the other hand, the regularized LS solutions were unable to accurately reconstruct the SLF, as depicted in Figs.~\ref{fig:Ridge_LS_syn} and~\ref{fig:TV_LS_syn}.

\begin{figure}[t!]
	\begin{minipage}[h]{.49\linewidth}
		\centering
		\includegraphics[width=\linewidth]{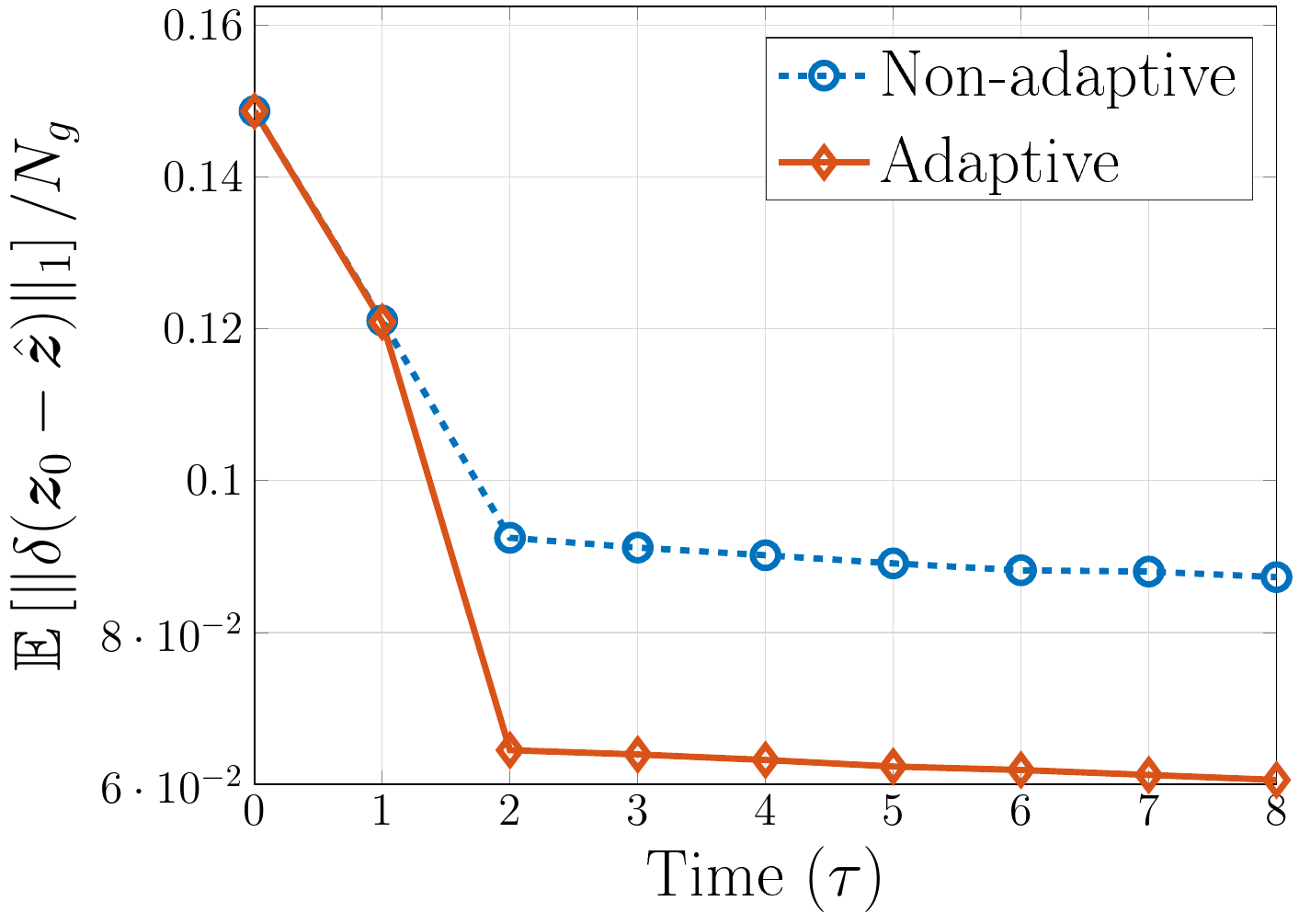}
		\vspace{-10pt}
		\subcaption{}
		\label{fig:ReconErrSynthetic}
	\end{minipage}  
	\begin{minipage}[h]{.49\linewidth}
		\centering
		\includegraphics[width=\linewidth]{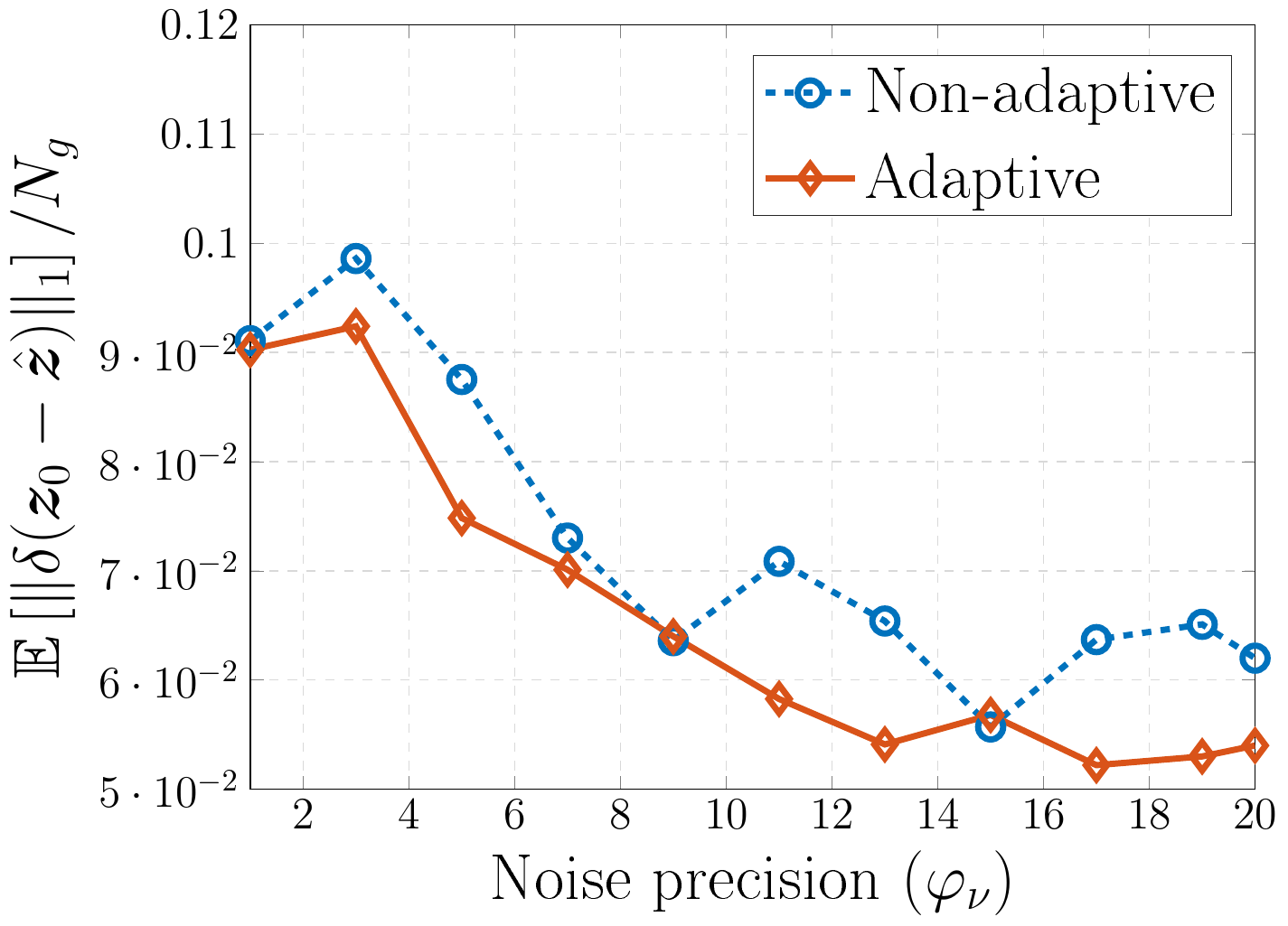}
		\vspace{-10pt}
		\subcaption{}
		\label{fig:ReconErrSyntheticRobust}
	\end{minipage}  
	\caption{Progression of estimation error of $\vZfield$ versus (a) time $\tau$; and (b) noise precision $\precNVar$, averaged over 20 MC runs.}\label{fig:ReconErrs}
\end{figure}

To test the proposed sensor selection method, $\hat{\mfield}$ and $\hat\mZfield$ found using the non-adaptive VB algorithm are depicted in Figs.~\ref{fig:VB_random_syn} and~\ref{fig:VB_random_hidden_syn}. Visual comparison of Figs.~\ref{fig:VB_adap_syn} and~\ref{fig:VB_random_syn} reveals that the reconstruction performance for $\mfield$ can be improved with the same number of measurements by adaptively selecting pairs of sensors. Accuracy of $\hat{\vZfield}$ was also quantitatively measured by the
labeling-error, defined using the entrywise Kronecker delta $\delta(\cdot)$, as $\| \delta(\vZfield_0 - \hat{\vZfield}) \|_{1}/N_g$. Progression of the labeling error averaged over $20$ Monte Carlo (MC) runs is 
displayed in Fig.~\ref{fig:ReconErrSynthetic}, where the proposed method consistently outperforms the non-adaptive one. This shows that informative measurements adaptively collected to decrease uncertainty of $\vfield$ given a current estimate of $\hParameter$ improve accuracy of $\hat \vfield$ and $\hat \vZfield$ in the next time slot. As a result, the SLF reconstruction accuracy of Alg.~\ref{alg:AdapBayesCG} improves accordingly with fewer measurements, as confirmed by comparing Figs.~\ref{fig:VB_adap_syn} 
and~\ref{fig:VB_full_syn}. 

The next experiment tests robustness of the proposed algorithms against measurement noise $\noise_{\tau}$. 
We adopted the labeling-error for $\vZfield$ averaged over sensor locations and realizations of $\{\noise_{\tau}\}_{\tau}^{t}$ to quantify the reconstruction performance. Fig.~\ref{fig:ReconErrSyntheticRobust} shows the progression of the labeling 
error at $\tau = 8$ as a function of the noise precision $\precNVar$ averaged over 20 MC runs. 
Note that Figs.~\ref{fig:VB_adap_hidden_syn} and~\ref{fig:VB_random_hidden_syn} correspond to the rightmost point of the x-axis of Fig.~\ref{fig:ReconErrSyntheticRobust}. Clearly, the reconstruction performance does not severely decrease as $\precNVar$ decreases, or equivalently $\nVar$ increases. This confirms that the proposed algorithm is reasonably robust against measurement noise.

Averaged estimates of $\hParameter$ and associated standard deviation denoted with $\pm$ are listed in Table~\ref{tab:parameter_est_syn}. Together with Fig.~\ref{fig:est_Synthetic_SLFs}, the high estimation accuracy of hyperparameters implies that the proposed method can effectively reveal patterns of objects in $\mathcal{A}$
by correctly inferring the underlying statistical properties of each piecewise homogeneous region
in the SLF. Note that $\vfPrec$ entries are overestimated in Table~\ref{tab:parameter_est_syn}. 
This can be intuitively understood in the sense that minimizing the KL divergence in~\eqref{eq:KL} 
leads to $q(\hParameter_f)$ avoiding regions in which $p(\vfield|\vZfield,\hParameter_f)p(\hParameter_f)$ 
is small by setting each $\fprec{k}$ to a large value $\forall k$, which corroborates the result in~\cite[p.~468]{bishop2006}. 

\begin{table}[t]
	\centering \caption{Hyper-parameters of $\hParameter_f$ for synthetic data tests.}
	\begin{tabular}{|c|c|c|c|c|c|c|c|}
		\hline
		$\fMeanMean{1}$ & $\fMeanMean{2}$ & $\fMeanMean{3}$ & $\fMeanMean{4}$ & $\fVarVar{1}$ & $\fVarVar{2}$ 
		& $\fVarVar{3}$ & $\fVarVar{4}$ \\ \hline
		$0$ & $0.9$  & $2.7$ & $5.3$ & $10^{-4}$ & $10^{-4}$ & $10^{-4}$ & $10^{-4}$   \\ \hline \hline
		$a_{1}$ & $a_{2}$ & $a_{3}$ & $a_{4}$ & $b_{1}$ & $b_{2}$ 
		& $b_{3}$ & $b_{4}$ \\ \hline
		$0.8$ & $0.8$  & $0.8$ & $0.8$ & $1$ & $1$ & $0.5$ & $0.5$   \\ \hline
	\end{tabular}%
	\label{tab:parameter_setup_synthetic}
\end{table}

\begin{table}[t!]
	\centering \caption{True $\hParameter$ and estimated $\hat{\hParameter}$ via Alg.~\ref{alg:AdapBayesCG} 
		(setting of Fig.~\ref{fig:VB_adap_syn}); 
		and non-adaptive VB algorithm (setting of Fig.~\ref{fig:VB_random_syn}) 
		averaged over 20 independent MC runs.}
	\begin{tabular}{|c|c|c|c|}
		\hline
		$\hParameter$ & True & Est. (Alg.~\ref{alg:AdapBayesCG}) & Est. (non-adaptive)  \\ \hline 
		$\precNVar$ & $20$   & $18.329 \pm 6\times10^{-3}$   & $18.461 \pm 4.6\times10^{-3}$\\ \hline
		$\fMean{1}$ & $0$   & $0.022\pm 1.2\times10^{-2} $   & $0.018 \pm 1.9\times10^{-2}$ \\ \hline
		$\fMean{2}$ & $1$   & $0.957 \pm 1.7\times10^{-2}$    & $0.962 \pm 1.6\times10^{-2}$ \\ \hline		
		$\fMean{3}$ & $2.5$   & $2.573\pm 1.7\times10^{-2} $   & $2.578 \pm 2.6\times10^{-2}$ \\ \hline
		$\fMean{4}$ & $5.5$   & $5.399 \pm 2.7\times10^{-2}$    & $5.374 \pm 7\times10^{-3}$ \\ \hline		
		$\fprec{1}$ & $10$   & $40.178 \pm 3\times10^{-3}$    & $42.352 \pm 2\times10^{-3}$ \\ \hline
		$\fprec{2}$ & $10$   & $14.634 \pm 1.4\times 10^{-2}$  & $15.845 \pm 1.2\times10^{-2}$ \\ \hline
		$\fprec{3}$ & $2$   & $7.712 \pm 2.7\times10^{-2}$    & $7.493 \pm 2.2\times10^{-2}$ \\ \hline
		$\fprec{4}$ & $2$   & $4.620 \pm 6.1\times 10^{-2}$  & $5.451 \pm 4\times10^{-2}$ \\ \hline
	\end{tabular}%
	\label{tab:parameter_est_syn}
\end{table}

Next, we will validate the efficacy of the proposed algorithms for channel-gain cartography using the setup of Fig.~\ref{fig:est_Synthetic_SLFs}. From the estimate $\hat {\bm f}_{\textrm{MMSE}}$ obtained through Alg.~\ref{alg:AdapBayesCG}, we found the shadowing attenuation $\hat{s}(\xgen,\xgen')$ between two arbitrary points $\xgen$ and $\xgen'$ in $\interestRegion$ using~\eqref{eq:aproxSLFmodel} after replacing ${\bm f}$ with $\hat {\bm f}_{\textrm{MMSE}}$. Subsequently, we obtained the estimated channel-gain $\hat{g}(\xgen,\xgen')$ after substituting $\hat{s}(\xgen,\xgen')$ into~\eqref{eq:cg}.

Since  $\uGain$ and $\pathlossexp$ are known, obtaining $s({\xgen, \xgen'})$ is equivalent to finding
$g(\xgen, \xgen')$; cf.~\eqref{eq:cg}. This suggests adopting a
performance metric quantifying the mismatch between $s(\xgen,\xgen')$ and
$\hat{s}(\xgen,\xgen')$, using the normalized mean-square error 
\begin{equation}
\textrm{NMSE}:= \frac{ \mathbb{E}\left[ \int_{\interestRegion}\big(s(\xgen,\xgen')
	- \hat{s}(\xgen,\xgen')\big)^{2}d\xgen d\xgen' \right]}
{\mathbb{E}\left[ \int_\interestRegion s^2(\xgen,\xgen') d\xgen
	d\xgen'\right]} \label{eq:NMSE}
\end{equation}
where the expectation is over the set
$\{\xgen_{n}\}_{n=1}^N$ of sensor locations and realizations of
$\{\noise_\tau\}_{\tau}$. The integrals are approximated by averaging the
integrand over $500$ pairs of $(\xgen,\xgen')$ chosen independently and
uniformly at random on the boundary of $\interestRegion$. The expectations are
estimated by averaging simulated deviates over $20$ MC runs. 

\begin{figure}[t!]
	\centering
	\includegraphics[width=0.6\linewidth]{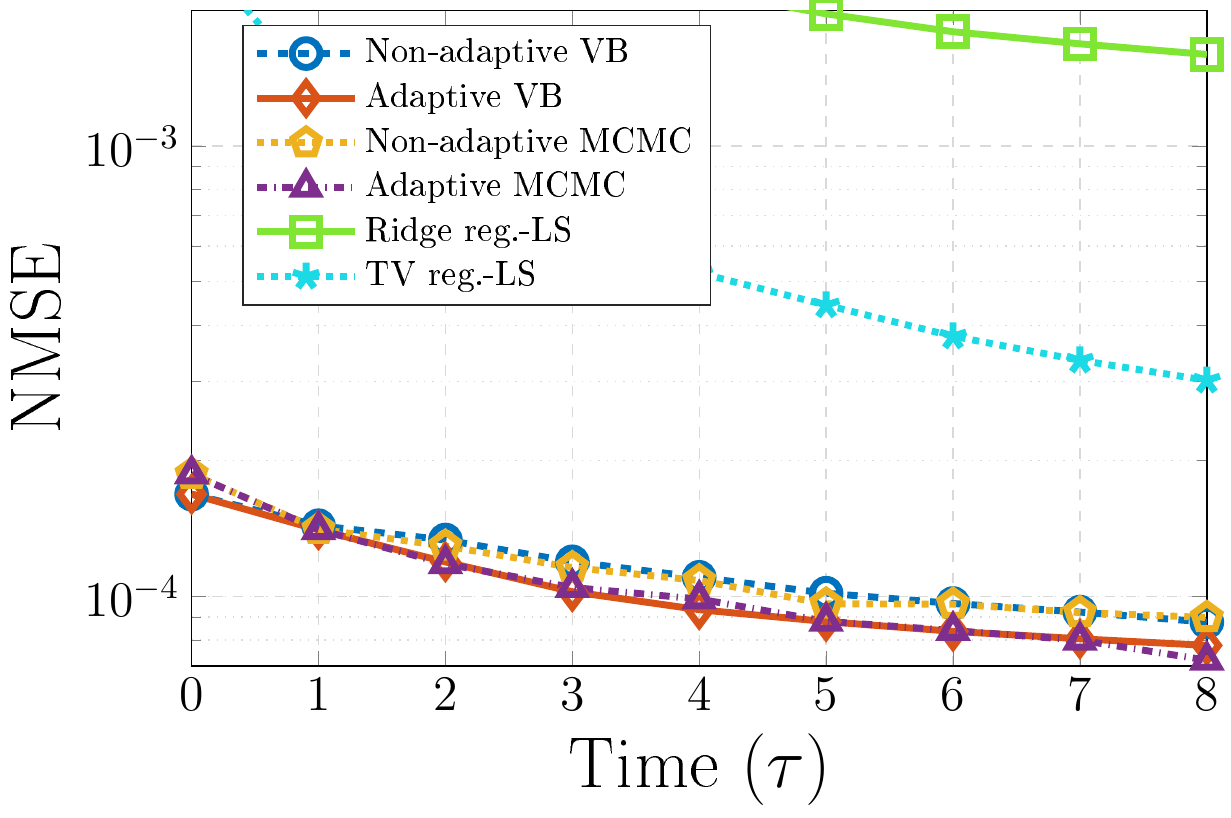}
	\caption{Progression of channel-gain estimation error.}\label{fig:CGErrSynthetic}
\end{figure}

Fig.~\ref{fig:CGErrSynthetic} depicts the NMSE of the proposed method and those of competing alternatives. 
Clearly, the approximate Bayesian inference methods outperform the regularized LS solutions. Furthermore, the performance of the VB methods is comparable to those of the MCMC methods. Noticeably, the adaptive VB method consistently exhibits lower NMSE than both non-adaptive ones, which highlights the efficacy in estimating channel-gain via the data-adaptive sensor selection. This suggests that the proposed VB framework is a 
viable solution for both radio tomography and channel-gain cartography, while enjoying low computational complexity.     

\subsection{Test with real data}
\label{sec:real_test}

\begin{figure}[t!]
	\centering
	\includegraphics[width=.45\linewidth]{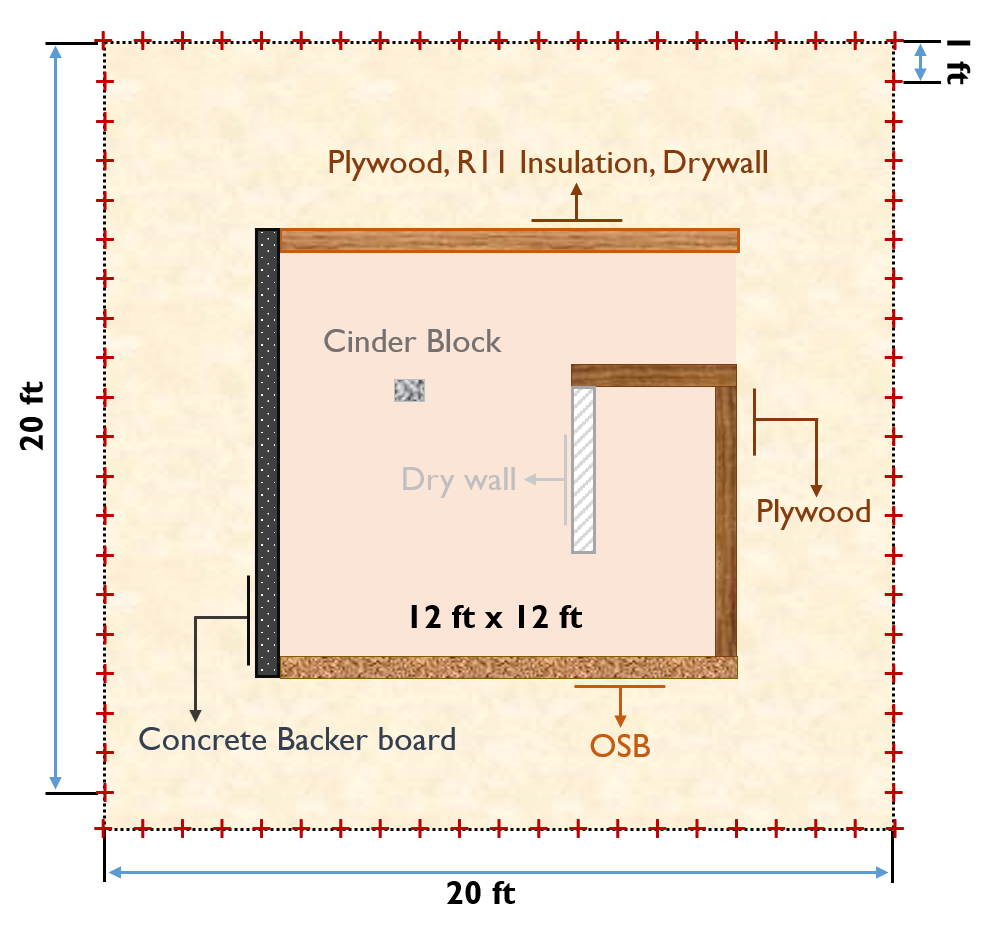}
	\caption{Configuration of the testbed with $N=80$ sensor locations marked with red crosses.}\label{fig:testbed}
\end{figure}

This section validates the proposed method using the real
dataset in~\cite{hamilton2014modeling}. The test setup is depicted in Fig.~\ref{fig:testbed}, where
$\interestRegion=[0.5,20.5]\times[0.5,20.5]$ is a square with sides of 20 feet (ft), over which a grid $\{\xgrd_i\}_{i=1}^{3,721}:=\{1,\ldots,61\}^2$
of $N_g = 3,721$ points is defined. A collection of $N=80$ sensors  
measure the RSS at $2.425$~GHz between pairs of sensor positions, marked with the
$N=80$ crosses in Fig.~\ref{fig:testbed}. To estimate $\uGain$ and $\pathlossexp$ 
using the approach in~\cite{hamilton2014modeling}, a
first set of $2,400$ measurements was obtained before placing objects.
Estimates $\hat{g}_0=54.6$ (dB) and $\hat\pathlossexp=0.276$ were obtained during the 
calibration phase. Afterwards, the structure comprising one pillar and six walls of 
different materials was assembled as shown in Fig.~\ref{fig:testbed}, and $T=2,380$ measurements
$\{\check{g}_{\tau'}\}_{\tau'=1}^T$ were collected. Calibrated measurements $\{\nshd_{\tau'}\}_{\tau'=1}^T$
were then obtained from $\{\check{g}_{\tau'}\}_{\tau'=1}^T$ after substituting $\hat{g}_0$ and 
$\hat{\pathlossexp}$ into~\eqref{eq:calibratedMea}. The weights $\{\vWeight_{\tau'}^{(n,n')}\}_{\tau'=1}^{T}$ 
were constructed with $w$ in~\eqref{eq:ellipse_model} by using known 
locations of sensor pairs. Note that $\tau'$ is introduced to distinguish indices 
of the real data from $\tau$ used to index time slots in numerical tests.

We randomly selected $1,380$ measurements
from $\{\nshd_{\tau'}\}_{\tau'=1}^T$ to initialize $\VnshdIni$ and $\mWeight^{(0)}$, and used
the remaining $1,000$ measurements to run the proposed algorithm under the mini-batch operation for $\tau=0,1,\ldots,5$.
At each time slot $\tau$, $\mSet_{\tau+1}$ was formed by sensors corresponding to  
$|\mSet_{\tau+1}| = 200$ weight vectors uniformly selected at random from $\{\vWeight_{\tau'}^{(n,n')}\}_{\tau'}$ 
associated with the remaining $1,000$ measurements
without replacement. Then, $\numBatch=100$ measurements were chosen from $\{\nshd_{\tau'}\}_{\tau'}$
associated with $\mSet_{\tau+1}$.

Simulation parameters were set to $\NumStep=3,000$, $\xi = 10^{-6}$, and $K=3$; and
hyper-hyper parameters of $\hParameter$ were set to $a_{\nu}=b_{\nu}=10^{-3}$, 
$[m_1,m_2,m_3]^{\transpose}=[0,0.035,0.05]^{\transpose}$,
$\fVarVar{k}=10^{-4}~\forall k$, and $a_k=b_k=0.1~\forall k$, respectively. 
To execute Alg.~\ref{alg:VB}, variational parameters of $q^{(0)}(\vfield,\vZfield,\hParameter)$ 
were initialized as follows: $\left\{\VIfmeanItr{k}{i}{0}\right\}_{i=1}^{N_g}~\forall k$, $\VIbNVarItr{0}$, $\left\{\VIfMeanVarItr{k}{0}\right\}_{k=1}^{3}$, and $\left\{\VIfprecaItr{k}{0},\VIfprecbItr{k}{0}\right\}_{k=1}^{3}$ 
were drawn from the uniform distribution $\mathcal{U}(0,1)$,
while $\VIfMeanMeanItr{k}{0} = m_k~\forall k$ and $\VIzfieldProbItr{k}{\xgrd_{i}}{0}=1/3~\forall i,k$.

Following~\cite{AgP09,hamilton2014modeling}, a spatial covariance matrix was used for $\bbC_{\field}$ of the ridge-regularized LS estimator, which models the similarity between points $\xgrd_{i}$, and $\xgrd_{j}$ as $\big[\bbC_{\field}\big]_{ij} = {\sigma^{2}_{s}}\exp[-{\|\xgrd_{i} - \xgrd_{j}\|_{2}}/{\kappa}]$~\cite{AgP09} with $\sigma_{s}^{2}=\kappa=1$, and $\RegWeight=0.015$ found with the L-curve~\cite[Chapter 26]{lawson74Lcurev}. For the TV-regularized LS estimator, it was set to $\RegWeight=6$ found through the 
generalized cross validation~\cite{golub79gcv}. To assess the efficacy of our Bayesian model
with the $K$-ary hidden label field, we tested the adaptive MCMC method in~\cite{dbg18adaptiveRT} 
with $K=2$. 

\begin{figure}[t!]
	\begin{minipage}[h]{.24\linewidth}
		\centering
		\includegraphics[width=\linewidth]{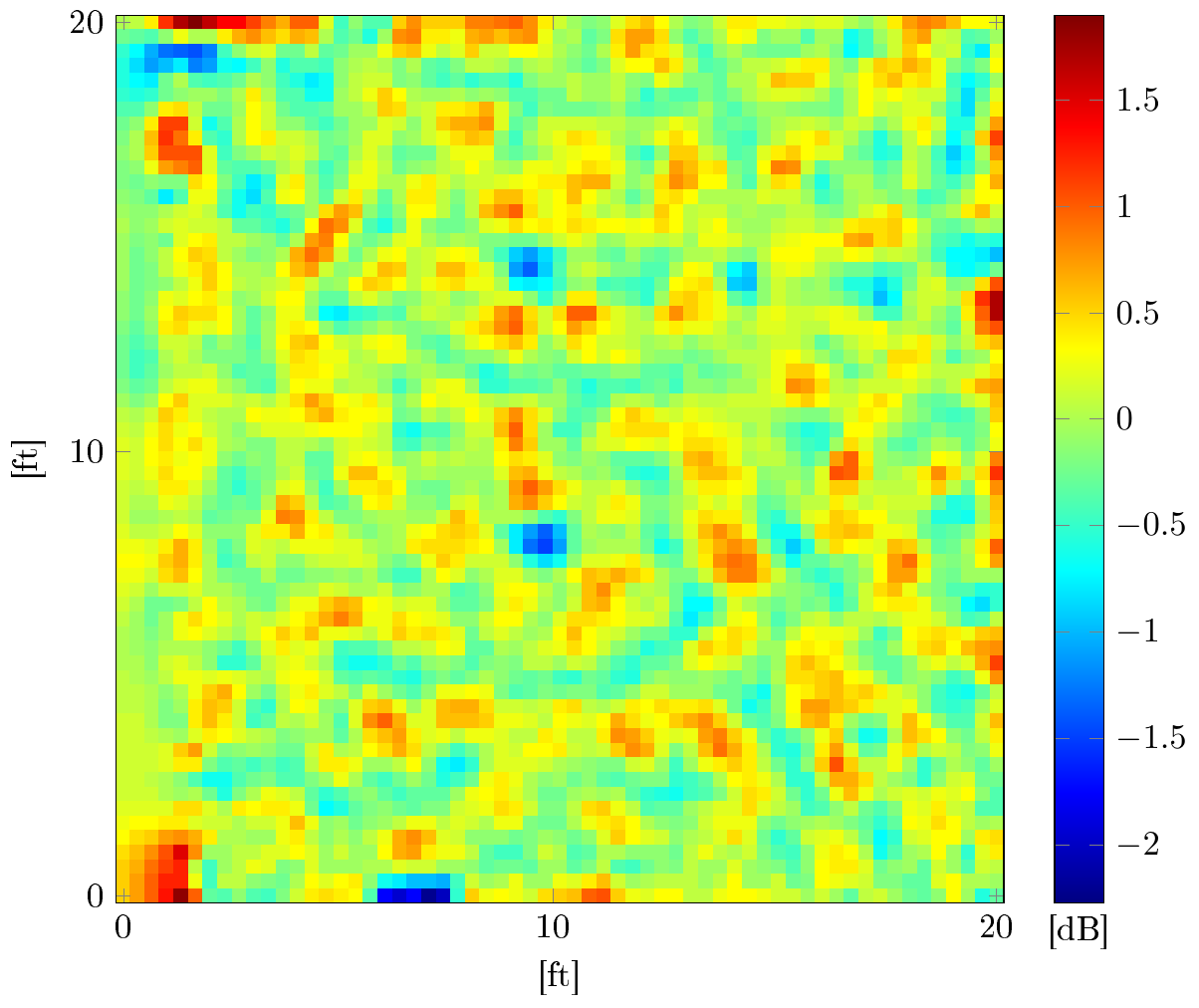}
		\vspace{-10pt}
		\subcaption{}
		\label{fig:Ridge_LS_real}
	\end{minipage}  
	\begin{minipage}[h]{.24\linewidth}
		\centering
		\includegraphics[width=\linewidth]{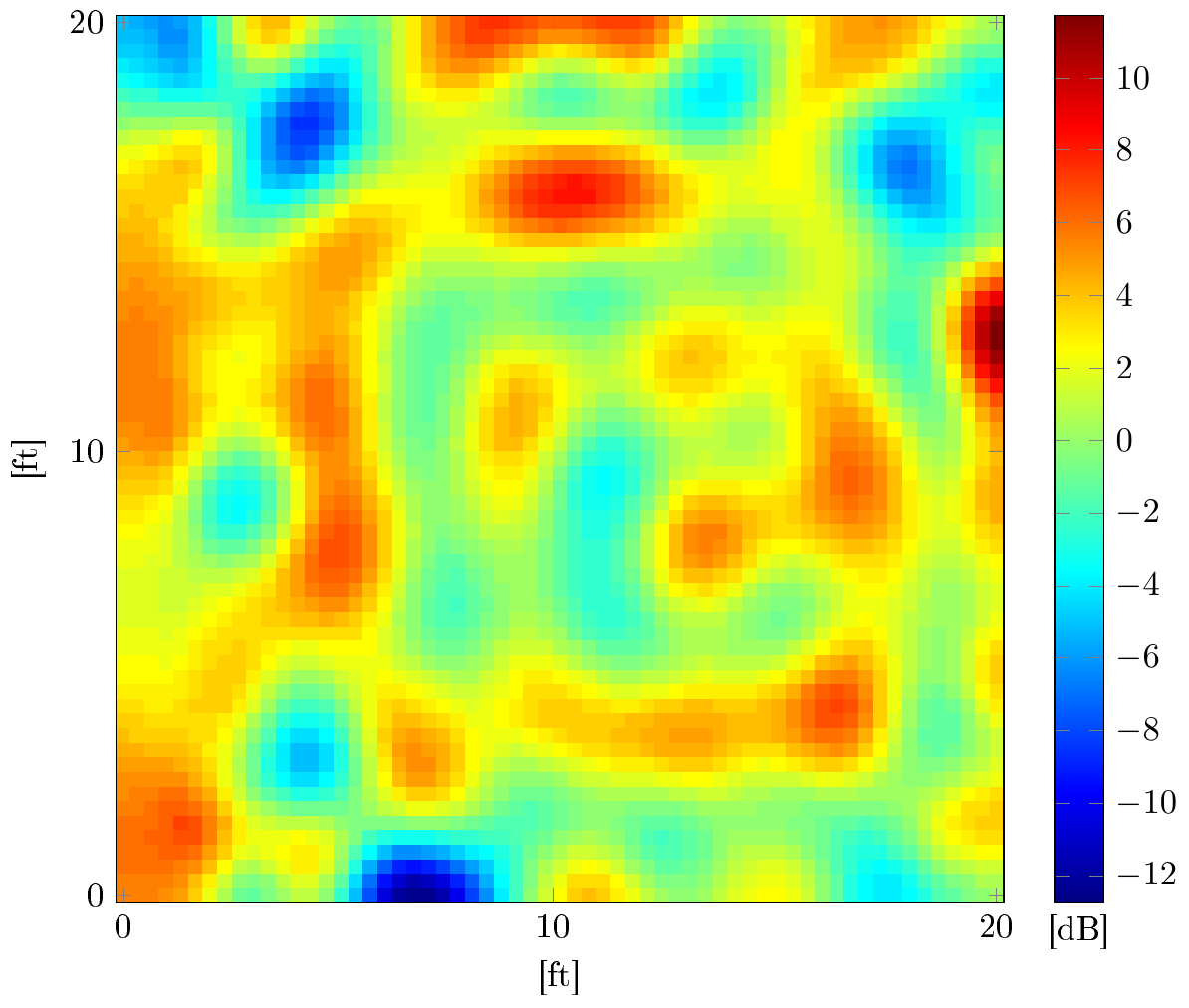}
		\vspace{-10pt}
		\subcaption{}
		\label{fig:TV_LS_real}
	\end{minipage} 
	\begin{minipage}[h]{.24\linewidth}
		\centering
		\includegraphics[width=\linewidth]{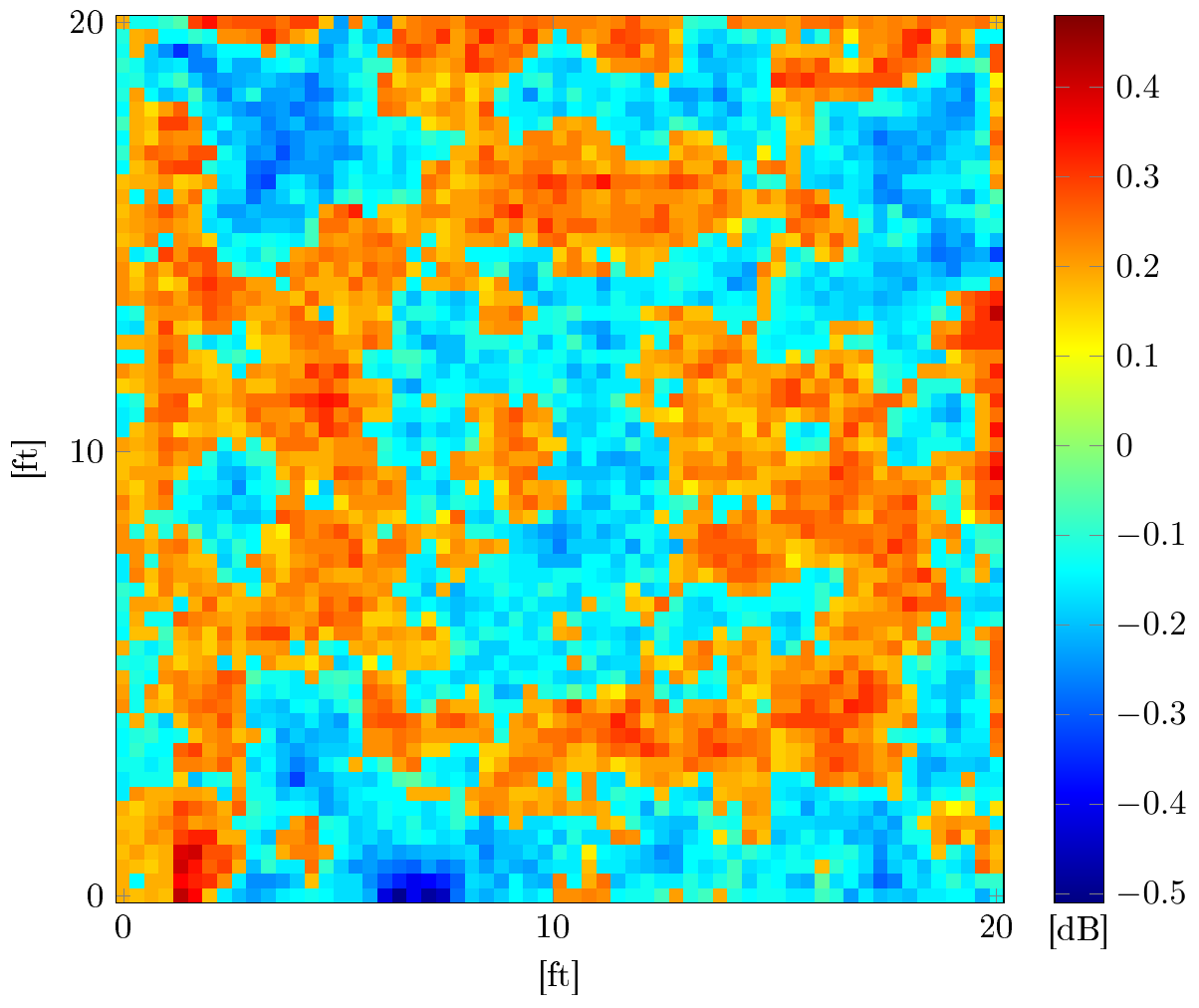}
		\vspace{-10pt}
		\subcaption{}
		\label{fig:MCMC_adap_real}
	\end{minipage}  
	\begin{minipage}[h]{.24\linewidth}
		\centering
		\includegraphics[width=\linewidth]{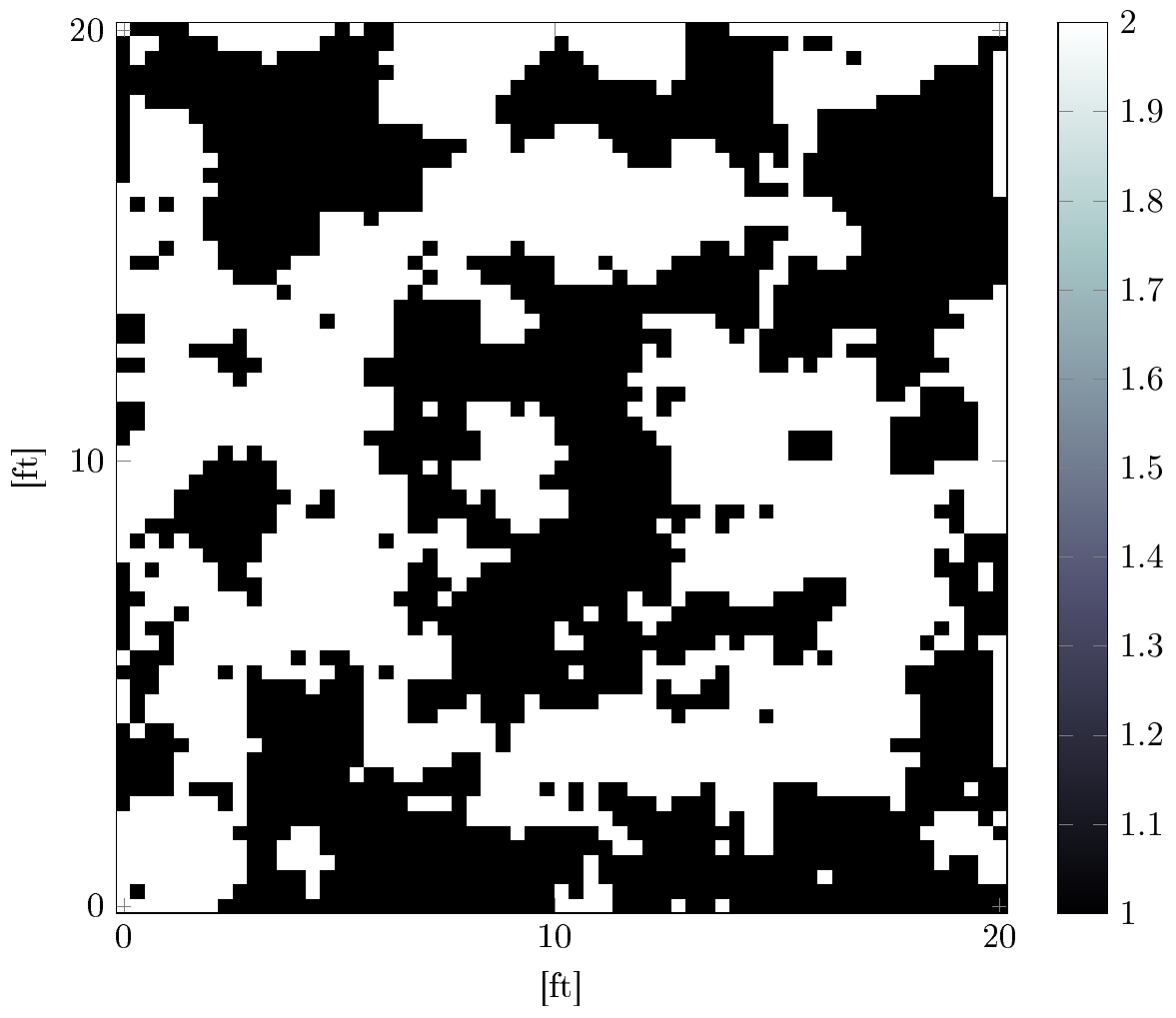}
		\vspace{-10pt}
		\subcaption{}
		\label{fig:MCMC_adap_hidden_real}
	\end{minipage} 
	
	\begin{minipage}[h]{.24\linewidth}
		\centering
		\includegraphics[width=\linewidth]{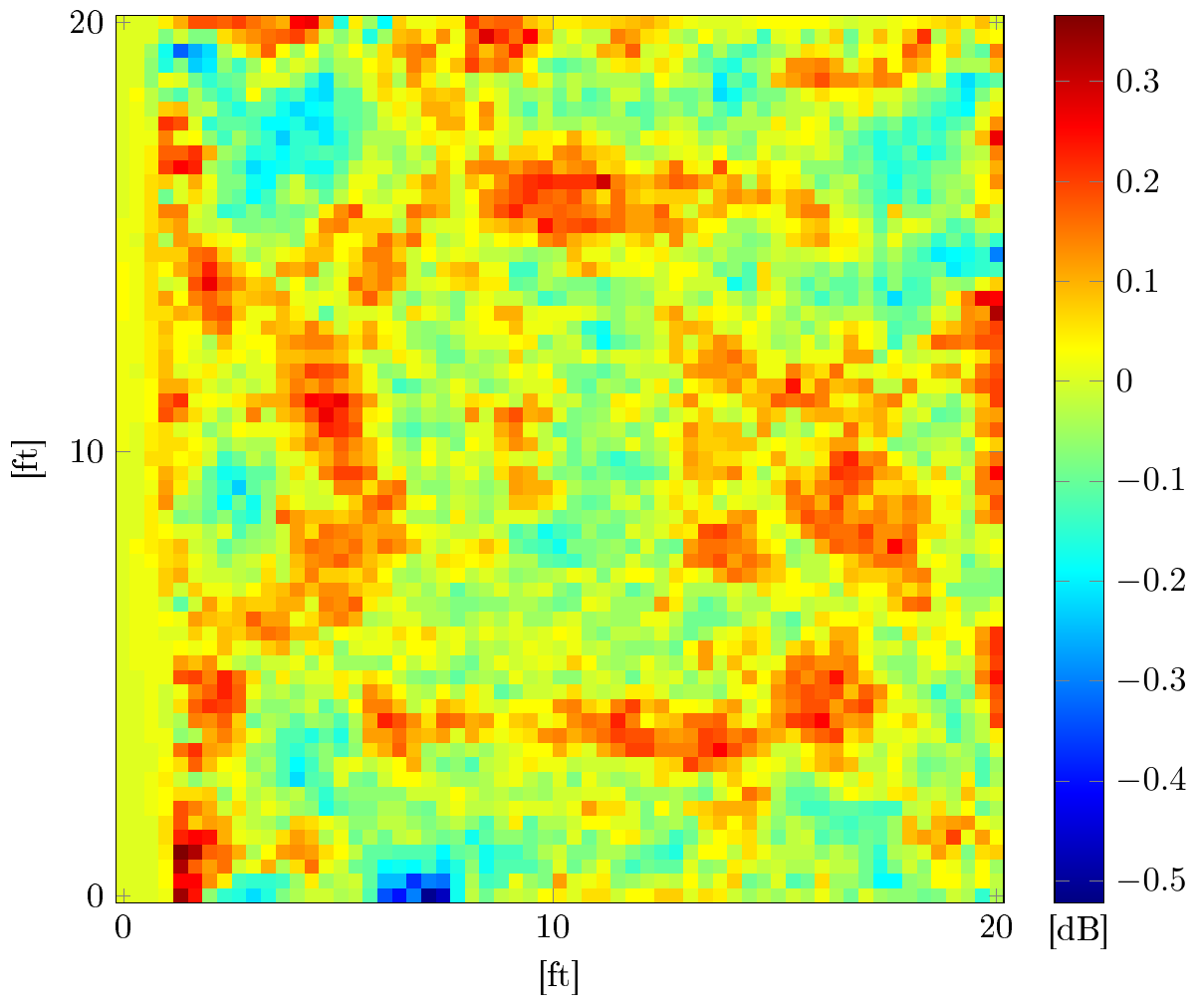}
		\vspace{-10pt}
		\subcaption{}
		\label{fig:VB_adap_real}
	\end{minipage}  
	\begin{minipage}[h]{.24\linewidth}
		\centering
		\includegraphics[width=\linewidth]{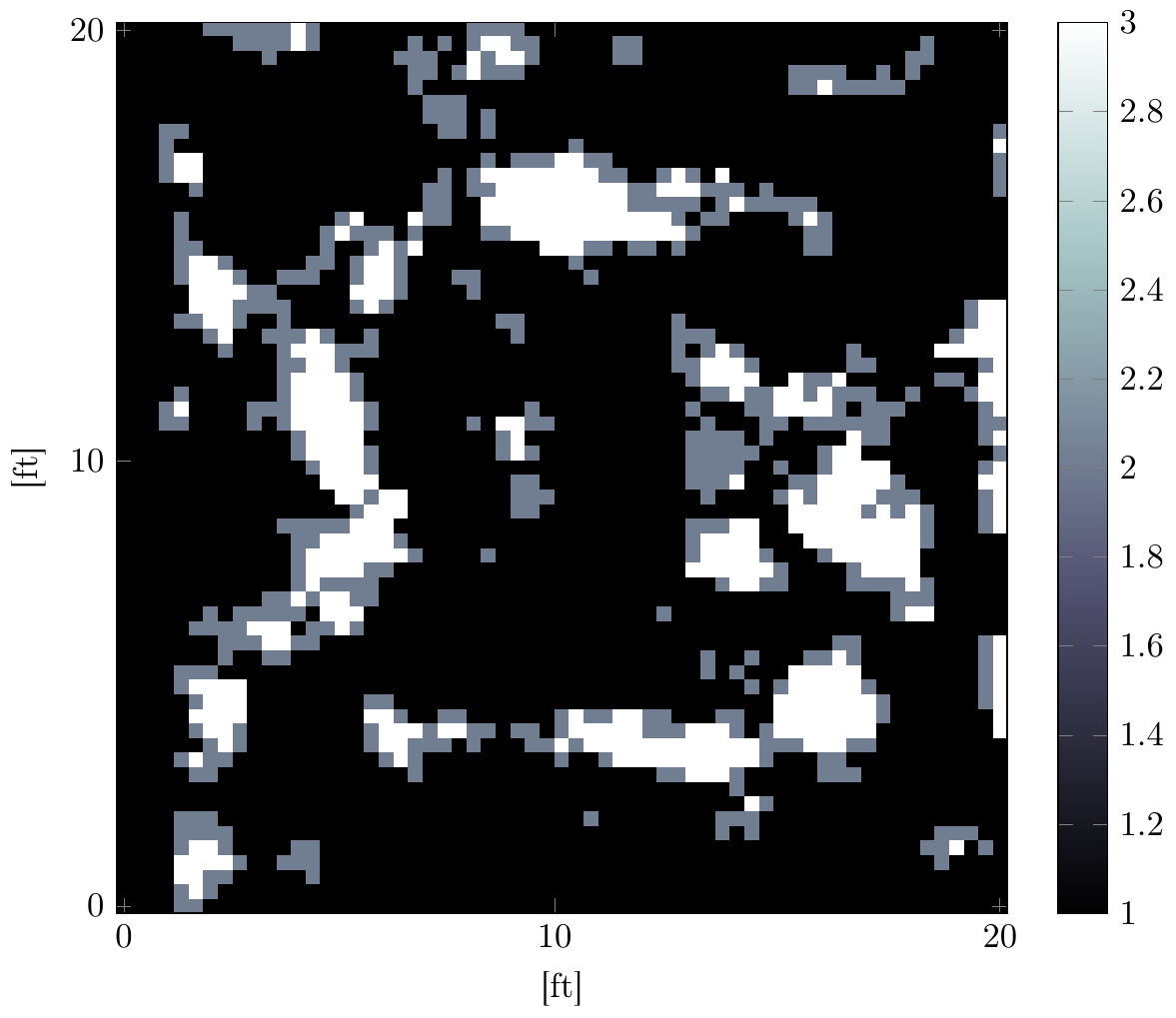}
		\vspace{-10pt}
		\subcaption{}
		\label{fig:VB_adap_hidden_real}
	\end{minipage}
	\begin{minipage}[h]{.24\linewidth}
		\centering
		\includegraphics[width=\linewidth]{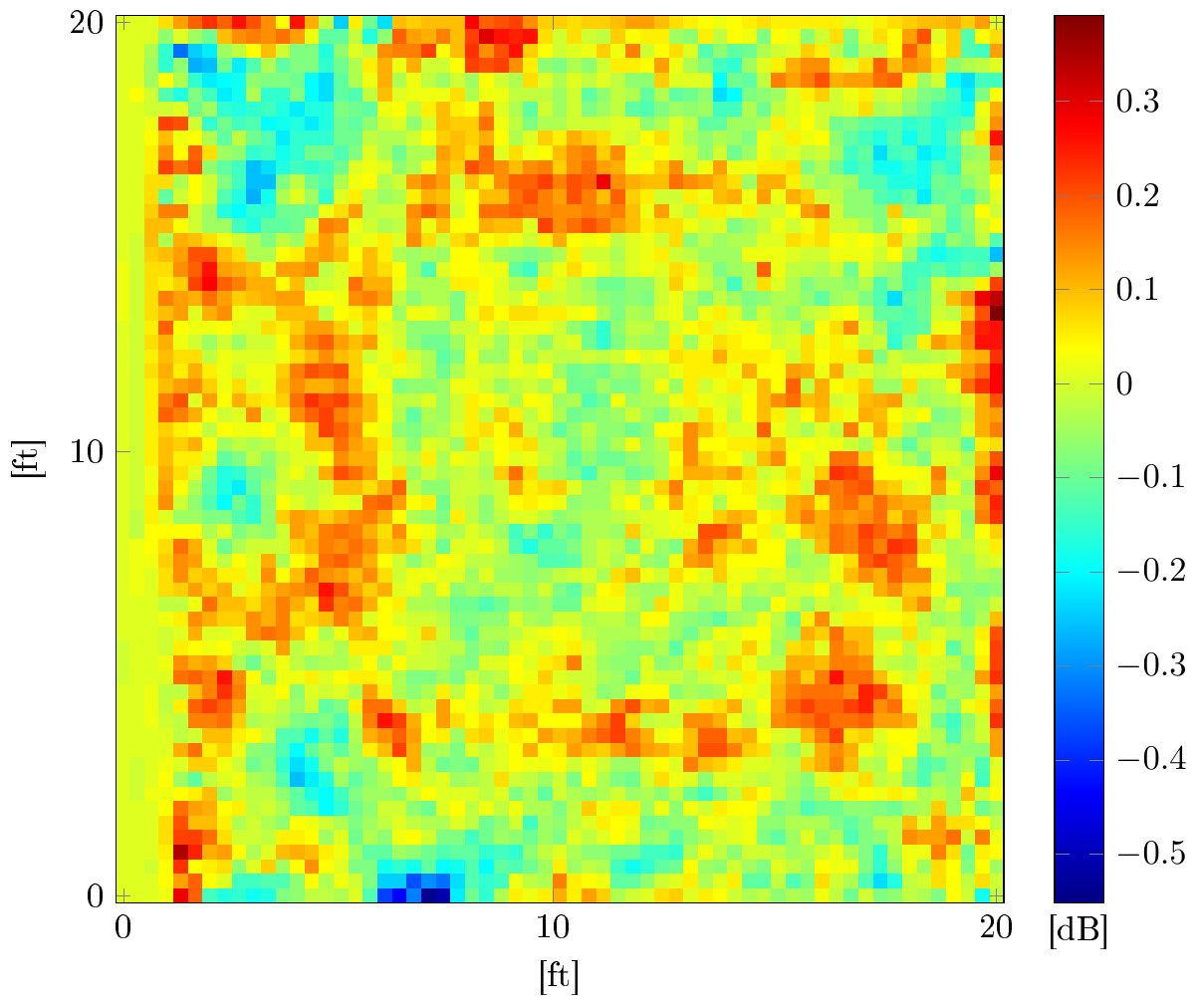}
		\vspace{-10pt}
		\subcaption{}
		\label{fig:VB_random_real}
	\end{minipage}  
	\begin{minipage}[h]{.24\linewidth}
		\centering
		\includegraphics[width=\linewidth]{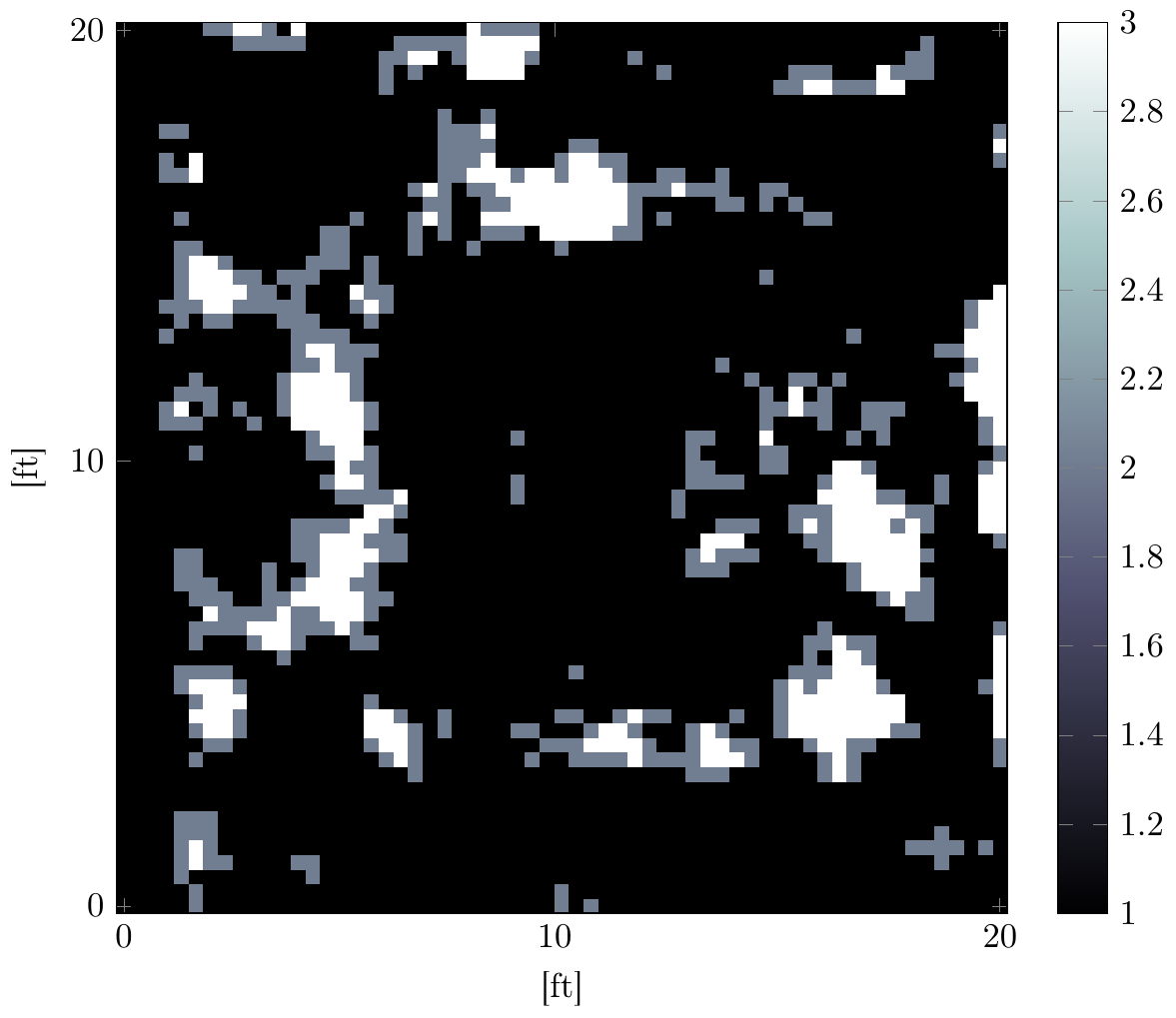}
		\vspace{-10pt}
		\subcaption{}
		\label{fig:VB_random_hidden_real}
	\end{minipage} 
	
	\begin{minipage}[h]{.24\linewidth}
		\centering
		\includegraphics[width=\linewidth]{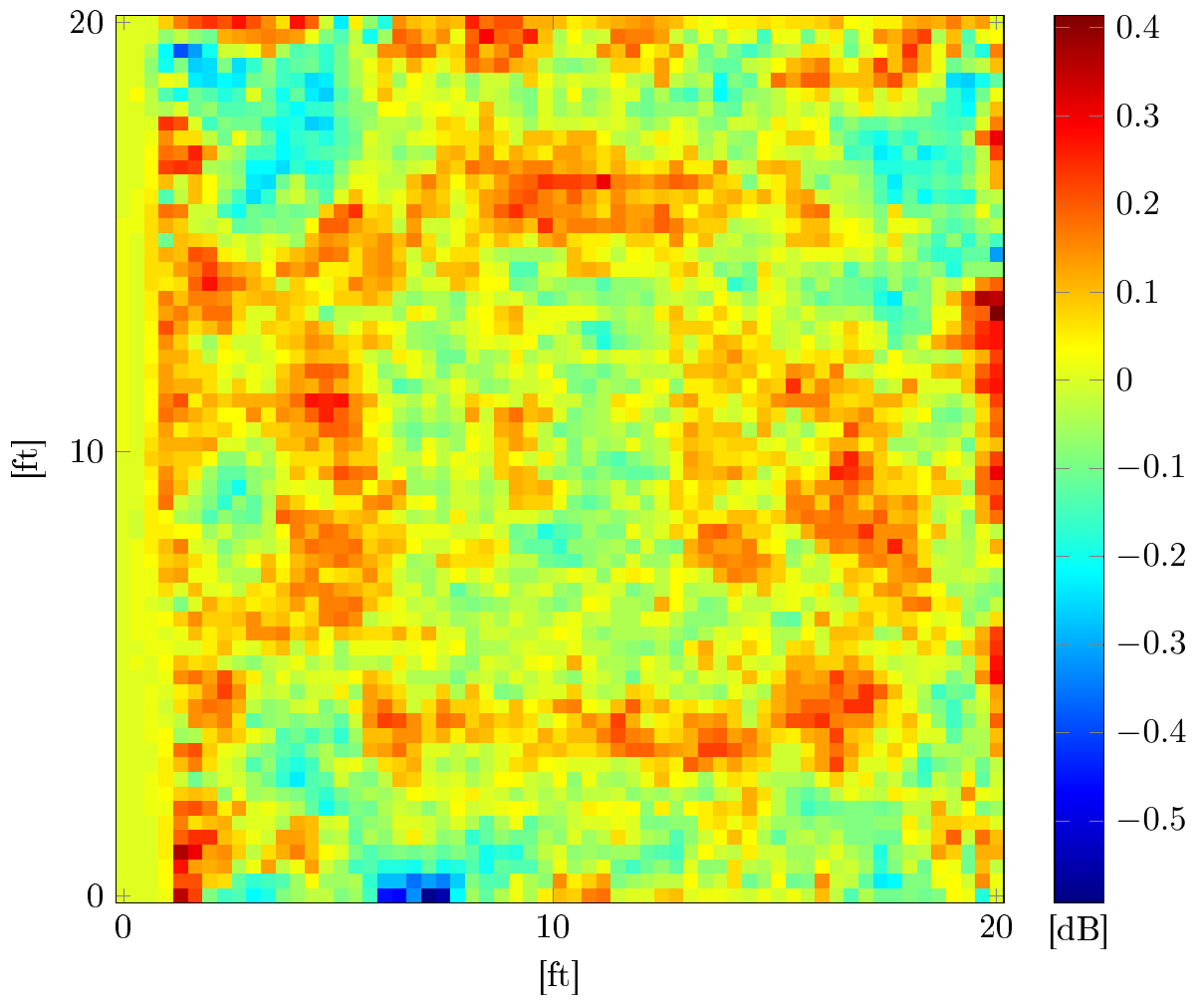}
		\vspace{-10pt}
		\subcaption{}
		\label{fig:VB_full_real}
	\end{minipage}  
	\begin{minipage}[h]{.24\linewidth}
		\centering
		\includegraphics[width=\linewidth]{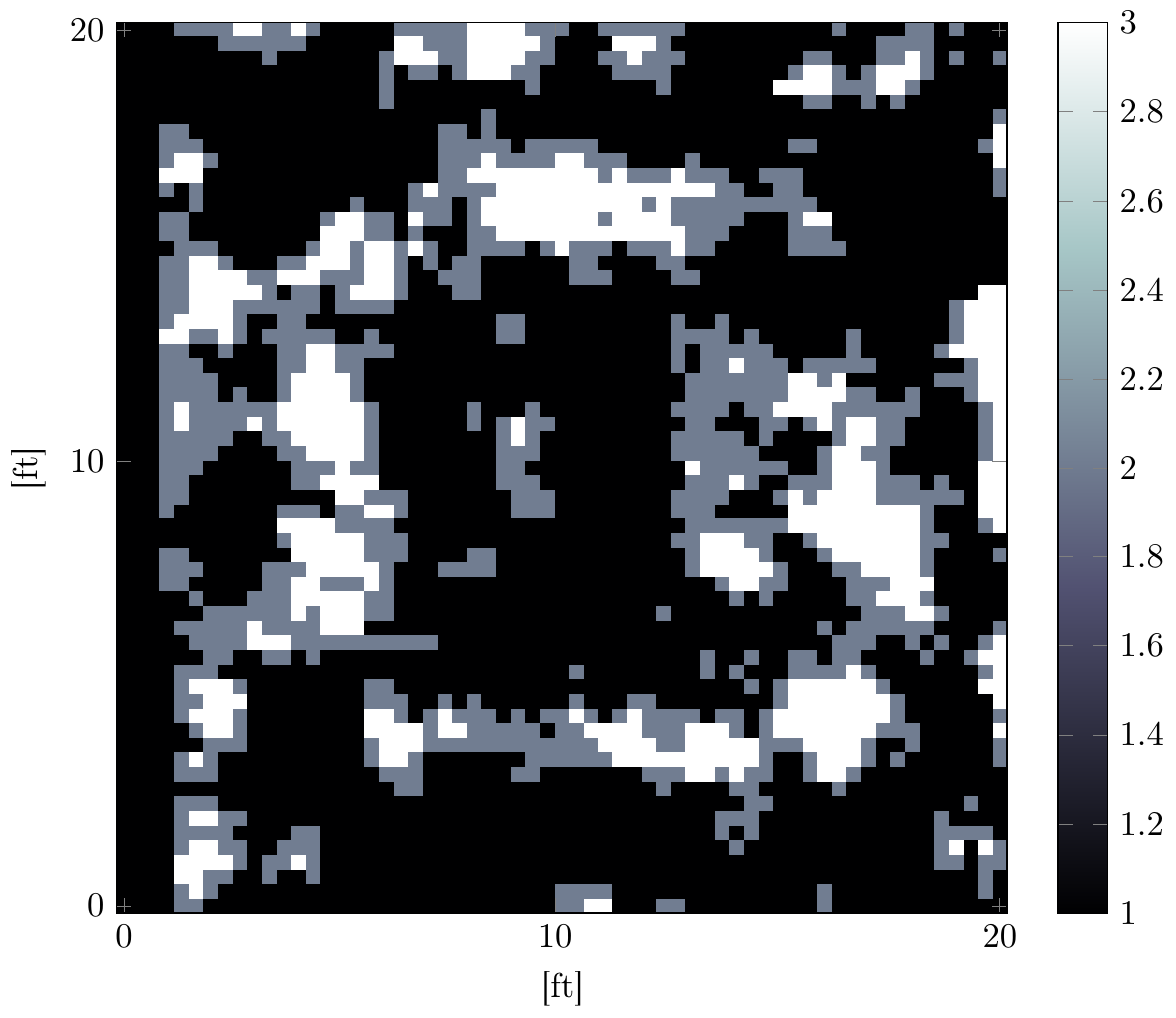}
		\vspace{-10pt}
		\subcaption{}
		\label{fig:VB_full_hidden_real}
	\end{minipage}  
	\caption{SLF estimates $\hat{\mfield}$ at $\tau=5$ (with $1,880$ measurements) via; 
		(a) ridge-regularized LS ($\RegWeight=0.015$ and
		$\bbC_{\field}=\bbI_{3,600}$); (b) TV-regularized LS ($\RegWeight=6$);
		(c) adaptive MCMC algorithm in~\cite{dbg18adaptiveRT} with $K=2$ through 
		(d) estimated hidden field $\hat{\mZfield}$;
		(e) Alg.~\ref{alg:AdapBayesCG} through (f) $\hat{\mZfield}$;
		(g) non-adaptive VB algorithm through (h) $\hat{\mZfield}$; 
		and (i) $\hat{\mfield}_{\textrm{full}}$ and (j) $\hat{\mZfield}_{\textrm{full}}$ 
		obtained by using the full data 
		(with $2,380$ measurements).} \label{fig:est_Real_SLFs}
\end{figure}

Figs.~\ref{fig:Ridge_LS_real}--\ref{fig:VB_random_hidden_real} depict SLF estimates 
$\hat{\mfield}$ and associated hidden fields $\hat{\mZfield}$ at $\tau=5$ obtained 
via the proposed algorithms and competing alternatives. As a benchmark, one-shot 
estimates of the SLF $\hat{\mfield}_{\textrm{full}}$ and associated hidden field 
$\hat{\mZfield}_{\textrm{full}}$ are also displayed in Figs.~\ref{fig:VB_full_real}
and~\ref{fig:VB_full_hidden_real} obtained via Alg.~\ref{alg:AdapBayesCG} by using 
the entire set of $2,380$ measurements. Comparing Figs.~\ref{fig:VB_adap_real} 
and~\ref{fig:VB_full_real} (or Figs.~\ref{fig:VB_adap_hidden_real} and~\ref{fig:VB_full_hidden_real}) 
shows that the proposed method accurately reveals the structural pattern of the testbed 
by using fewer number of measurements; e.g., the cinder block in the testbed was not 
captured by the SLF in Fig.~\ref{fig:VB_random_real}, but that in Fig.~\ref{fig:VB_adap_real}.  
For competing alternatives, the testbed structure was not captured through the SLFs in 
Figs.~\ref{fig:Ridge_LS_real} and~\ref{fig:TV_LS_real} estimated via both regularized LS methods. 
On the other hand, the MCMC method reveals the structure through $\hat{\mfield}$ and 
$\hat{\mZfield}$ in Figs.~\ref{fig:MCMC_adap_real} and~\ref{fig:MCMC_adap_hidden_real}, 
although they are less accurately delineated than those from the proposed method. 
This illustrates the benefits of considering a general Bayesian model with $K \geq 2$ 
addressing a richer class of spatial heterogeneity.  

\begin{figure}[t!]
	\centering
	\includegraphics[width=0.5\linewidth]{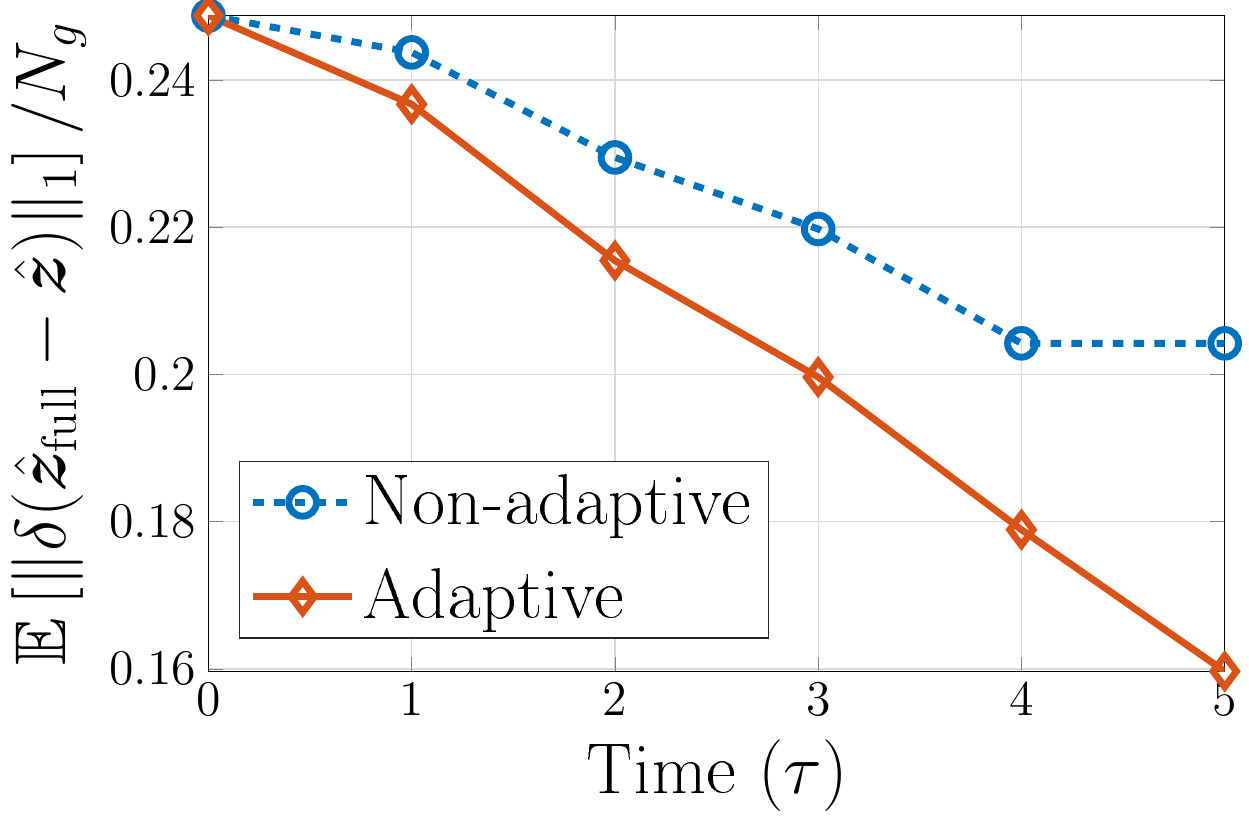}
	\caption{Progression of a mismatch between $\hat\vZfield$ and $\hat{\vZfield}_{\textrm{full}}$.}\label{fig:ReconErrReal}
\end{figure}

Efficacy of the data-driven sensor selection scheme is further analyzed. Specifically, the accuracy of $\hat{\vZfield}$ measured by the labeling error $\| \delta(\hat{\vZfield}_{\textrm{full}} - \hat{\vZfield}) \|_{1}/N_g$ with $\hat{\vZfield}_{\textrm{full}}:=\vectorize(\hat{\mZfield}_{\textrm{full}})$ 
was used as performance metric. Progression of the labeling error for Alg.~\ref{alg:AdapBayesCG} is depicted in Fig.~\ref{fig:ReconErrReal}
with that for the non-adaptive VB algorithm, where the proposed method consistently outperforms
the non-adaptive one for every $\tau$. This implies that the proposed sensor selection 
strategy helps to reveal object patterns more accurately while reducing data collection costs.

To corroborate the hyperparameter estimation capability of the proposed algorithm, estimates of $\hParameter$ averaged over 20 MC runs are listed in Table~\ref{tab:parameter_est_real}. Estimated $\hat{\hParameter}$
obtained by using the full data was considered as a benchmark, to demonstrate that Alg.~\ref{alg:AdapBayesCG} yields estimates $\hParameter$ closer to the benchmark than its non-adaptive counterpart (except $\precNVar$). Note that the level of measurement noise is high since $\widehat{\nVar}=\widehat{\precNVar}^{-1}\approx 15$. This can be justified because the testbed structure was accurately revealed in $\hat\mfield$ and $\hat\mZfield$ from the proposed method by incorporating imperfect calibration effects in the measurement noise. 

The last simulation assesses performance of the proposed algorithms for channel-gain cartography.
The set of shadowing measurements and setup was the one used in the first simulated tests of this 
section. A channel-gain map is constructed to portray the channel-gain between every point in the map $\xgen$, and a fixed receiver location $\xgen_{\textrm{rx}}$. Specifically, Alg.~\ref{alg:AdapBayesCG} is executed and estimates $\{\hat{\shd}(\xgrd_i,\xgen_{\textrm{rx}})\}_{i=1}^{N_g}$ are obtained by substituting $\hat{\vfield}$ and $w$ into~\eqref{eq:aproxSLFmodel}. Subsequently, $\{\hat{g}(\xgrd_i,\xgen_{\textrm{rx}})\}_{i=1}^{N_g}$ are obtained by substituting $\{\hat{\shd}(\xgrd_i,\xgen_{\textrm{rx}})\}_{i=1}^{N_g}$ into~\eqref{eq:cg} 
with $\hat{g}_0$ and $\hat\pathlossexp$. Upon defining $\hat{\bm{g}}:=[\hat{g}(\xgrd_1,\xgen_{\textrm{rx}}),\ldots,
\hat{g}(\xgrd_{N_g},\xgen_{\textrm{rx}})]^{\transpose}\in\mathbb{R}^{N_g}$, we construct the channel-gain map $\hat{\mCGmap}:=\unvec(\hat{\bm{g}})$ with the receiver located at $\xgen_{\textrm{rx}}$.  

\begin{figure}[t!]
	\begin{minipage}[h]{.24\linewidth}
		\centering
		\includegraphics[width=\linewidth]{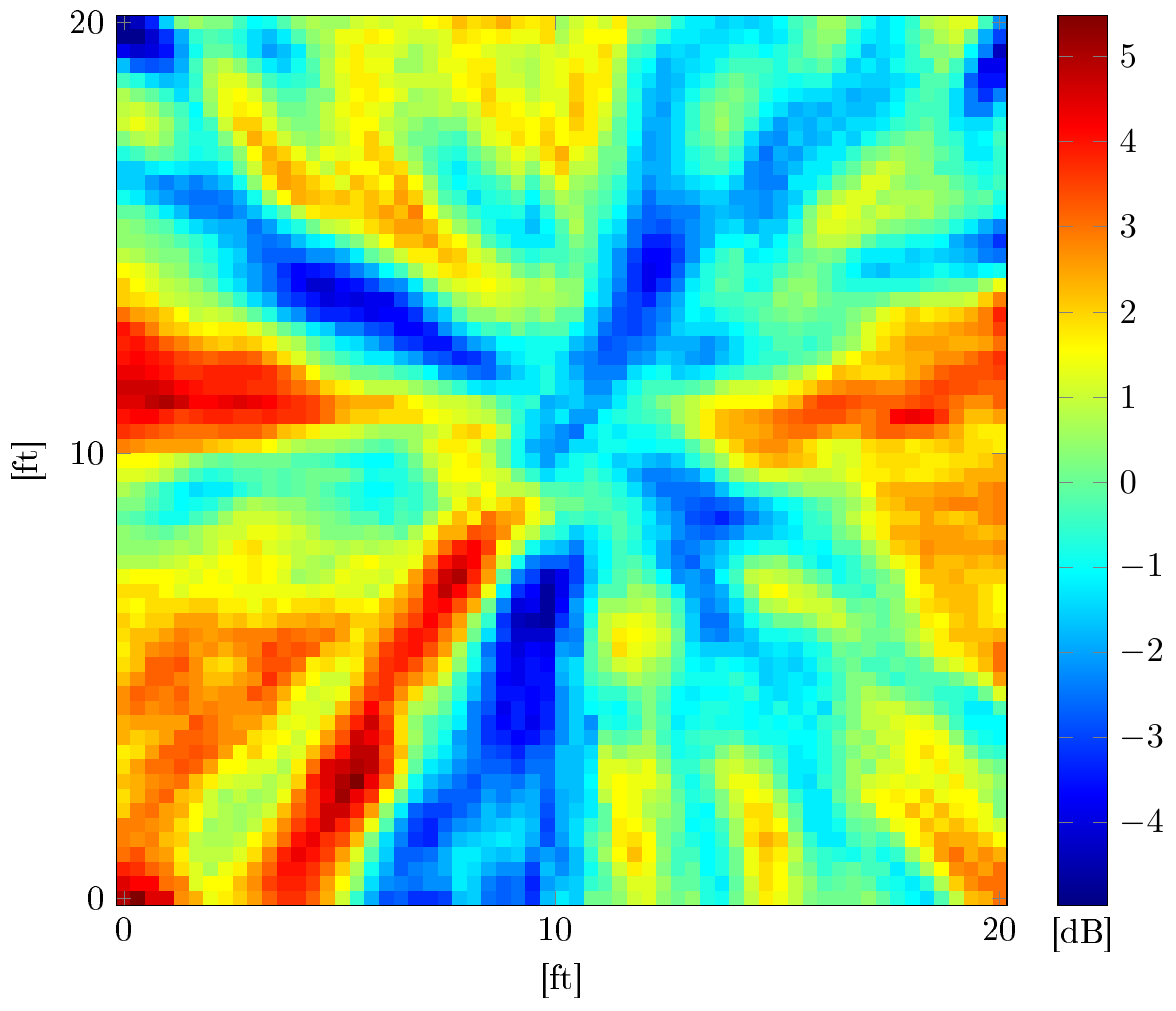}
		\vspace{-10pt}
		\subcaption{}
		\label{fig:est_shadow_Tik_real}
	\end{minipage}  
	\begin{minipage}[h]{.24\linewidth}
		\centering
		\includegraphics[width=\linewidth]{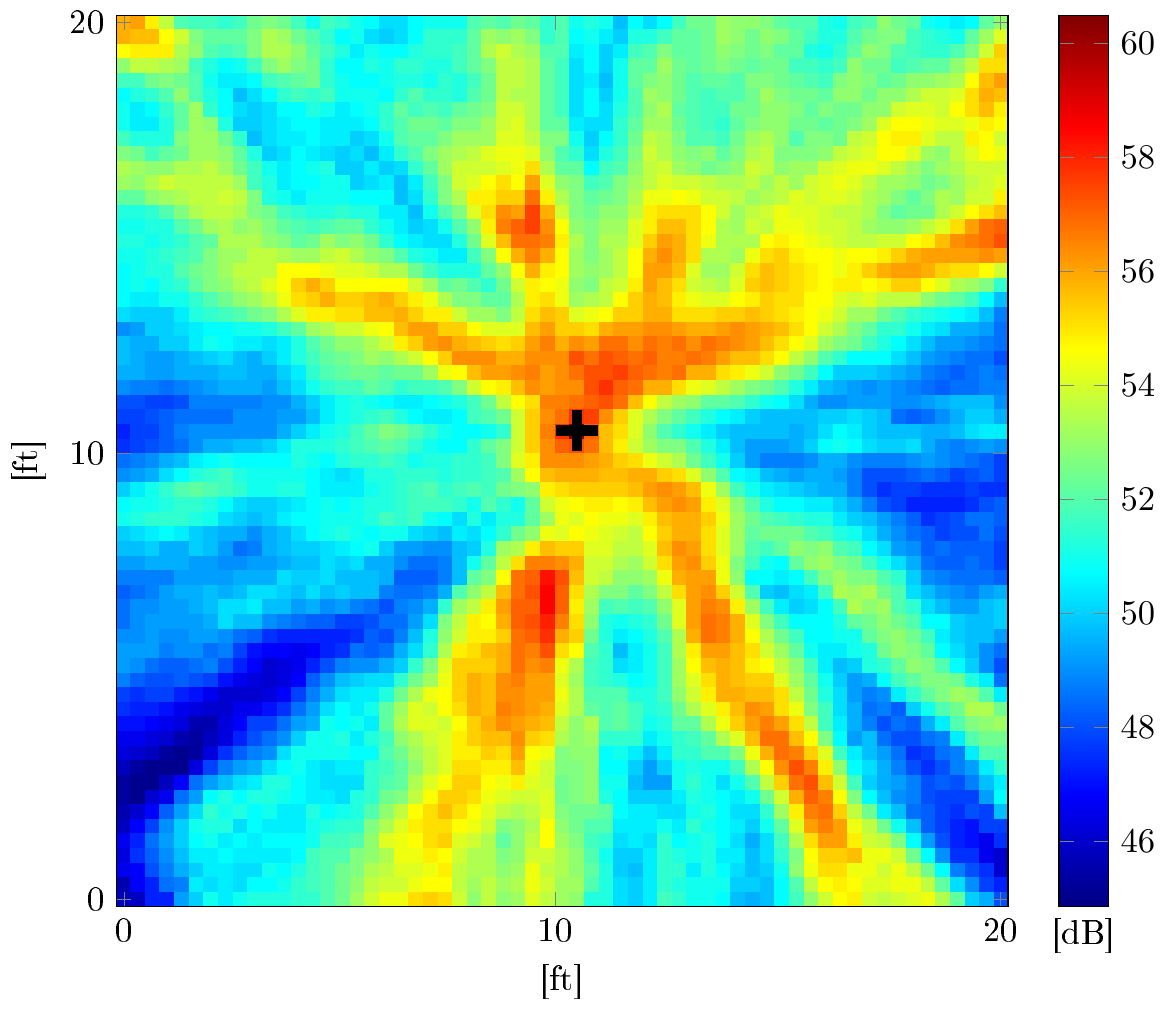}
		\vspace{-10pt}
		\subcaption{}
		\label{fig:est_CG_Tik_real}
	\end{minipage} 	
	\begin{minipage}[h]{.24\linewidth}
		\centering
		\includegraphics[width=\linewidth]{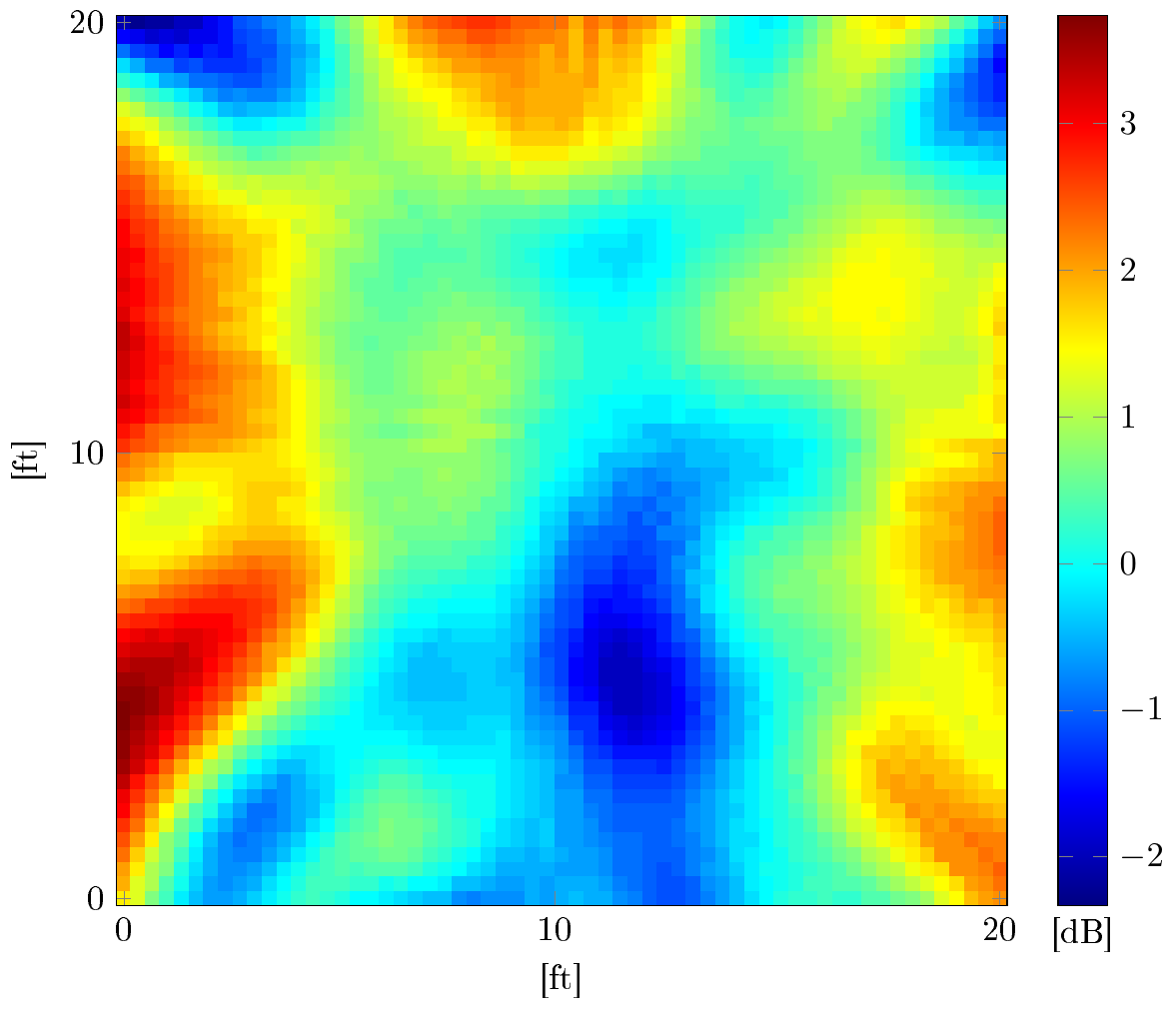}
		\vspace{-10pt}
		\subcaption{}
		\label{fig:est_shadow_TV_real}
	\end{minipage}  
	\begin{minipage}[h]{.24\linewidth}
		\centering
		\includegraphics[width=\linewidth]{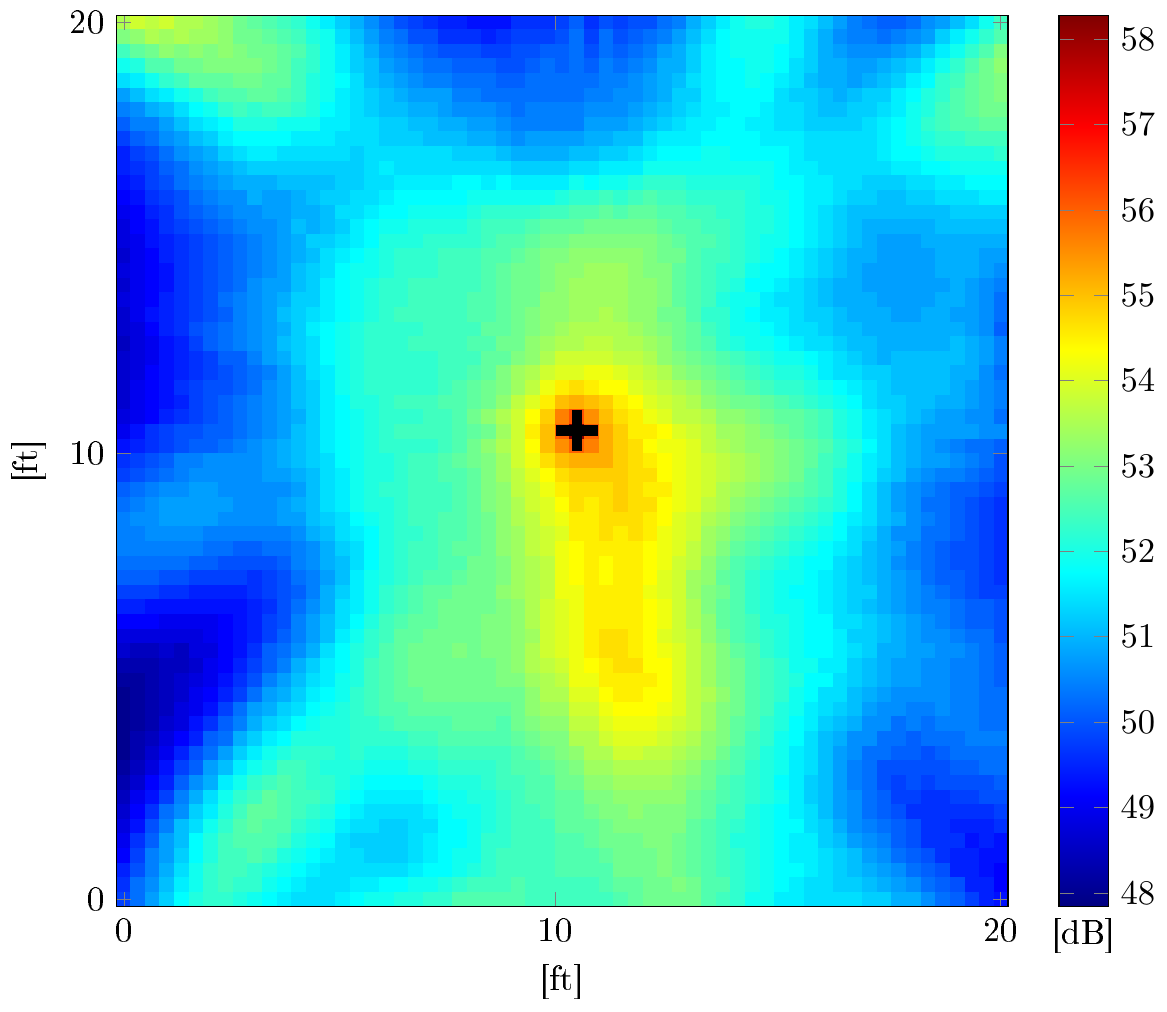}
		\vspace{-10pt}
		\subcaption{}
		\label{fig:est_CG_TV_real}
	\end{minipage} 
	
	\begin{minipage}[h]{.24\linewidth}
		\centering
		\includegraphics[width=\linewidth]{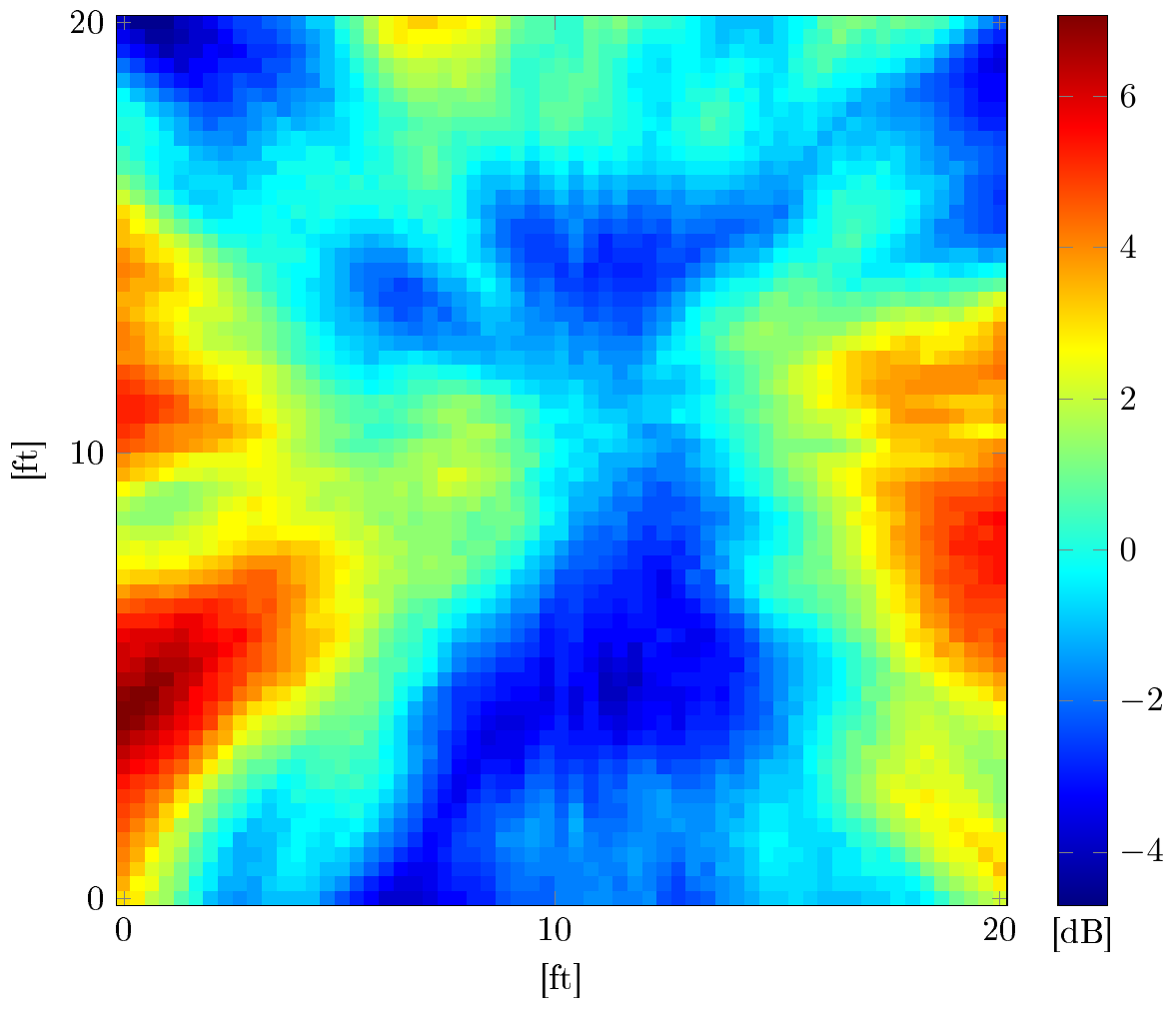}
		\vspace{-10pt}
		\subcaption{}
		\label{fig:est_shadow_MCMC_adap_real}
	\end{minipage}  
	\begin{minipage}[h]{.24\linewidth}
		\centering
		\includegraphics[width=\linewidth]{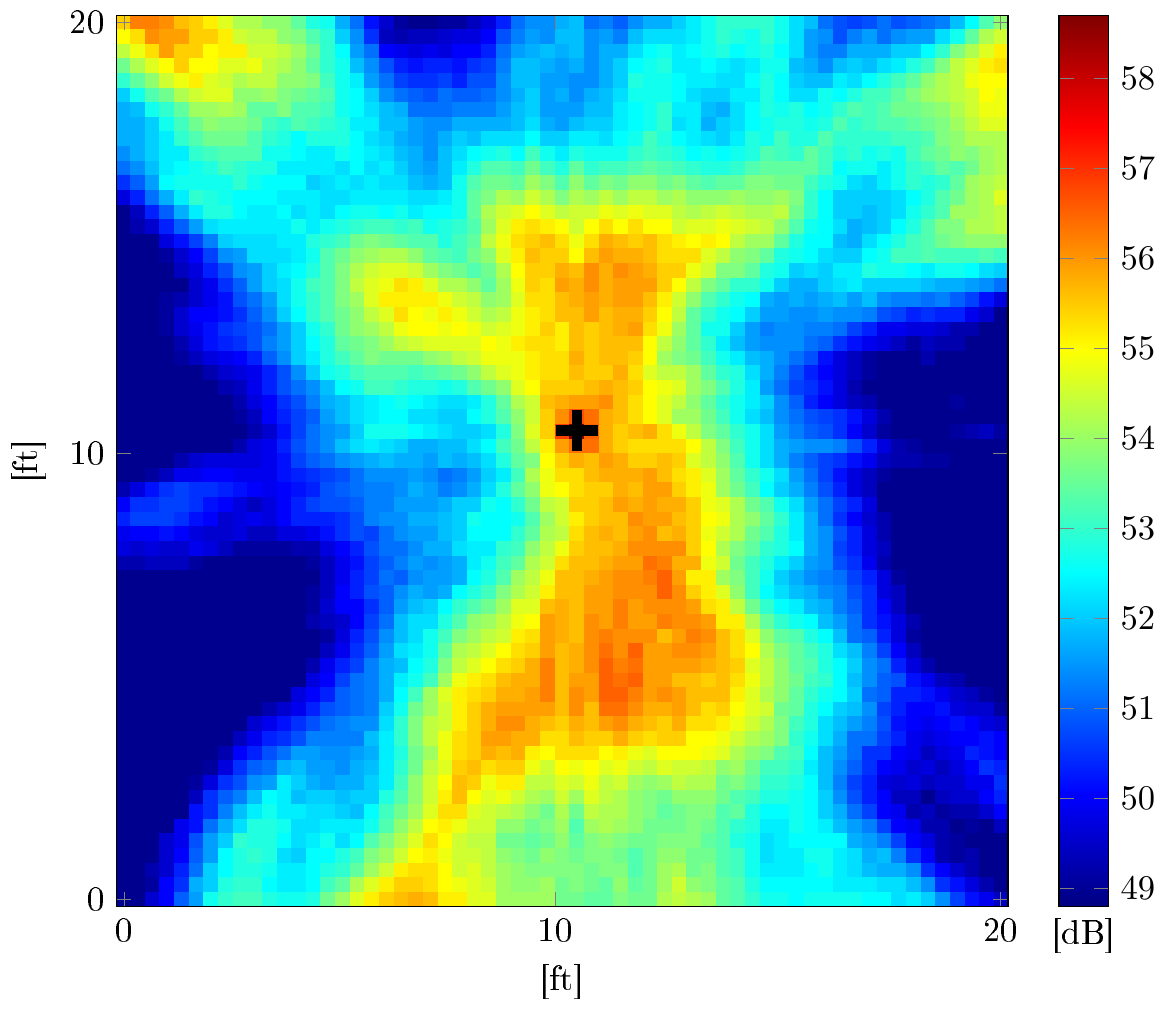}
		\vspace{-10pt}
		\subcaption{}
		\label{fig:est_CG_Bayes_MCMC_adap_real}
	\end{minipage} 
	\begin{minipage}[h]{.24\linewidth}
		\centering
		\includegraphics[width=\linewidth]{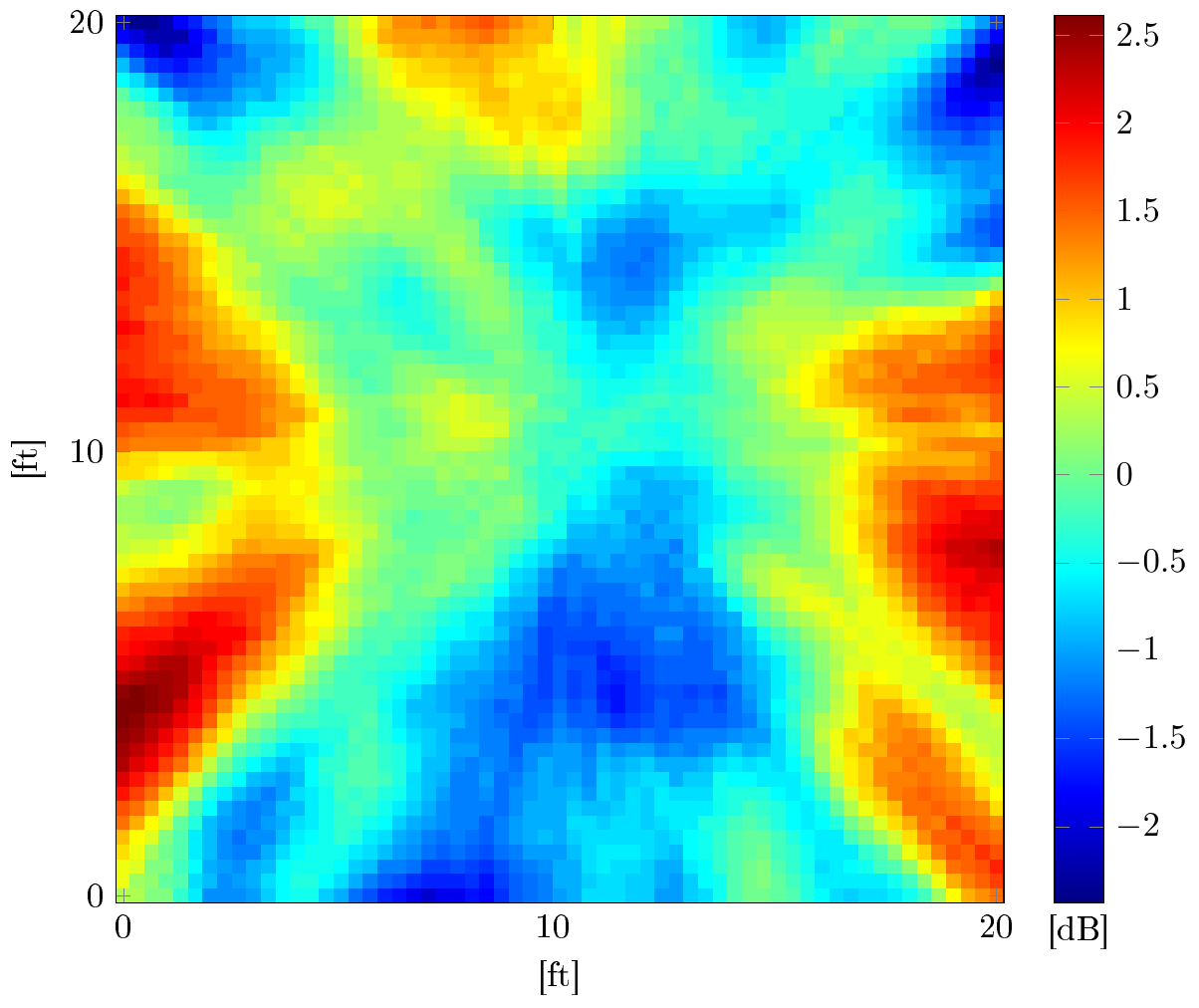}
		\vspace{-10pt}
		\subcaption{}
		\label{fig:est_shadow_VB_adap_real}
	\end{minipage}  
	\begin{minipage}[h]{.24\linewidth}
		\centering
		\includegraphics[width=\linewidth]{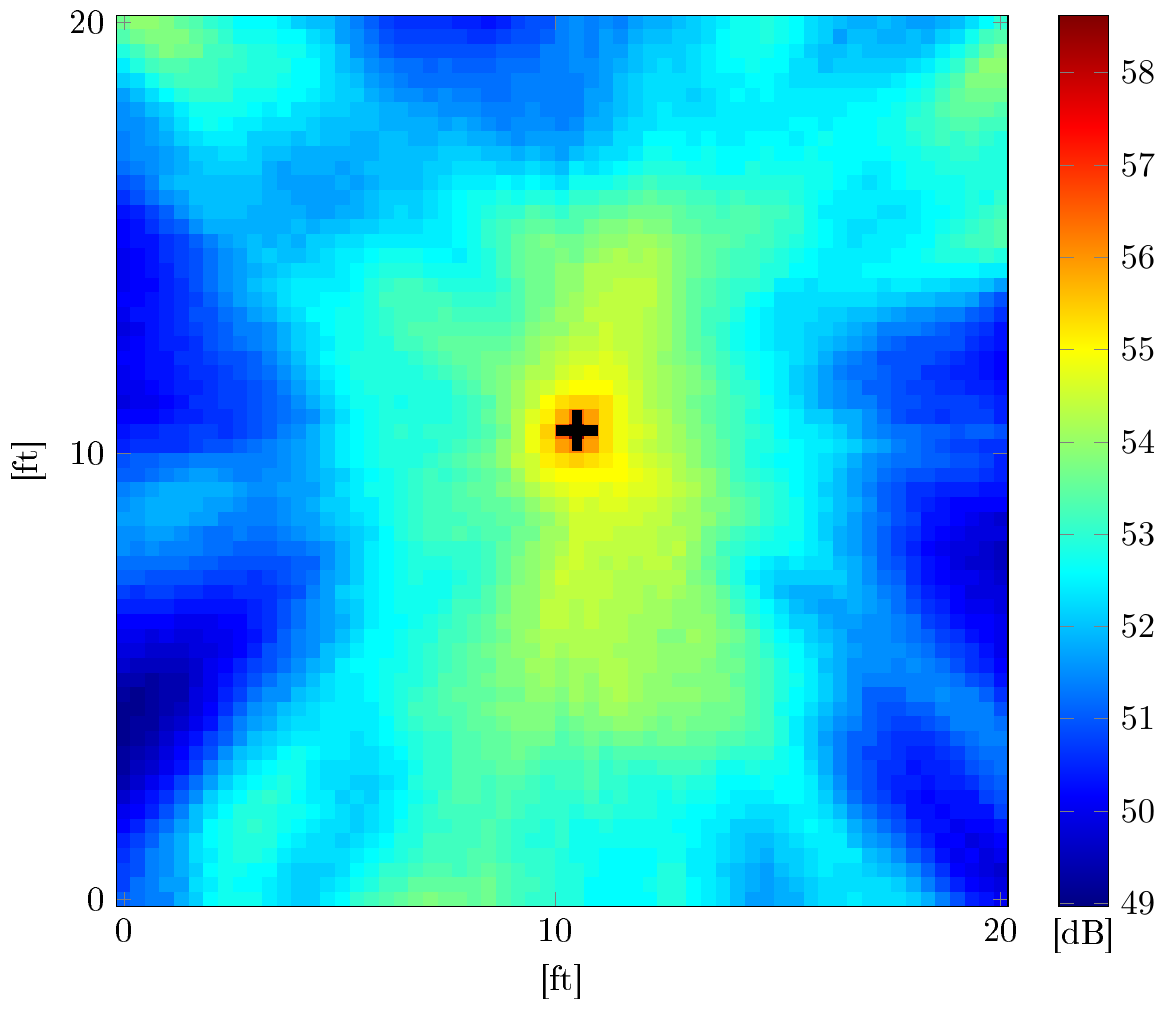}
		\vspace{-10pt}
		\subcaption{}
		\label{fig:est_CG_VB_adap_real}
	\end{minipage} 
	
	\begin{minipage}[h]{.24\linewidth}
		\centering
		\includegraphics[width=\linewidth]{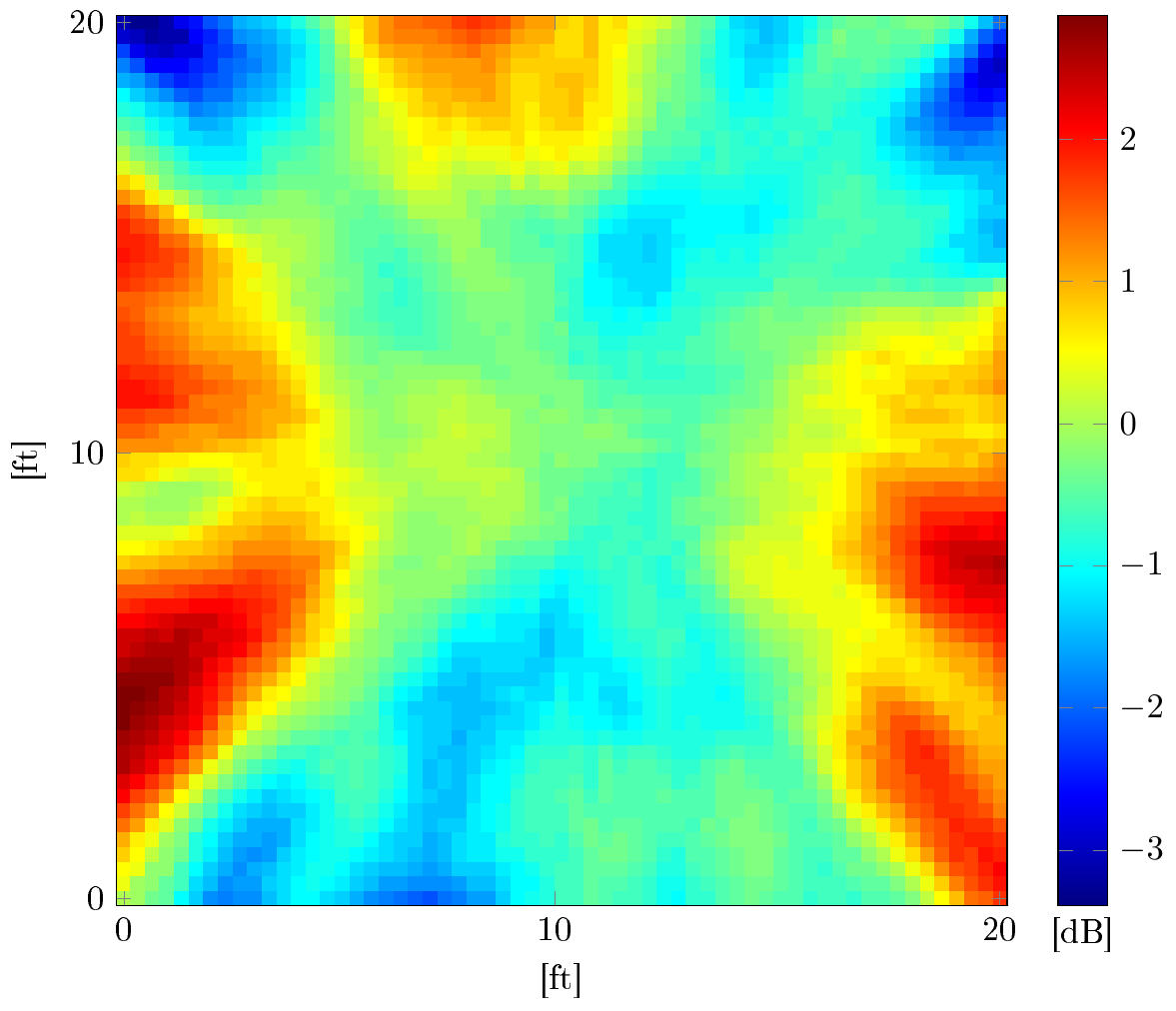}
		\vspace{-10pt}
		\subcaption{}
		\label{fig:est_shadow_VB_random_real}
	\end{minipage}  
	\begin{minipage}[h]{.24\linewidth}
		\centering
		\includegraphics[width=\linewidth]{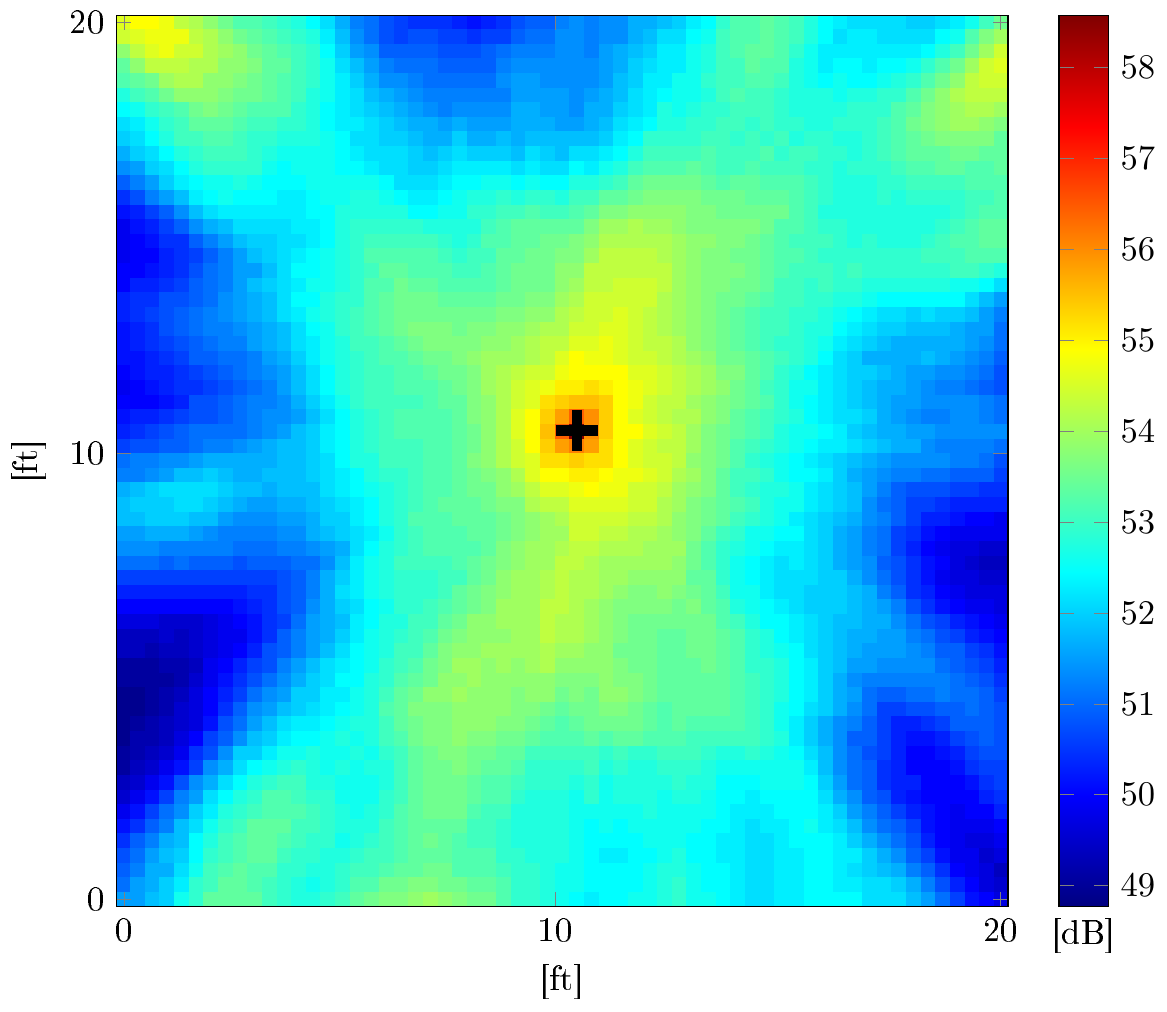}
		\vspace{-10pt}
		\subcaption{}
		\label{fig:est_CG_VB_random_real}
	\end{minipage} 
	\begin{minipage}[h]{.24\linewidth}
		\centering
		\includegraphics[width=\linewidth]{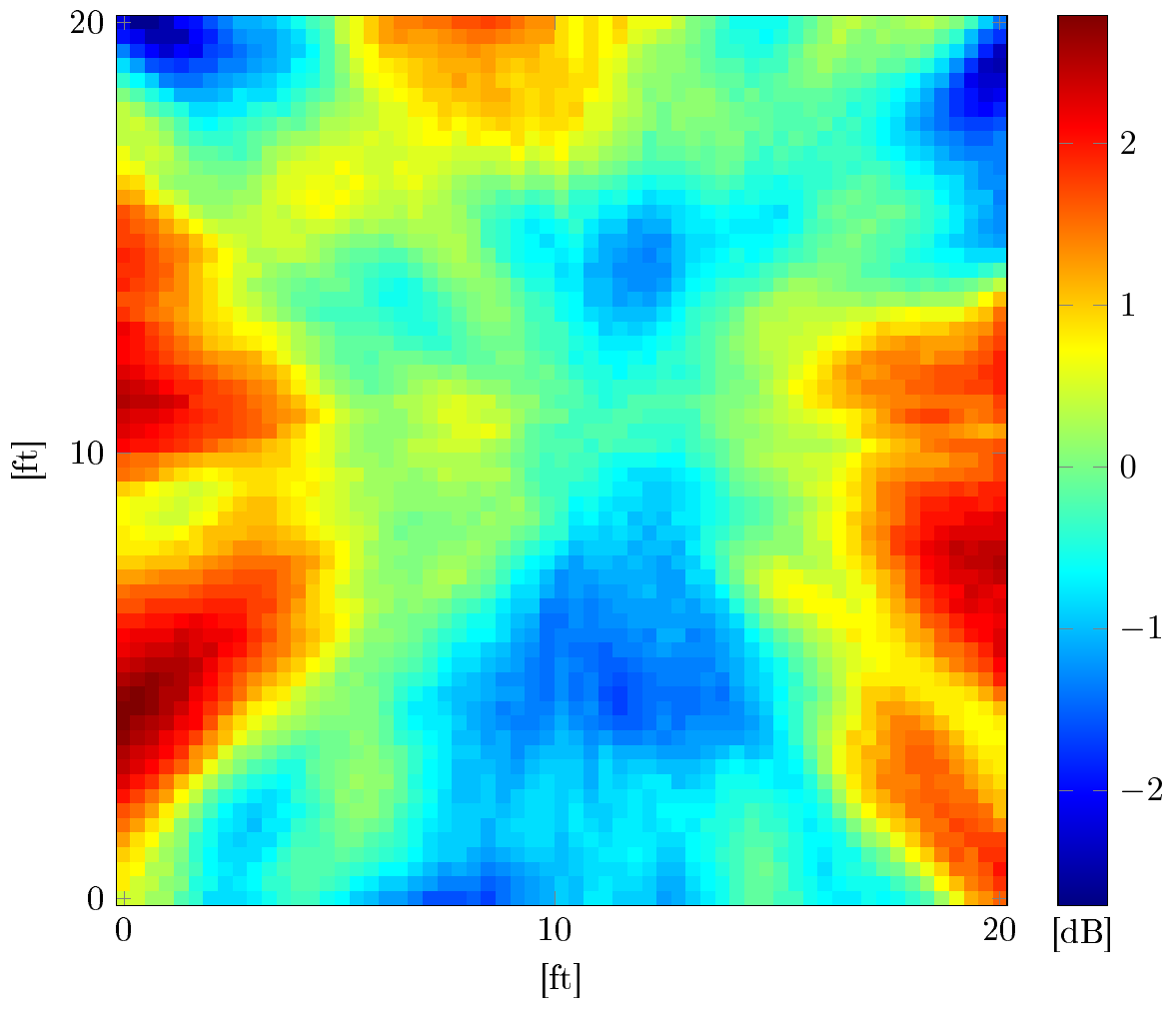}
		\vspace{-10pt}
		\subcaption{}
		\label{fig:est_shadow_VB_full_real}
	\end{minipage}  
	\begin{minipage}[h]{.24\linewidth}
		\centering
		\includegraphics[width=\linewidth]{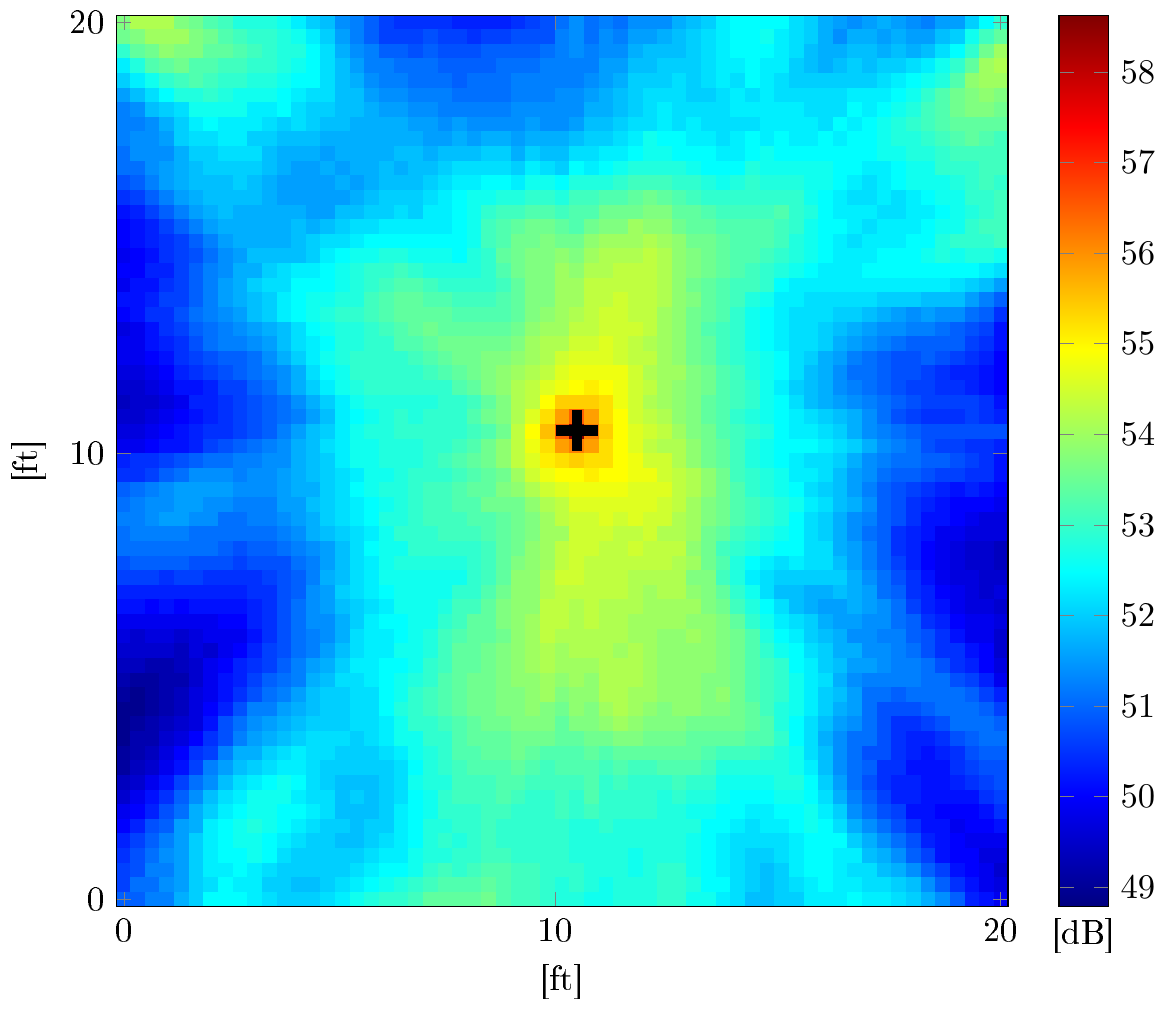}
		\vspace{-10pt}
		\subcaption{}
		\label{fig:est_CG_VB_full_real}
	\end{minipage}

	\caption{Estimated shadowing maps $\hat{\mShadowMap}$ and corresponding 
		channel-gain maps $\hat{\mCGmap}$ at $\tau=5$ via (a)-(b)
		ridge-regularized LS (setting of Fig.~\ref{fig:Ridge_LS_real}); 
		(c)-(d) TV-regularized LS (setting of Fig.~\ref{fig:TV_LS_real});
		(e)-(f) adaptive MCMC algorithm in~\cite{dbg18adaptiveRT} with $K=2$ 
		(setting of Fig.~\ref{fig:MCMC_adap_real}); 
		(g)-(h) Alg.~\ref{alg:AdapBayesCG} (setting of Fig.~\ref{fig:VB_adap_real}); 
		and (i)-(j) non-adaptive VB algorithm (setting of Fig.~\ref{fig:VB_random_real}); 
		and (k)--(l) benchmark algorithm (setting of Fig.~\ref{fig:VB_full_real}), 
		with the receiver location at $\xgen_{\textrm{rx}}=(10.3, 10.7)$ 
		(ft) marked with the black cross.} \label{fig:real_CGmaps}
\end{figure}

Let $\hat{\mShadowMap}:=\unvec(\hat{\bm{s}})$ denote a shadowing map with $\hat{\bm{s} }:=[\hat{s}(\xgrd_1,\xgen_{\textrm{rx}}),\ldots,\hat{s}(\xgrd_{N_g},\xgen_{\textrm{rx}})]^{\transpose}\in\mathbb{R}^{N_g}$. 
Fig.~\ref{fig:real_CGmaps} displays estimated shadowing maps and corresponding 
channel-gain maps constructed via Alg.~\ref{alg:AdapBayesCG} and the competing 
alternatives, when the receiver is located at $\xgen_{\textrm{rx}}=(10.3, 10.7)$ 
(ft) marked by the cross. In every channel-gain map of Fig.~\ref{fig:real_CGmaps}, 
stronger attenuation is observed when signals propagate through either more building materials 
(bottom-right side of $\hat{\mCGmap}$), or the concrete wall (left side of $\hat{\mCGmap}$). 
On the other hand, only the channel-gain maps in Figs.~\ref{fig:est_CG_Bayes_MCMC_adap_real},~\ref{fig:est_CG_VB_adap_real},~\ref{fig:est_CG_VB_random_real}, 
and~\ref{fig:est_CG_VB_full_real} constructed by the approximate Bayesian inference methods 
exhibit less attenuation along the entrance of the structure (top-right side of $\hat{\mCGmap}$); 
this cannot be seen through the channel-gain maps in Figs.~\ref{fig:Ridge_LS_real} 
and~\ref{fig:TV_LS_real} constructed by both regularized LS methods. The reason is that free space and objects are more distinctively delineated in $\hat{\mfield}$ by the proposed method. All in all, the simulation results confirm that our approach could provide more site-specific information of the propagation medium, and thus endows the operation of cognitive radio networks with more accurate interference management.

\begin{table}[t!]
	\centering \caption{Estimated $\hat{\hParameter}$ via benchmark algorithm (setting of Fig.~\ref{fig:VB_full_real}); Alg.~\ref{alg:AdapBayesCG} (setting of Fig.~\ref{fig:VB_adap_real}); and non-adaptive VB algorithm (setting of Fig.~\ref{fig:VB_random_real}), averaged over 20 independent MC runs.} 
	\begin{tabular}{|c|c|c|c|}
		\hline
		$\hParameter$ & Benchmark            & Est. (Alg.~\ref{alg:AdapBayesCG})             & Est. (non-adaptive)   \\ \hline 
		$\precNVar$ & $0.075 \pm 10^{-16}$   & $0.068 \pm 0.13$   & $0.071 \pm 0.24$\\ \hline
		$\fMean{1}$ & $-0.001\pm 10^{-17} $   & $-0.001\pm 10^{-7} $   & $-0.001\pm 10^{-7} $ \\ \hline
		$\fMean{2}$ & $0.032 \pm 10^{-17}$   & $0.032 \pm 10^{-8}$    & $0.032 \pm 10^{-8}$ \\ \hline		
		$\fMean{3}$ & $0.045 \pm 10^{-17} $   & $0.046\pm 10^{-8} $   & $0.046\pm 10^{-8} $ \\ \hline
		$\fprec{1}$ & $5.524 \pm 10^{-18}$   & $4.951 \pm 10^{-3}$    & $4.789 \pm 1.9\times10^{-3}$ \\ \hline
		$\fprec{2}$ & $5.524 \pm 10^{-18}$   & $4.942 \pm 10^{-3}$  & $4.782 \pm 1.7\times10^{-3}$ \\ \hline
		$\fprec{3}$ & $5.524 \pm 10^{-18}$   & $4.935 \pm 10^{-3}$    & $4.775 \pm  1.7\times10^{-3}$ \\ \hline
	\end{tabular}%
	\label{tab:parameter_est_real}
\end{table}

\section{Conclusions}
\label{sec:conc}
This paper developed a variational Bayes approach to adaptive radio tomography, which estimates the spatial loss field of the tomographic model at affordable complexity by using measurements collected from sensing radio pairs that are adaptively chosen with an uncertainty sampling criterion. Extensive synthetic and real data tests corroborated the efficacy of the proposed novel algorithm for imaging and channel-gain cartography applications. Future research will include an online approach to radio tomography for streaming data.

\newpage 

\appendix
\section{Appendix}
Here we derive the variational distributions in~\eqref{eq:factPosterior}. Terms not related to a target variable will be lumped in a generic constant $c$. The iteration index $\ell$ will be omitted for simplicity.

\subsection{Variational distribution of the SLF  $q(\vfield|\vZfield)$}
\label{sec:CondMeanVar4SLF}
Recall that the conditional posterior obeys $p(\vfield,\vZfield,\hParameter|\Vnshd) \propto p(\Vnshd|\vfield,\precNVar)p(\vfield|\vZfield,\hParameter_f)$. The first factor in~\eqref{eq:factPosterior}, is expressed as
\begin{align}
	\ln q(\vfield|\vZfield)=\sum_{i=1}^{N_g}\ln q(\field_{i}|\Zfield_{i}) 
	=\sum_{k=1}^{K}\sum_{i:\xgrd_{i}\in \Region{k}}\ln q(\field_{i}|\Zfield_{i}=k)
\end{align}
where $\ln q(\field_{i}|\Zfield_{i}=k)$ can be written as 
\begin{align}
	\ln q(\field_{i}|\Zfield_{i}=k) &= \EXPwo{\ln p(\vfield,\vZfield,\hParameter|\Vnshd)}
	{-q(\field_{i}|\Zfield_{i}=k)} + c\nonumber\\
	&= \EXPwo{\ln p(\Vnshd|\vfield,\precNVar)}{-q(\field_{i}|\Zfield_{i}=k)} \nonumber \\
	&+ \EXPwo{\ln p(\vfield|\vZfield,\hParameter_f)}{-q(\field_{i}|\Zfield_{i}=k)} + c. \label{eq:logPosField}
\end{align}
Each term on the RHS in~\eqref{eq:logPosField} is thus given by 
\begin{align}
&\EXPwo{\ln p(\Vnshd|\vfield,\precNVar)}{-q(\field_{i}|\Zfield_{i}=k)} \nonumber \\
&\leftrightarrow \frac{\MeanprecNVar}{2} \sum_{\tau=1}^{t} \bigg[w_{\tau,i}^{2} f_i^{2}-2\bigg(\nshd_{\tau}-\sum_{j\neq i} 
w_{\tau,j}\Meanfield{j}\bigg)w_{\tau,i} f_i\bigg]  \label{eq:MeanDataLikelihood}
\end{align}
where $\Meanfield{j}:=\sum_{k=1}^{K} \VIzfieldProb{k}{\xgrd_{j}}\VIfmean{k}{j}$, and
\begin{align}
&\EXPwo{\ln p(\vfield|\vZfield,\hParameter_f)}{-q(\field_{i}|\Zfield_{i}=k)} \nonumber \\
&\leftrightarrow \EXPwo{\frac{\fprec{k}}{2}\big(f_i^2 - 2 \fMean{k}f_i\big)}{-q(\field_{i}|\Zfield_{i}=k)} 
= \frac{\Meanfprec{k}}{2}\big(f_i^2 - 2 \VIfMeanMean{k}f_i\big). \label{eq:MeanfieldConditional}
\end{align}
After substituting~\eqref{eq:MeanDataLikelihood} and~\eqref{eq:MeanfieldConditional} into~\eqref{eq:logPosField}, the pdf $q(\field_{i}|\Zfield_{i}=k)$ can be expressed as
\begin{align}
	q(\field_{i}|\Zfield_{i}=k) &\propto \exp\bigg\{-\frac{1}{2} \left({\MeanprecNVar} 
	\sum_{\tau=1}^{t}w_{\tau,i}^{2} + \Meanfprec{k}\right)f_i^2 \nonumber \\
	&\hspace{-1.2cm}+\bigg[\MeanprecNVar\sum_{\tau=1}^{t} \bigg(\nshd_{\tau}-\sum_{j\neq i} 
	w_{\tau,j}\Meanfield{j}\bigg)w_{\tau,i} + \Meanfprec{k}\VIfMeanMean{k}
	\bigg]f_i\bigg\}. \label{eq:VIposField}
\end{align}
By completing the square, it can be readily verified that $q(\field_{i}|\Zfield_{i}=k) = \NormDist{\VIfmean{k}{i}}{\VIfvar{k}{i}}~\forall k$, where
\begin{align}
	\VIfvar{k}{i} &= \bigg({\MeanprecNVar} \sum_{\tau=1}^{t}w_{\tau,i}^{2} 
	+ \Meanfprec{k}\bigg)^{-1} \label{eq:VIfVar}\\
	\VIfmean{k}{i} &= \VIfvar{k}{i} \label{eq:VIfmean}\\
	&\times\bigg[\MeanprecNVar\sum_{\tau=1}^{t} \bigg(\nshd_{\tau}-\sum_{j\neq i} 
	w_{\tau,j}\Meanfield{j}\bigg)w_{\tau,i} + \Meanfprec{k}\VIfMeanMean{k}
	\bigg]. \nonumber
\end{align}
Upon defining $\Meanshd{\tau}:=\sum_{i=1}^{N_g} w_{\tau,i}\Meanfield{i}$,
$\VIfmean{k}{i}$ in~\eqref{eq:VIfmean}, it follows that   
\begin{align}
	\VIfmean{k}{i} &= \Meanfield{i} + \VIfvar{k}{i} \\
	&\times\bigg[\big(\VIfMeanMean{k} - \Meanfield{i} \big)\Meanfprec{k}
	+\MeanprecNVar\sum_{\tau=1}^{t} w_{\tau,i}\big(\nshd_{\tau}-\Meanshd{\tau}\big)\bigg].~~\qquad\blacksquare \nonumber 
\end{align}

\subsection{Variational distribution of the hidden label field $q(\vZfield)$}
\label{sec:VIhiddenFieldPosterior}
Since $q(\vZfield) = \prod_{i=1}^{N_g}q(\Zfield_i)$ in~\eqref{eq:factPosterior} because  
$\Zfield_i$ and $\Zfield_j~\forall i\neq j$ are independent, we focus on the derivation of $q(\Zfield_i)$.
By proportionality of the conditional posterior
$p(\vfield,\vZfield,\hParameter|\Vnshd) \propto p(\vfield|\vZfield,\hParameter_f)p(\vZfield;\granPara)$ 
wrt $\vZfield$, after singling out the terms that involve $q(\Zfield_i)$, we arrive at  
\begin{align}
	&\ln q(\Zfield_i) = \EXPwo{\ln p(\vfield,\vZfield,\hParameter|\Vnshd)}{-q(\Zfield_{i})} + c \nonumber \\
	&\leftrightarrow \EXPwo{\ln p(\vfield|\vZfield,\hParameter_{f})}{-q(\Zfield_{i})} + \EXPwo{\ln p(\Zfield_i|\vZfield_{-i};\granPara)}{-q(\Zfield_{i})}. \label{eq:logPosHiddenField}
\end{align}
For $z_i = k$, each term on the RHS in~\eqref{eq:logPosHiddenField} becomes
\begin{align}
	&\EXPwo{\ln p(\vfield|\vZfield,\hParameter_{f})}{-q(\Zfield_{i})}  
	\leftrightarrow \frac{1}{2} \EXPwo{\ln\fprec{k} - \fprec{k}(f_i - \fMean{k})^2}{-q(\Zfield_{i})} \nonumber\\
	&=-\frac{\Meanfprec{k}}{2}\bigg[\underbrace{\EXPwo{f_i^2}{-q(\Zfield_{i})}}_{=\VIfvar{k}{i}+\breve{\mu}_{f_k}^2(\xgrd_{i})} - 2\VIfMeanMean{k}\VIfmean{k}{i} + \underbrace{\EXPwo{\fMean{k}^2}{-q(\Zfield_{i})}}
	_{=\VIfMeanVar{k}+\VIfMeanMean{k}^2}\bigg]\nonumber \\
	&+ \frac{1}{2}\underbrace{\EXPwo{\ln\fprec{k}}{-q(\Zfield_{i})}}_{=\diGamma{\VIfpreca{k}}+\ln \VIfprecb{k}} 
\end{align}
and
\begin{align}
&\EXPwo{\ln p(\Zfield_i=k|\vZfield_{-i},\granPara)}{-q(\Zfield_{i})} \nonumber \\
&\leftrightarrow \EXPwo{\granPara \sum_{j \in \neighbor{\xgrd_{i}}}  \delta(\Zfield_j-k)}{-q(\Zfield_{i})}
= \granPara \sum_{j \in \neighbor{\xgrd_{i}}}  \VIzfieldProb{k}{\xgrd_{j}}. 
\end{align}
All in all, the variational pdf $q(\Zfield_i=k)$ becomes
\begin{align}
	&\hspace{-0.2cm}q(\Zfield_i=k) \propto \exp\bigg\{-\frac{\Meanfprec{k}}{2}\bigg[\VIfvar{k}{i}+\breve{\mu}_{f_k}^2(\xgrd_{i}) - 2\VIfMeanMean{k}\VIfmean{k}{i} \nonumber \\
	&\hspace{-0.2cm}+\VIfMeanVar{k}+\VIfMeanMean{k}^2 \bigg]
	+ \frac{1}{2}(\diGamma{\VIfpreca{k}}+\ln \VIfprecb{k}) +\sum_{j \in \neighbor{\xgrd_{i}}} 
	\granPara  \VIzfieldProb{k}{\xgrd_{j}} \bigg\}.
\end{align}
which leads to the update rule of $q(\Zfield_i=k)$ in~\eqref{eq:VP_hidden}.\hspace*{\fill} $\blacksquare$

\subsection{Variational distribution of the noise precision $q(\precNVar)$}
\label{sec:VINPrecPosterior}
As the conditional posterior $p(\vfield,\vZfield,\hParameter|\Vnshd)$ is proportional to  $p(\Vnshd|\vfield,\precNVar)p(\precNVar)$ wrt~$\precNVar$, we can write 
\begin{align}
&\ln q(\precNVar) = \EXPwo{\ln p(\vfield,\vZfield,\hParameter|\Vnshd)}{-q(\precNVar)} + c \nonumber \\
&\leftrightarrow \EXPwo{\ln p(\Vnshd|\vfield,\precNVar)}{-q(\precNVar)} + \EXPwo{\ln p(\precNVar)}
{-q(\precNVar)} \label{eq:logPosNPrec}
\end{align}
where 
\begin{align}
&\EXPwo{\ln p(\Vnshd|\vfield,\precNVar)}{-q_{\precNVar}} \leftrightarrow \frac{t}{2}\ln \precNVar 
- \frac{\precNVar}{2} \|\Vnshd - \mWeight_{t}^{\transpose}\vfield \|_{2}^2 \nonumber \\
&=\frac{t}{2}\ln \precNVar - \frac{\precNVar}{2} \sum_{\tau=1}^{t}\nshd_{\tau}^2-2\nshd_{\tau}\Meanshd{\tau} 
+ \EXPwo{({\vWeight_{\tau}^{(n,n')}}^{\transpose}\vfield)^2}{-q(\precNVar)} \label{eq:VIprecLikelihood}
\end{align}
and 
\begin{align}
\EXPwo{\ln p(\precNVar)}{-q(\precNVar)} \leftrightarrow (a_{\nu} - 1)\ln \precNVar - \frac{\precNVar}{b_{\nu}}.
\label{eq:VIprecPrior}
\end{align}
After substituting~\eqref{eq:VIprecLikelihood} and~\eqref{eq:VIprecPrior} into~\eqref{eq:logPosNPrec}, we can easily see that $q(\precNVar)=\G{\VIaNVar}{\VIbNVar}$ with $\VIaNVar := a_{\nu} + {t}/{2}$, and 
\begin{equation}
	\VIbNVar := \left(\frac{1}{b_{\nu}} + \frac{1}{2}\sum_{\tau=1}^{t} \nshd_{\tau}^2-2\nshd_{\tau}\Meanshd{\tau}
	+ \EXPwo{({\vWeight_{\tau}^{(n,n')}}^{\transpose}\vfield)^2}{-q(\precNVar)}\right)^{-1}
\end{equation}
where 
\begin{align}
&\EXPwo{({\vWeight_{\tau}^{(n,n')}}^{\transpose}\vfield)^2}{-q(\precNVar)} \nonumber \\
&= \textrm{Var}\left[ {\vWeight_{\tau}^{(n,n')}}^{\transpose}\vfield \right] + \left(\EXPwo{{\vWeight_{\tau}^{(n,n')}}^{\transpose}\vfield}{-q(\precNVar)}\right)^2\\
&= \sum_{i=1}^{N_g}w_{\tau,i}^2 \left[\sum_{k=1}^{K}\VIzfieldProb{k}{\xgrd_{i}}\left( {\VIfvar{k}{i}} +
\breve{\mu}_{f_k}^2(\xgrd_{i}) \right) - \Meanfield{i}^2\right] + \Meanshd{\tau}^2 \label{eq:LTV}
\end{align}
by the law of total variance on $\textrm{Var}\left[ {\vWeight_{\tau}^{(n,n')}}^{\transpose}\vfield \right]$
\cite[p.~401]{bj14lawofTotalVariance}.
\hspace*{\fill} $\blacksquare$

\subsection{Variational distribution of the field means $q(\vfMean)$}
\label{sec:VIClassMeansPosterior}
Since the conditional posterior $p(\vfield,\vZfield,\hParameter|\Vnshd)$ is proportional to  $p(\vfield|\vZfield,\hParameter_f)p(\vfMean)$ wrt~$\vfMean$, the entries of $\vfMean$ are iid, 
we have   
\begin{align}
	&\ln q(\vfMean) = \EXPwo{\ln p(\vfield,\vZfield,\hParameter|\Vnshd)}{-q(\vfMean)} + c \nonumber \\
	&\leftrightarrow \EXPwo{\ln p(\vfield|\vZfield,\hParameter_f)}{-q(\vfMean)} + \EXPwo{\sum_{k=1}^{K} 
		\ln p(\fMean{k})}{-q(\vfMean)} \label{eq:logPosfMean}
\end{align}
where
\begin{align}
	&\EXPwo{\ln p(\vfield|\vZfield,\hParameter_f)}{-q(\vfMean)} \nonumber \\
	&\hspace{1.4cm} \leftrightarrow \sum_{k=1}^{K}\sum_{i=1}^{N_g}\VIzfieldProb{k}{\xgrd_{i}}
	\Meanfprec{k}\bigg(\fMean{k}^2 - 2 \VIfmean{k}{i}\fMean{k}\bigg) \label{eq:VImeanLikelihood}
\end{align}
and
\begin{align}
	\EXPwo{ \sum_{k=1}^{K}\ln p(\fMean{k})}{-q(\vfMean)} \leftrightarrow \sum_{k=1}^{K}
	\frac{1}{\sigma_{k}^2}(\fMean{k}^2 - 2 \fMean{k}m_{k}).\label{eq:VImeanPrior}
\end{align}
Together with~\eqref{eq:VImeanLikelihood} and~\eqref{eq:VImeanPrior}, $\ln q(\vfMean)$ becomes 
\begin{align}
\ln q(\vfMean) &\leftrightarrow \sum_{k=1}^{K}\bigg[\bigg(\frac{1}{\sigma_{k}^2} + \sum_{i=1}^{N_g}\VIzfieldProb{k}{\xgrd_{i}}\Meanfprec{k}\bigg)\fMean{k}^2 \nonumber \\
& - 2\bigg(\frac{m_k}{\sigma_{k}^2} + \sum_{i=1}^{N_g}\VIzfieldProb{k}{\xgrd_{i}}\Meanfprec{k}\VIfmean{k}{i}\bigg)\fMean{k}\bigg].\label{eq:VIlogfMeanPost}
\end{align}
After completing the square of the summand in~\eqref{eq:VIlogfMeanPost}, we find
$q(\fMean{k})=\NormDist{\VIfMeanMean{k}}{\VIfMeanVar{k}}$ $\forall k$
with
\begin{align}
	&\VIfMeanVar{k} := \bigg(\frac{1}{\sigma_{k}^2} + \sum_{i=1}^{N_g}\VIzfieldProb{k}{\xgrd_{i}}
	\Meanfprec{k}\bigg)^{-1}\\
	&\VIfMeanMean{k} := \VIfMeanVar{k}\bigg(\frac{m_k}{\sigma_{k}^2} + \sum_{i=1}^{N_g}\VIzfieldProb{k}{\xgrd_{i}}\Meanfprec{k}\VIfmean{k}{i}\bigg)
\end{align}
by inspection since $q(\vfMean) = \prod_{k=1}^{K}q(\fMean{k})$, as in~\eqref{eq:factPosterior}.
\hspace*{\fill} $\blacksquare$
 
\subsection{Variational distribution of the field precisions $q(\vfPrec)$}
\label{sec:VIClassPrecsPosterior}
Similar to $q(\vfMean)$, the pdf $q(\vfPrec)$ can be expressed as
\begin{align}
	&\ln q(\vfPrec) = \EXPwo{\ln p(\vfield,\vZfield,\hParameter|\Vnshd)}{-q(\vfPrec)} + c \nonumber \\
	&\leftrightarrow \EXPwo{\ln p(\vfield|\vZfield,\hParameter_f)}{-q(\vfPrec)} + \EXPwo{\sum_{k=1}^{K}
		\ln p(\fprec{k})}{-q(\vfPrec)} \label{eq:logPosfPrec} 
\end{align}
by appealing to the proportionality of the conditional posterior
$p(\vfield,\vZfield,\hParameter|\Vnshd) \propto p(\vfield|\vZfield,\hParameter_f)p(\vfPrec)$ w.r.t.~$\vfPrec$.
Each term on the RHS in~\eqref{eq:logPosfPrec} can be thus expressed as
\begin{align}
	&\EXPwo{\ln p(\vfield|\vZfield,\hParameter_f)}{-q(\vfPrec)} \nonumber \\
	&= \frac{1}{2}\sum_{k=1}^{K}\sum_{i=1}^{N_g} {\VIzfieldProb{k}{\xgrd_{i}}}\bigg[
	\ln\fprec{k} - \fprec{k}\EXPwo{(f_i - \fMean{k})^2}{-q(\Zfield_i)}\bigg] + c \label{eq:VIfVarLikelihood}
\end{align}
where 
\begin{align}
	\EXPwo{(f_i - \fMean{k})^2}{-q(\Zfield_i)} &= \VIfvar{k}{i}+\breve{\mu}_{f_k}^2(\xgrd_{i}) 
	- 2\VIfMeanMean{k}\VIfmean{k}{i} \nonumber \\
	&+\VIfMeanVar{k}+\VIfMeanMean{k}^2,
\end{align}
and  
\begin{align}
	\EXPwo{\sum_{k=1}^{K}\ln p(\fprec{k})}{-q(\vfPrec)} 
	= \sum_{k=1}^{K}\bigg[(a_k - 1)\ln\fprec{k} - \frac{\fprec{k}}{b_k}\bigg] + c \label{eq:VIfVarPrior}. 
\end{align}
After substituting~\eqref{eq:VIfVarLikelihood} and~\eqref{eq:VIfVarPrior} into~\eqref{eq:logPosfPrec}, 
$\vfPrec$ can be shown to follow $q(\fprec{k})=\G{\VIfpreca{k}}{\VIfprecb{k}}$ $\forall k$ with
\begin{align}
	\VIfpreca{k} &:= a_k + \frac{1}{2}\sum_{i=1}^{N_g} \VIzfieldProb{k}{\xgrd_{i}}\\
	\VIfprecb{k} &:= \bigg[\frac{1}{b_k} + \frac{1}{2}\sum_{i=1}^{N_g} \VIzfieldProb{k}{\xgrd_{i}}  \\
	&\times \bigg(\VIfvar{k}{i} + \breve{\mu}_{f_k}^2(\xgrd_{i}) -2\VIfmean{k}{i}\VIfMeanMean{k} 
	+ \VIfMeanVar{k}+\VIfMeanMean{k}^2 \bigg) \bigg]^{-1} \nonumber
\end{align}
where we used that $q(\vfPrec) = \prod_{k=1}^{K}q(\fprec{k})$, as in~\eqref{eq:factPosterior}.
\hspace*{\fill} $\blacksquare$

\subsection{Derivation of the cross-entropy $H(\vfield|\vZfield,\Vnshdat{\tau+1};\hat{\hParameter}_{\tau})$}
\label{sec:approxEntropyDerivation}
To establish the expression for $H(\vfield|\vZfield,\Vnshdat{\tau+1};\hat{\hParameter}_{\tau})$ 
in~\eqref{eq:ApproxCondEntropyAtNextTimeSlot}, consider that at time slot $\tau+1$. Similar to
\eqref{eq:condEntropyGivenParametersVI}, we have
\begin{align}
& H(\vfield|\vZfield=\vZfield',\Vnshdat{\tau+1}=\Vnshdat{\tau+1}';\hat{\hParameter}_{\tau}) \nonumber \\
& \hspace{1cm}\approx \frac{1}{2}\ln \left| \mCondCovFVIwith{z',\Vnshdat{\tau+1}';\hat{\hParameter}_{\tau}}  
\right| + \frac{N_g}{2}\bigg(1 + \ln 2\pi\bigg). \label{eq:condEntropyGivenParametersVINextTimslot}
\end{align}
With $\mDiagWeight{\tau+1} :=\textrm{diag}\left(\vWeight_{\tau+1}^{(n,n')} \circ \vWeight_{\tau+1}^{(n,n')}\right)$, and using the construction of $\VIfvar{k}{i}$ in~\eqref{eq:VIfVar}, we can write 
\begin{align}
	\mCondCovFVIwith{z',\Vnshdat{\tau+1}';\hat{\hParameter}_{\tau}} 
	= \left[\mCondCovFVIwith{z',\Vnshdat{\tau}';\hat{\hParameter}_{\tau}}^{-1} + 
	\MeanprecNVar\mDiagWeight{\tau+1}\right]^{-1}
\end{align}
from which we deduce that 
\begin{align}
\bigg|\mCondCovFVIwith{z',\Vnshdat{\tau+1}';\hat{\hParameter}_{\tau}}\bigg|^{-1}
= \bigg|\bbI_{N_g}+\MeanprecNVar\mDiagWeight{\tau+1}\mCondCovFVIwith{z',\Vnshdat{\tau}';\hat{\hParameter}_{\tau}}\bigg| \bigg|\mCondCovFVIwith{z',\Vnshdat{\tau}';\hat{\hParameter}_{\tau}}^{-1}\bigg| \label{eq:approxCondEntropy_v2}
\end{align}
by using the matrix determinant identity lemma~\cite[Chapter 18]{harville97MatrixDeterminantLemma}.
Further substituting~\eqref{eq:approxCondEntropy_v2} into~\eqref{eq:condEntropyGivenParametersVINextTimslot}, leads to  
\begin{align}
H(\vfield|\vZfield=\vZfield',\Vnshdat{\tau+1}=\Vnshdat{\tau+1}';\hat{\hParameter}_{\tau}) &\approx  H(\vfield|\vZfield=\vZfield',\Vnshdat{\tau}=\Vnshdat{\tau}';\hat{\hParameter}_{\tau}) \nonumber \\
& \hspace{-2cm} - \frac{1}{2}\ln \bigg|\bbI_{N_g}+\MeanprecNVar\mDiagWeight{\tau+1}\mCondCovFVIwith{z',\Vnshdat{\tau}';\hat{\hParameter}_{\tau}}\bigg|. \label{eq:condEntropyGivenParametersVINextTimslot_v2}
\end{align}
It follows from the conditional entropy definition in~\eqref{eq:conditionalEntropy} that 
\begin{align}
&H(\vfield|\vZfield,\Vnshdat{\tau+1};\hat{\hParameter}_{\tau})  \nonumber \\
&\approx \sum_{\vZfield'\in\zSet}\int  p(\vZfield',\Vnshdat{\tau+1}';\hat{\hParameter}_{\tau}) \bigg( H(\vfield|\vZfield=\vZfield',\Vnshdat{\tau}=\Vnshdat{\tau}';\hat{\hParameter}_{\tau}) \nonumber \\
&\hspace{2.4cm}-\frac{1}{2}\ln\bigg|\bbI_{N_g}+\MeanprecNVar
\mDiagWeight{\tau+1}\mCondCovFVIwith{z',\Vnshdat{\tau}';\hat{\hParameter}_{\tau}}\bigg|\bigg)
d\Vnshdat{\tau+1}' \nonumber \\
&\stackrel{(e1)}{=} H(\vfield|\vZfield,\Vnshdat{\tau};\hat{\hParameter}_{\tau})    \label{eq:condEntropyVI} \\
&- \sum_{\vZfield'\in\zSet}\int  p(\vZfield',\Vnshdat{\tau}';\hat{\hParameter}_{\tau})
\frac{1}{2}\ln\bigg|\bbI_{N_g}+\MeanprecNVar
\mDiagWeight{\tau+1}\mCondCovFVIwith{z',\Vnshdat{\tau}';\hat{\hParameter}_{\tau}}\bigg|d\Vnshdat{\tau}', \nonumber
\end{align}
where (e1) is obtained after marginalizing out $\nshd_{\tau+1}$ from $p(\vZfield',\Vnshdat{\tau+1}';\hat{\hParameter}_{\tau})$ as the RHS 
of~\eqref{eq:condEntropyGivenParametersVINextTimslot_v2} is not a function of $\nshd_{\tau+1}$.
\hspace*{\fill} $\blacksquare$

\bibliographystyle{IEEEtranS}
\bibliography{myBibliography_DH,my_bibliography}

\end{document}

%% file: mysymbol.tex
\newcounter{lemma} 

\newcounter{corollary} 

\newcounter{proposition} 

\newcounter{remarkjs} 

\def\ccalA{{\ensuremath{\mathcal A}}}

\def\ccalN{{\ensuremath{\mathcal N}}}

\def\ccalR{{\ensuremath{\mathcal R}}}

\def\ccal0{{\ensuremath{\mathcal 0}}}
\def\bbC{{\ensuremath{\mathbf C}}}

\def\bbI{{\ensuremath{\mathbf I}}}

\def\bbW{{\ensuremath{\mathbf W}}}

\def\bbw{{\ensuremath{\mathbf w}}}

\def\bbs{{\ensuremath{\mathbf s}}}

\def\bbx{{\ensuremath{\mathbf x}}}

\def\bb0{{\ensuremath{\mathbf 0}}}
\def\bbDelta{{\mbox{\boldmath $\Delta$}}}

\def\bbPhi{{\mbox{\boldmath $\Phi$}}}